\documentclass[acmtog,screen]{acmart}

\acmSubmissionID{1246}

\renewcommand{\eqref}[1]{Eq.~(\ref{#1})}
\newcommand{\figref}[1]{Fig.~\ref{#1}}
\newcommand{\secref}[1]{Sec.~\ref{#1}}
\newcommand{\tabref}[1]{Tab.~\ref{#1}}
\newcommand{\algref}[1]{Alg.~\ref{#1}}
\newcommand{\appref}[1]{App.~\ref{#1}}
\newcommand{\cosangle}[2]{\langle #1,#2\rangle}

\usepackage[table]{xcolor}
\usepackage{booktabs}
\usepackage{graphicx}
\usepackage{adjustbox}
\usepackage{enumitem} 
\usepackage{multicol} 
\usepackage{multirow} 
\usepackage{subcaption} 

\usepackage{caption}
\usepackage{lipsum}

\usepackage[ruled,linesnumbered]{algorithm2e} 

\SetAlFnt{\footnotesize}
\SetAlCapFnt{\small}
\SetAlCapNameFnt{\small}
\SetAlCapHSkip{0pt}
\DontPrintSemicolon
\SetAlgoNlRelativeSize{-1}
\SetNlSty{textcolor{black!55}}{}{}

\SetCommentSty{algcommentstyle}

\setcopyright{none}
\acmJournal{TOG}

\begin{document}
\title{Occlusion-Point Reuse for Ray-Traced Ambient Occlusion and Shadow}


\author{Haojie Jin}
\orcid{0009-0007-7899-3603}
\affiliation{%
  \institution{Peking University}
  \city{Beijing}
  \country{China}}
\email{2201111642@pku.edu.cn}

\author{Fujia Su}
\affiliation{%
  \institution{Central Media Technology Institution, Huawei}
  \city{Shenzhen}
  \country{China}}
\email{sufujia@pku.edu.cn}

\author{Zehui Lin}
\affiliation{%
  \institution{Central Media Technology Institution, Huawei}
  \city{Shenzhen}
  \country{China}}
\email{zehui@pku.edu.cn}

\author{Chenxiao Hu}
\orcid{0009-0000-3306-6653}
\affiliation{%
  \institution{Peking University}
  \city{Beijing}
  \country{China}}
\email{hineven@pku.edu.cn}

\author{Jierui Ren}
\orcid{0009-0000-3556-4402}
\affiliation{%
  \institution{Peking University}
  \city{Beijing}
  \country{China}}
\email{2201112464@stu.pku.edu.cn}

\author{Yuqing Yuan}
\affiliation{%
  \institution{Peking University}
  \city{Beijing}
  \country{China}}
\email{yuanyq@stu.pku.edu.cn}

\author{Yanchen Zhang}
\orcid{0009-0007-2333-6294}
\affiliation{%
  \institution{Peking University}
  \city{Beijing}
  \country{China}}
\email{xy_zml@foxmail.com}

\author{Zhongtao Wang}
\orcid{0009-0006-1396-7474}
\affiliation{%
  \institution{Peking University}
  \city{Beijing}
  \country{China}}
\email{wangzhongtao@stu.pku.edu.cn}

\author{Yisong Chen}
\orcid{0000-0002-3406-7751}
\affiliation{%
  \institution{Peking University}
  \city{Beijing}
  \country{China}}
\email{chenyisong@pku.edu.cn}

\author{Kangying Cai}
\affiliation{%
  \institution{Central Media Technology Institution, Huawei}
  \city{Beijing}
  \country{China}}
\email{caikangying@huawei.com}

\author{Guoping Wang}
\orcid{0000-0001-7819-0076}
\affiliation{%
  \institution{Peking University}
  \city{Beijing}
  \country{China}}
\email{wgp@pku.edu.cn}

\author{Sheng Li}
\orcid{0000-0002-8901-2184}
\authornote{Corresponding author.}
\affiliation{%
  \institution{Peking University}
  \city{Beijing}
  \country{China}}
\email{lisheng@pku.edu.cn}


\begin{abstract}
    Ambient occlusion (AO) and soft shadows are critical visibility cues for spatial perception in real-time rendering. Hardware ray tracing provides a direct way to evaluate these effects, enabling ray-traced AO and area-light shadows that avoid many limitations of screen-space AO and shadow mapping. However, real-time budgets allow only a few rays per pixel, leaving raw ray-traced estimates noisy and expensive. We present an occlusion-point reuse framework that reuses traced samples in the domain of first-hit occlusion points instead of directly reusing final shading values or light samples. This provides a ray-reuse formulation for AO, rather than merely filtering or reusing completed AO values. The key idea is to transform AO and area-light shadow estimators into occluder-domain integrals, then combine neighboring occluder samples with a multiple-importance-sampling (MIS) formulation. For both AO and shadows, we derive unbiased estimators that validate convergence to the transformed integrals, as well as biased estimators designed for practical real-time execution. The biased variants assume local first-hit occluder consistency; for shadows, this occluder-based assumption better matches local visibility geometry than the visibility-consistency assumption commonly used when reusing light samples. Experiments show higher AO and shadow quality than non-reuse ray-traced baselines, and better shadow quality than light-sample reuse at comparable cost.
\end{abstract}

%
%


%
%



\begin{teaserfigure}
	\centering
	\includegraphics[width=1\textwidth]{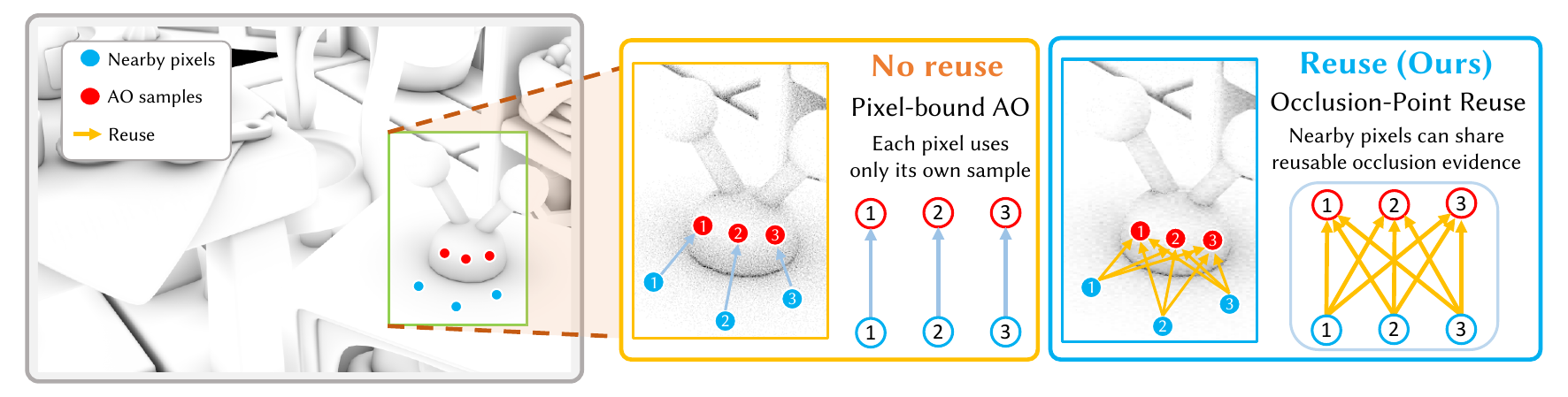}
	\caption{
		\textbf{Principle illustration of occlusion-point reuse for ray-traced ambient occlusion (RTAO).}
		Left: AO reference with a highlighted local region.
		Middle (No reuse): each pixel uses only its own sample (one-to-one correspondence), leading to higher Monte Carlo noise.
		Right (Ours): neighboring pixels share reusable occlusion evidence (many-to-many correspondence), effectively increasing usable samples and substantially reducing noise.
	}
	\label{fig:teaser}
	\Description{Teaser}
\end{teaserfigure}

\maketitle

\section{Introduction}
\label{sec:introduction}

Ambient occlusion (AO) and shadows are fundamental visibility cues in real-time rendering. Hardware ray tracing provides a direct way to evaluate them, enabling ray-traced AO and area-light shadows that avoid many limitations of screen-space AO~\cite{SSAO08,HBAO08} and shadow maps~\cite{SM78,PCF87,PCSS05,VSM06,MSM15}. In practice, however, real-time budgets allow only a few rays per pixel, leaving the resulting visibility estimates noisy and making sample reuse an attractive way to improve quality.

Existing reuse strategies are not a natural fit for this problem. Modern frameworks such as ReSTIR~\cite{ReSTIR20,GRIS22} are built around reusing light samples or light transport paths. For AO, there is no shared emitter domain from which neighboring pixels can directly reuse samples. For area-light shadows, such a domain does exist, but practical biased reuse often assumes that neighboring receivers have similar visibility to a reused light sample. This assumption can fail near penumbrae, thin occluders, and depth discontinuities, where binary visibility changes rapidly across pixels.

We address these limitations with \emph{occlusion-point reuse}, a framework built around first-hit occlusion points. Instead of treating a traced sample as only a final visibility value or a light sample, we reinterpret a blocked ray as a reusable geometric event: its first-hit occlusion point. We reformulate both AO and area-light shadow visibility as integrals over first-hit occluders, so neighboring receivers can reweight shared occlusion points according to their own geometry and support. This provides a reusable domain for AO and, for shadows, shifts the practical approximation from light-sample visibility consistency to local occluder consistency.

Our formulation yields both unbiased and practical biased estimators. The unbiased variants re-evaluate whether each reused occlusion point is also the first hit for the target receiver, establishing convergence to the transformed occluder-domain integral. The biased variants remove these cross-pixel first-hit validation rays by assuming local first-hit consistency, making reuse practical at real-time cost.

For AO, this gives, to our knowledge, the first practical formulation that reuses traced AO rays through first-hit occlusion points instead of filtering completed AO estimates. For shadows, the same framework uses a more geometry-consistent biased assumption than directly reusing the visibility of sampled light points. Together, these observations suggest that first-hit occlusion points form a more suitable reuse object for low-spp visibility effects than emitter samples alone.
In summary, our contributions are:
\begin{itemize}[leftmargin=*]
	\item We introduce occlusion-point reuse, a unified first-hit occluder reuse framework for ray-traced AO and area-light shadows.
	\item We reformulate both effects as occluder-domain integrals and derive corresponding reuse estimators, including unbiased reference variants and practical biased variants.
	\item We present the first practical AO ray-reuse formulation based on first-hit occlusion points, and show that its shadow variant adopts a more geometry-consistent biased assumption than conventional light-sample reuse, improving quality at modest overhead.
\end{itemize}

\section{Related Work}
\label{sec:related-work}

\subsection{Ambient Occlusion}
Ambient occlusion (AO) darkens regions where nearby geometry blocks much of the surrounding environment.
Early obscurance models introduced AO as a practical approximation of local visibility~\cite{AO98}. Ray-based AO can produce high-quality results by casting many short rays~\cite{LaineK10}, but the cost is high for real-time rendering. Most real-time methods therefore approximate AO in image space. SSAO~\cite{SSAO08}, HBAO~\cite{HBAO08}, Scalable AO~\cite{SAO12}, and GTAO~\cite{GTAO16} estimate occlusion from depth and normal buffers, while stochastic-depth methods add additional depth information to reduce missing-geometry artifacts~\cite{SDAO21,RTSDAO24}. Neural methods learn AO or related screen-space shading effects from geometric buffers~\cite{NNAO16,DeepShading17}. These methods are efficient, but they are limited to screen-space information and learned priors.

Ray-traced ambient occlusion (RTAO) directly queries scene visibility, avoiding many failure cases of screen-space methods, but limited ray budgets make raw estimates noisy. In contrast to filtering final AO values, our method reformulates the AO estimator as a first-hit occluder-domain integral and reuses traced occlusion points, providing a different reuse strategy at controllable cost.


\subsection{Shadow}

Shadow mapping~\cite{SM78} efficiently produces real-time hard shadows via light-space depth comparison, but extending it to area-light soft shadows often suffers from resolution- and bias-related artifacts such as aliasing and peter-panning.
Many shadow-mapping extensions improve filtering or approximate soft shadows, including PCF~\cite{PCF87}, PCSS~\cite{PCSS05}, and moment/distribution-based methods such as VSM~\cite{VSM06}, VSSM~\cite{VSSM10}, ESSM~\cite{ESSM13}, and MSM~\cite{MSM15}, though these approaches remain biased approximations dependent on depth-distribution and filtering assumptions.

Hardware ray tracing enables ray-traced shadows (RTShadow) by directly tracing shadow rays for visibility evaluation, naturally supporting area lights but suffering from noise under limited ray budgets. Unlike prior reuse methods based on light samples or filtered shadow-map values, our method reformulates the estimator in the first-hit occluder domain and reuses occlusion points across neighboring shading points.


\subsection{Reuse Techniques}
Sample reuse and spatiotemporal reconstruction are central to real-time ray tracing, where only a few samples can be traced per pixel. A common strategy is to reuse or reconstruct final radiance in image space. Temporal accumulation and TAA-style methods~\cite{TAAU20}, as well as spatiotemporal denoisers such as SVGF~\cite{SVGF17}, improve the stability of low-spp rendering by accumulating history and filtering noisy estimates. These methods are effective and general, but they operate after sampling and therefore depend on history validation and reconstruction heuristics near disocclusions, shadow boundaries, and geometric edges.

Another line of work reuses samples before reconstruction. ReSTIR DI~\cite{ReSTIR20}, ReSTIR GI~\cite{ReSTIRGI21}, and GRIS~\cite{GRIS22} reuse light/path samples through reservoir-based resampling, significantly improving sampling efficiency. However, these methods operate in the light/path-sample domain, where practical biased reuse often assumes visibility consistency across neighboring pixels, which can fail near penumbrae, thin occluders, and depth discontinuities. Related shadow-map approaches such as ReSTIR SM~\cite{RESTIRSM25} also use ReSTIR to select important lights, but still rely on shadow-map-based visibility approximations.

Our method instead reuses first-hit occlusion points produced by AO or shadow rays. By expressing AO and shadow estimation in the occluder domain and combining reused samples with MIS-style weighting~\cite{MIS95}, our approach provides a different reuse strategy from light/path-sample reuse, particularly for AO where no natural emitter-sampling domain exists.


\section{Preliminary}
\label{sec:preliminary}

\subsection{Notation}

We denote a 3D shading position by $\mathrm{x}$ and its normal by $\mathbf{n}_{\mathrm{x}}$. Directions are denoted by $\omega$, and $\Omega(\mathrm{x})$ is the normal hemisphere centered around $\mathbf{n}_{\mathrm{x}}$. For area-light shadows, $\mathcal{A}$ denotes the area-emitter sampling domain, $\mathrm{y}\in\mathcal{A}$ denotes an emitter-surface sample, and $\mathrm{d}\mu(\cdot)$ denotes surface measure. We write $\cosangle{\mathbf{a}}{\mathbf{b}}$ for the clamped cosine between two unit direction or normal vectors. For example, $\cosangle{\mathbf{n}_{\mathrm{x}}}{\omega}$ uses the normal at $\mathrm{x}$ and direction $\omega$. We use $V$ for binary visibility, with $V=1$ indicating an unoccluded direction or segment.

\subsection{Ray-Traced Ambient Occlusion}

For a shading point $\mathrm{x}$, radius-limited ray-traced AO can be written as a cosine-weighted visibility integral over the normal hemisphere:
\begin{equation}
	\mathrm{AO}(\mathrm{x})
	=
	\frac{1}{\pi}
	\int_{\Omega(\mathrm{x})}
	V(\mathrm{x},\omega,R)
	\,\cosangle{\mathbf{n}_{\mathrm{x}}}{\omega}
	\,\mathrm{d}\omega,
	\label{eq:ao-hemisphere}
\end{equation}
where $V(\mathrm{x},\omega,R)$ tests whether the ray from $\mathrm{x}$ along $\omega$ remains unoccluded up to radius $R$.
Standard RTAO samples directions using the cosine-weighted hemisphere density
\begin{equation}
	p_{\Omega}(\omega\mid\mathrm{x})=\frac{\cosangle{\mathbf{n}_{\mathrm{x}}}{\omega}}{\pi}.
	\label{eq:ao-direction-pdf}
\end{equation}
With samples $\{\omega_j\}_{j=1}^{M}$ drawn from \eqref{eq:ao-direction-pdf}, the corresponding Monte Carlo estimator is
\begin{equation}
	\left\langle \mathrm{AO}(\mathrm{x}) \right\rangle_M
	=
	\frac{1}{M}
	\sum_{j=1}^{M}
	V(\mathrm{x},\omega_j,R).
	\label{eq:ao-mc-estimator}
\end{equation}
This baseline operates entirely in the local directional domain, and its binary visibility samples have high variance at real-time ray budgets.

\subsection{Ray-Traced Area-Light Shadows}

For area-light shadows, we use the normalized visibility over the area-emitter domain
\begin{equation}
	\mathrm{Shadow}(\mathrm{x})
	=
	\frac{1}{\mathrm{Area}(\mathcal{A})}
	\int_{\mathcal{A}}
	V(\mathrm{x},\mathrm{y})
	\,\mathrm{d}\mu(\mathrm{y}),
	\label{eq:shadow-area}
\end{equation}
where $V(\mathrm{x},\mathrm{y})=1$ if the segment from $\mathrm{x}$ to $\mathrm{y}$ is unoccluded. This term can be obtained from the direct-lighting integral by separating the average visibility over the emitter domain; we provide the derivation in \appref{app:shadow-visibility}. We use uniform area sampling over $\mathcal{A}$,
\begin{equation}
	p_{\mathcal{A}}(\mathrm{y})=\frac{1}{\mathrm{Area}(\mathcal{A})},
	\label{eq:shadow-light-pdf}
\end{equation}
and trace a shadow ray toward each sampled point. With samples $\{\mathrm{y}_j\}_{j=1}^{M}$ drawn from \eqref{eq:shadow-light-pdf}, the corresponding Monte Carlo estimator is
\begin{equation}
	\left\langle \mathrm{Shadow}(\mathrm{x}) \right\rangle_M
	=
	\frac{1}{M}
	\sum_{j=1}^{M}
	V(\mathrm{x},\mathrm{y}_j).
	\label{eq:shadow-mc-estimator}
\end{equation}
Area-light shadows face the same low-sample visibility problem, but their samples are points on the area-emitter domain rather than directions on the receiver hemisphere.

\subsection{Sample Reuse Problem}



RTAO samples receiver-local hemisphere directions governed by \eqref{eq:ao-direction-pdf}, but these samples become visibility queries only after being attached to a specific shading point and thus do not form a naturally reusable cross-pixel domain. Reusing only final AO values may reduce noise, but cannot provide unbiased ray reuse and often breaks near geometric discontinuities. In contrast, area-light shadows possess an explicit shared emitter-sampling domain sampled by \eqref{eq:shadow-light-pdf}, enabling unbiased light-sample reuse through re-evaluation at neighboring receivers. Practical biased variants, however, often assume shared visibility for reused light samples, effectively reducing reuse to the final binary visibility result in \eqref{eq:shadow-mc-estimator}, which becomes fragile near penumbrae.

These observations suggest that existing reuse strategies are fundamentally tied either to final visibility values or to emitter/path samples, both of which become unstable when visibility changes rapidly across neighboring receivers. A more suitable reuse representation should instead capture the underlying geometric occlusion event itself, rather than only its final visibility outcome. This motivates considering first-hit occlusion points as a reusable domain shared by both AO and shadow rays, where local occluder consistency may provide a more stable approximation than final-value or light-sample visibility consistency.


\section{Occlusion-Point Reuse}
\label{sec:methods}

\subsection{Overview}

Our key idea is to realize ray reuse through first-hit occlusion points rather than final visibility values. Each shading point $\mathrm{x}_j$ still samples rays in the original domain: cosine-weighted hemisphere rays for AO or shadow rays toward sampled light points. When a ray hits geometry within the valid support, its first intersection $\mathrm{z}_j$ becomes a reusable occluder candidate for nearby receivers. Instead of directly sampling scene surfaces---which rarely yields valid first-hit occluders---we reuse the naturally generated first-hit points from the original ray samplers and express the estimator in the occluder domain with induced sampling densities.

The exact formulation remains unbiased by validating first-hit visibility for every reused candidate and receiver pair, but this is expensive. Our practical estimator instead relies on a key observation: neighboring pixels often share locally consistent first-hit occluders, allowing cross-pixel reuse without additional validation rays.

Unlike reusing final AO or shadow values, the reused object is a geometric occlusion event with its own position, normal, and sampling density. Each receiver can therefore reweight the same occluder candidate according to its local geometry and support, enabling true ray reuse rather than value filtering.

\subsection{First-Hit Occluder Domain}

Let $\mathcal{S}$ denote the scene surface. For a shading point $\mathrm{x}$ and a candidate surface point $\mathrm{z}\in\mathcal{S}$, we define the first-hit predicate
\begin{equation}
	\mathrm{Occ}(\mathrm{x},\mathrm{z})=1
	\iff
	\mathrm{z}\ \text{is the first intersection along}\ \overrightarrow{\mathrm{x}\mathrm{z}}.
	\label{eq:occ-first-hit}
\end{equation}
This predicate encodes whether a candidate is the first hit for the receiver being evaluated. In the exact induced density $p(\mathrm{z}\mid\mathrm{x})$, it appears together with effect-specific sampling support, such as the AO radius or whether a shadow candidate corresponds to a ray that intersects the emitter domain.

\begin{figure}[tb]
	\centering
	\includegraphics[width=0.9\linewidth]{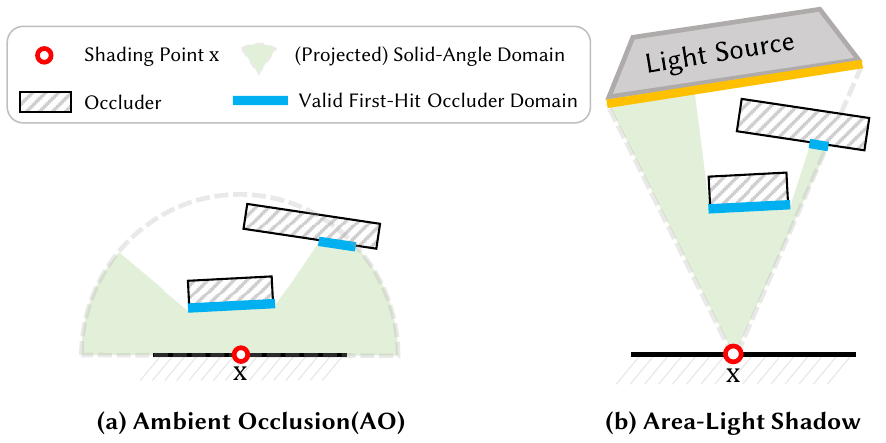}
	\caption{\textbf{First-Hit Occluder Domain.}
		AO and area-light shadow have different original sampling domains, but both can be interpreted in a unified first-hit occluder view.
		AO is sampled directly in solid-angle space, whereas shadows are sampled on the emitter surface, which induces a corresponding solid-angle domain at the receiver.
	}
	\label{fig:first-hit-occluder-domain}
	\Description{First-hit occluder domain.}
\end{figure}

The occluder-domain reformulation lets AO and shadow share the same reuse structure. For a target point $\mathrm{x}_i$, let
\begin{equation}
	\mathcal{N}_i=\{\mathrm{x}_1,\ldots,\mathrm{x}_N\}
	\label{eq:reuse-group}
\end{equation}
be its local shared group. Each source point $\mathrm{x}_j$ contributes one candidate occluder $\mathrm{z}_j$ or an invalid miss. Consider an occluder-domain integral
\begin{equation}
	I(\mathrm{x}_i)
	=
	\int_{\mathcal{S}}
	\mathrm{Occ}(\mathrm{x}_i,\mathrm{z})
	F(\mathrm{x}_i,\mathrm{z})
	\,\mathrm{d}\mu(\mathrm{z}),
	\label{eq:generic-occluder-integral}
\end{equation}
where $F$ contains the effect-specific contribution after factoring out first-hit visibility. The density $p(\mathrm{z}\mid\mathrm{x})$ is induced by the original ray sampler and includes both the first-hit predicate and the support tests of that sampler. With one sample from each induced source density $p(\mathrm{z}\mid\mathrm{x}_j)$, a generic MIS estimator is
\begin{equation}
	\left\langle I(\mathrm{x}_i) \right\rangle
	=
	\sum_{j=1}^{N}
	w_j(\mathrm{z}_j)
	\frac{\mathrm{Occ}(\mathrm{x}_i,\mathrm{z}_j)F(\mathrm{x}_i,\mathrm{z}_j)}
	{p(\mathrm{z}_j\mid\mathrm{x}_j)}.
	\label{eq:generic-mis-estimator}
\end{equation}
We use the balance heuristic, whose weight is
\begin{equation}
	w_j(\mathrm{z})
	=
	\frac{p(\mathrm{z}\mid\mathrm{x}_j)}
	{\sum_{k=1}^{N}p(\mathrm{z}\mid\mathrm{x}_k)}.
	\label{eq:generic-mis-weight}
\end{equation}
Substituting \eqref{eq:generic-mis-weight} into \eqref{eq:generic-mis-estimator} gives the estimator used in our occluder-domain reuse:
\begin{equation}
	\left\langle I(\mathrm{x}_i) \right\rangle
	=
	\sum_{j=1}^{N}
	\frac{\mathrm{Occ}(\mathrm{x}_i,\mathrm{z}_j)F(\mathrm{x}_i,\mathrm{z}_j)}
	{\sum_{k=1}^{N}p(\mathrm{z}_j\mid\mathrm{x}_k)}.
	\label{eq:generic-mis-estimator-simplified}
\end{equation}
Here $p(\mathrm{z}\mid\mathrm{x})$ is not a directly sampled occluder distribution. It is induced by the original sampler after mapping a traced ray to its first-hit point. Evaluating it for a reused point therefore requires checking whether that point would also be the first hit from the evaluation receiver.
AO and shadow have different density expressions, derived in \appref{app:ao-occluder-derivation} and \appref{app:shadow-occluder-derivation}.
The same occluder-domain view can also be extended to full direct illumination; we provide the derivation in \appref{app:di-occluder-derivation}. However, this paper focuses its implementation and evaluation on AO and normalized area-light shadow visibility.
\figref{fig:first-hit-occluder-domain} provides an overview of the First-Hit Occluder Domain interpretation for both effects.

\subsection{Reuse for Ambient Occlusion}

Changing variables from the directional domain to the first-hit surface domain introduces the solid-angle to area-measure Jacobian
\begin{equation}
	J_{\Omega\!\to\!\mathcal{S}}(\mathrm{x},\mathrm{z})
	=
	\frac{\cosangle{\mathbf{n}_{\mathrm{z}}}{\omega_{\mathrm{z}\to\mathrm{x}}}}
	{\|\mathrm{x}-\mathrm{z}\|^2}.
	\label{eq:ao-jacobian}
\end{equation}
Here and below, $\mathbb{I}_{(\cdot)}$ denotes an indicator function that returns one when its condition is true and zero otherwise.
Using the first-hit predicate, the complement of AO can be written as an occluder-domain integral:
\begin{equation}
	\mathrm{AO}(\mathrm{x})
	=
	1-
	\frac{1}{\pi}
	\int_{\mathcal{S}}
	\mathrm{Occ}(\mathrm{x},\mathrm{z})
	\cosangle{\mathbf{n}_{\mathrm{x}}}{\omega_{\mathrm{x}\to\mathrm{z}}}
	J_{\Omega\!\to\!\mathcal{S}}(\mathrm{x},\mathrm{z})
	\mathbb{I}_{\|\mathrm{x}-\mathrm{z}\|\le R}
	\,\mathrm{d}\mu(\mathrm{z}).
	\label{eq:ao-occluder-domain}
\end{equation}
This expression maps cosine-weighted hemisphere samples to surface points while keeping first-hit visibility as the explicit predicate $\mathrm{Occ}$. The same change of variables also gives the induced finite-support AO density:
\begin{equation}
	p_{\mathrm{AO}}(\mathrm{z}\mid\mathrm{x})
	=
	\frac{1}{\pi}
	\cosangle{\mathbf{n}_{\mathrm{x}}}{\omega_{\mathrm{x}\to\mathrm{z}}}
	J_{\Omega\!\to\!\mathcal{S}}(\mathrm{x},\mathrm{z})
	\mathbb{I}_{\|\mathrm{x}-\mathrm{z}\|\le R}
	\mathrm{Occ}(\mathrm{x},\mathrm{z}),
	\label{eq:ao-pdf}
\end{equation}
We derive this density in \appref{app:ao-occluder-derivation}.

For AO, the numerator in \eqref{eq:generic-mis-estimator-simplified} first has the geometric contribution form
\begin{equation}
	\begin{aligned}
		     & \mathrm{Occ}(\mathrm{x}_i,\mathrm{z}_j)
		F_{\mathrm{AO}}(\mathrm{x}_i,\mathrm{z}_j)      \\
		= {} &
		\mathrm{Occ}(\mathrm{x}_i,\mathrm{z}_j)\,
		\frac{1}{\pi}
		\cosangle{\mathbf{n}_{\mathrm{x}_i}}{\omega_{\mathrm{x}_i\to\mathrm{z}_j}}
		J_{\Omega\!\to\!\mathcal{S}}(\mathrm{x}_i,\mathrm{z}_j)
		\mathbb{I}_{\|\mathrm{x}_i-\mathrm{z}_j\|\le R} \\
		= {} &
		p_{\mathrm{AO}}(\mathrm{z}_j\mid\mathrm{x}_i).
	\end{aligned}
	\label{eq:ao-generic-numerator}
\end{equation}
Substituting this relation into \eqref{eq:generic-mis-estimator-simplified} gives the unbiased AO estimator:
\begin{equation}
	\left\langle \mathrm{AO}(\mathrm{x}_i) \right\rangle_{\mathrm{unbiased}}
	=
	1-
	\sum_{j=1}^{N}
	\frac{p_{\mathrm{AO}}(\mathrm{z}_j\mid\mathrm{x}_i)}
	{\sum_{k=1}^{N}p_{\mathrm{AO}}(\mathrm{z}_j\mid\mathrm{x}_k)}.
	\label{eq:ao-unbiased-estimator}
\end{equation}
Here each exact density evaluation includes the AO radius, valid hemisphere support, and the first-hit predicate. This is why evaluating $p_{\mathrm{AO}}(\mathrm{z}_j\mid\mathrm{x}_k)$ for a reused candidate requires a first-hit validation from $\mathrm{x}_k$.
We provide the full pseudocode in \appref{app:ao-reuse-algorithm}.

\eqref{eq:ao-unbiased-estimator} establishes the convergence target, but it is too expensive for real-time use. Its overhead comes from two sources:
\begin{enumerate}[leftmargin=*]
	\item The primary AO ray must return the closest hit position and normal, rather than only a binary hit/miss result.
	\item Each exact density evaluation requires a first-hit test: evaluating $p_{\mathrm{AO}}(\mathrm{z}_j\mid\mathrm{x}_k)$ must verify whether $\mathrm{z}_j$ is first hit from $\mathrm{x}_k$, adding many validation rays within one shared group.
\end{enumerate}
The second cost dominates in practice, because it scales with the number of reused candidate--receiver pairs and requires additional shadow-ray-style visibility tests.

For the practical biased estimator, we exploit local first-hit occluder consistency. Since $\mathrm{z}_j$ is generated as the first-hit occluder of its source point $\mathrm{x}_j$, $\mathrm{Occ}(\mathrm{x}_j,\mathrm{z}_j)=1$. For neighboring points, we approximate $\mathrm{Occ}(\mathrm{x}_k,\mathrm{z}_j)\approx\mathrm{Occ}(\mathrm{x}_j,\mathrm{z}_j)$. Substituting this approximation into density evaluations removes cross-pixel first-hit validation and leaves only finite-radius and hemisphere support terms in $\widetilde{p}_{\mathrm{AO}}$. The resulting biased estimator is
\begin{equation}
	\left\langle \mathrm{AO}(\mathrm{x}_i) \right\rangle_{\mathrm{biased}}
	=
	1-
	\sum_{j=1}^{N}
	\frac{\widetilde{p}_{\mathrm{AO}}(\mathrm{z}_j\mid\mathrm{x}_i)}
	{\sum_{k=1}^{N}\widetilde{p}_{\mathrm{AO}}(\mathrm{z}_j\mid\mathrm{x}_k)}.
	\label{eq:ao-biased-estimator}
\end{equation}
Each reused occluder is still rejected if it fails the target hemisphere or radius support tests, since then $\widetilde{p}_{\mathrm{AO}}(\mathrm{z}_j\mid\mathrm{x}_i)=0$. Missed source rays provide no reusable occluder. By removing cross-pixel first-hit validation, the biased estimator avoids extra shadow-ray-style tests after the source closest-hit queries; the remaining work is density evaluation and MIS normalization.

\begin{figure}[tb]
	\centering
	\includegraphics[width=0.85\linewidth]{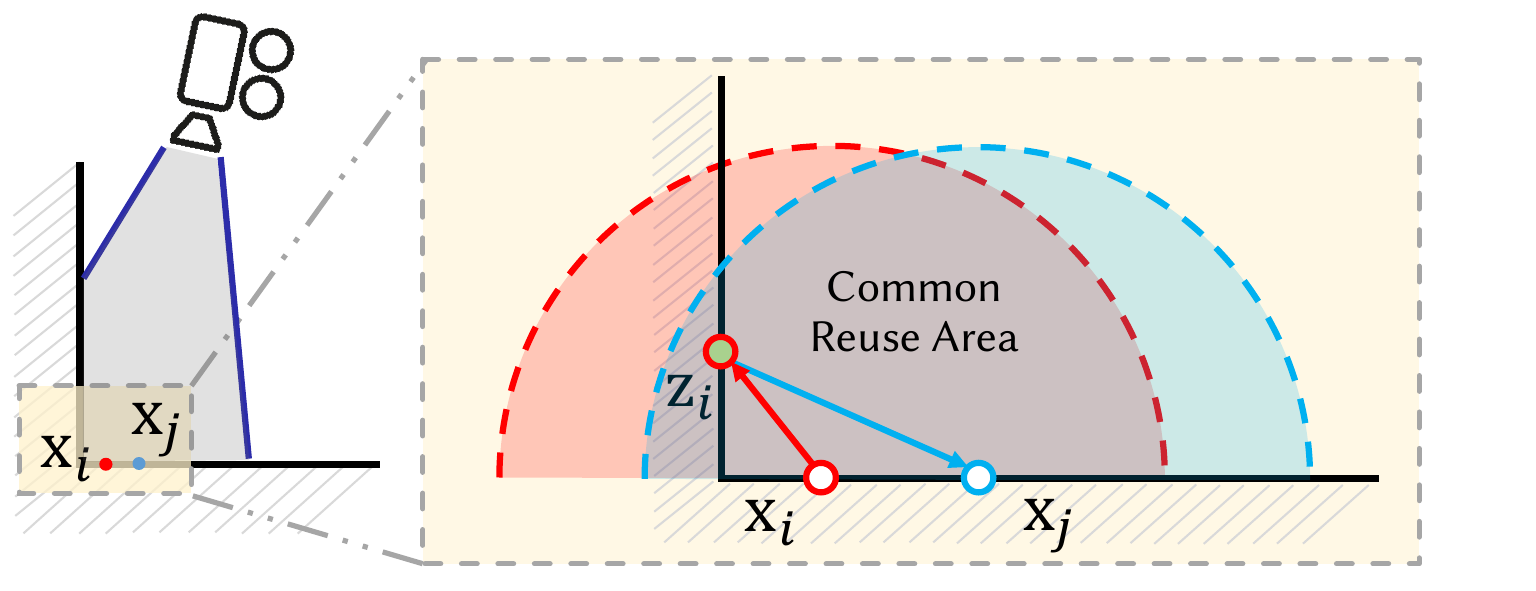}
	\caption{\textbf{RTAO Ray Reuse Example.}
		The right panel enlarges the local configuration shown on the left. A source shading point $\mathrm{x}_i$ samples a direction in its normal hemisphere and traces an AO ray. The first-hit occluder $\mathrm{z}_i$ is then reused by neighboring shading points $\mathrm{x}_j$, where it is reweighted according to the induced occluder-domain densities. Any miss ray or occlusion point outside the ``Common Reuse Area'' makes zero contribution.
	}
	\label{fig:rtao-reuse}
	\Description{RTAO ray reuse example.}
\end{figure}

\subsection{Reuse for Area-Light Shadows}
\label{sec:application-shadow}

For area-light shadows, \eqref{eq:shadow-area} integrates visibility over the area-emitter domain $\mathcal{A}$. For a blocked emitter point $\mathrm{y}$, let $\mathrm{z}$ be the first occluder along the segment from $\mathrm{x}$ to $\mathrm{y}$. We use $\chi_{\mathcal{A}}(\mathrm{x},\mathrm{z})$ to indicate whether the ray from $\mathrm{x}$ through $\mathrm{z}$ intersects a valid point on $\mathcal{A}$. Mapping blocked light samples to their first-hit occluders gives
\begin{equation}
	\mathrm{Shadow}(\mathrm{x})
	=
	1-
	\int_{\mathcal{S}}
	\mathrm{Occ}(\mathrm{x},\mathrm{z})
	\frac{1}{\mathrm{Area}(\mathcal{A})}
	J(\mathrm{x},\mathrm{z})
	\chi_{\mathcal{A}}(\mathrm{x},\mathrm{z})
	\,\mathrm{d}\mu(\mathrm{z}).
	\label{eq:shadow-occluder-domain}
\end{equation}
The Jacobian from the light area to the occluder area is
\begin{equation}
	J(\mathrm{x},\mathrm{z})
	=
	\frac{
	\cosangle{\mathbf{n}_{\mathrm{z}}}{\omega_{\mathrm{z}\to\mathrm{x}}}
	\|\mathrm{x}-\mathrm{y}\|^2
	}{
	\cosangle{\mathbf{n}_{\mathrm{y}}}{\omega_{\mathrm{y}\to\mathrm{x}}}
	\|\mathrm{x}-\mathrm{z}\|^2
	}.
	\label{eq:shadow-jacobian}
\end{equation}
Here $\mathrm{y}$ is the emitter point paired with $\mathrm{z}$ in the domain conversion. We provide the detailed derivation of \eqref{eq:shadow-occluder-domain} and \eqref{eq:shadow-jacobian} in \appref{app:shadow-occluder-derivation}.

For density evaluation during reuse, the original emitter point may no longer be the one seen by the current receiver. We therefore introduce $\mathrm{y}'(\mathrm{x},\mathrm{z})$, obtained by intersecting the ray from $\mathrm{x}$ through $\mathrm{z}$ with $\mathcal{A}$, and evaluate the same Jacobian using this light-intersection point. In density evaluations below, $J(\mathrm{x},\mathrm{z})$ uses $\mathrm{y}'(\mathrm{x},\mathrm{z})$ whenever the candidate is reused at a different receiver. For a general area sampler, the induced occluder density is the light-sample density multiplied by the Jacobian, support, and first-hit terms. With the uniform area sampler used in our estimator, this becomes
\begin{equation}
	p_{\mathrm{Sh}}(\mathrm{z}\mid\mathrm{x})
	=
	\frac{1}{\mathrm{Area}(\mathcal{A})}
	J(\mathrm{x},\mathrm{z})
	\chi_{\mathcal{A}}(\mathrm{x},\mathrm{z})
	\mathrm{Occ}(\mathrm{x},\mathrm{z}).
	\label{eq:shadow-pdf}
\end{equation}
It is zero when $\mathrm{z}$ is not the first hit from $\mathrm{x}$, when the ray through $\mathrm{z}$ does not intersect the emitter domain, or when the projected-area terms are invalid. We provide the full arbitrary emitter-area sampling form in \appref{app:shadow-occluder-derivation}.

\begin{figure}[htbp]
	\centering
	\includegraphics[width=0.9\linewidth]{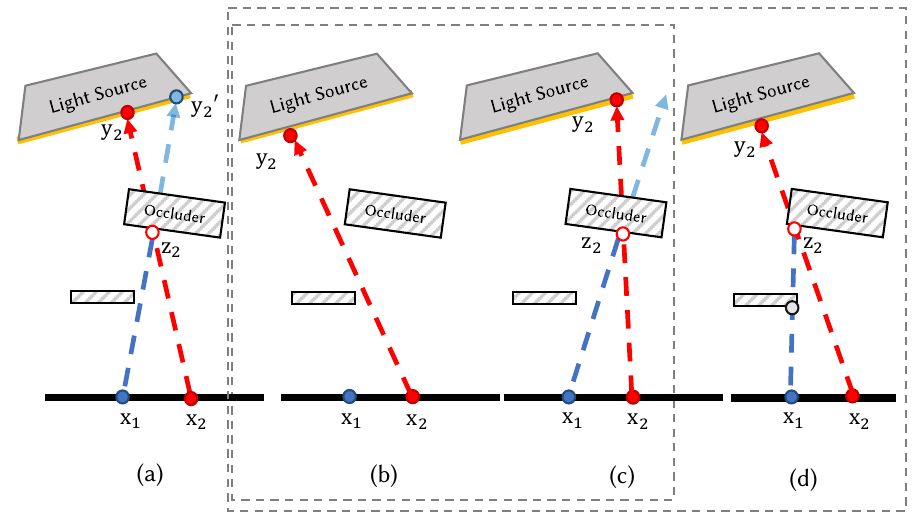}
	\caption{\textbf{RTShadow Ray Reuse Example.}
		Reusing a sample from $\mathrm{x}_2$ at $\mathrm{x}_1$ yields four outcomes.
		Under \eqref{eq:shadow-estimator}, only case (a) contributes to $K(\mathrm{x}_i)$.
		(a) valid reuse: $\mathrm{x}_2$ samples light point $\mathrm{y}_2$ and obtains occluder $\mathrm{z}_2$; for receiver $\mathrm{x}_1$, the ray through $\mathrm{z}_2$ intersects the emitter at $\mathrm{y}_2'$, and $\mathrm{z}_2$ is the first hit along $\mathrm{x}_1\!\to\!\mathrm{z}_2$, so contribution is evaluated using \eqref{eq:shadow-pdf}.
		(b) no occluder is found for the source sample, so the reused contribution is zero.
		(c) the ray from $\mathrm{x}_1$ through $\mathrm{z}_2$ does not hit the emitter, so $\chi_{\mathcal{A}}=0$.
		(d) $\mathrm{z}_2$ is not the first hit from $\mathrm{x}_1$ in that direction, so the exact contribution is zero.
		The practical biased estimator skips the first-hit validation in (d) under the local first-hit consistency approximation.
	}
	\label{fig:rtshadow-reuse}
	\Description{RTShadow ray reuse example.}
\end{figure}

For shadows, the numerator in \eqref{eq:generic-mis-estimator-simplified} has the blocked-emitter contribution form
\begin{equation}
	\begin{aligned}
		     & \mathrm{Occ}(\mathrm{x}_i,\mathrm{z}_j)
		F_{\mathrm{Sh}}(\mathrm{x}_i,\mathrm{z}_j)     \\
		= {} &
		\mathrm{Occ}(\mathrm{x}_i,\mathrm{z}_j)
		\frac{1}{\mathrm{Area}(\mathcal{A})}
		J(\mathrm{x}_i,\mathrm{z}_j)
		\chi_{\mathcal{A}}(\mathrm{x}_i,\mathrm{z}_j)  \\
		= {} &
		p_{\mathrm{Sh}}(\mathrm{z}_j\mid\mathrm{x}_i).
	\end{aligned}
	\label{eq:shadow-generic-numerator}
\end{equation}
Substituting this relation into \eqref{eq:generic-mis-estimator-simplified} gives the unbiased shadow estimator:
\begin{equation}
	\left\langle \mathrm{Shadow}(\mathrm{x}_i) \right\rangle_{\mathrm{unbiased}}
	=
	1-
	\sum_{j=1}^{N}
	\frac{p_{\mathrm{Sh}}(\mathrm{z}_j\mid\mathrm{x}_i)}
	{\sum_{k=1}^{N}p_{\mathrm{Sh}}(\mathrm{z}_j\mid\mathrm{x}_k)} = 1 - K(\mathrm{x}_i).
	\label{eq:shadow-estimator}
\end{equation}
The per-target evaluation mirrors the AO algorithm. Each thread in the shared group samples one emitter point and stores the first occluder if the segment is blocked. The group then exchanges candidate occluders, evaluates exact induced densities, and accumulates the MIS-normalized blocked area.
\figref{fig:rtshadow-reuse} visualizes typical validity cases when reusing a candidate generated at one receiver for another receiver.
We provide the full pseudocode in \appref{app:shadow-reuse-algorithm}.

As in AO, exact density evaluation is costly: each reused occluder must be validated as the first hit from every receiver that evaluates it. The practical biased estimator applies the same first-hit consistency approximation, $\mathrm{Occ}(\mathrm{x}_k,\mathrm{z}_j)\approx\mathrm{Occ}(\mathrm{x}_j,\mathrm{z}_j)$. It still keeps the emitter hit support $\chi_{\mathcal{A}}$ and the Jacobian terms, yielding
\begin{equation}
	\left\langle \mathrm{Shadow}(\mathrm{x}_i) \right\rangle_{\mathrm{biased}}
	=
	1-
	\sum_{j=1}^{N}
	\frac{\widetilde{p}_{\mathrm{Sh}}(\mathrm{z}_j\mid\mathrm{x}_i)}
	{\sum_{k=1}^{N}\widetilde{p}_{\mathrm{Sh}}(\mathrm{z}_j\mid\mathrm{x}_k)}.
	\label{eq:shadow-biased-estimator}
\end{equation}
Unlike light-sample reuse, this approximation is made on a physical occlusion point rather than on the binary visibility of a particular emitter sample. Since occluder geometry is often locally coherent, this assumption can be more stable across neighboring receivers than light-sample visibility consistency, especially near soft-shadow penumbrae.

In addition, the unbiased estimator can produce bright outliers in very dark regions precisely in case~(c) of \figref{fig:rtshadow-reuse}, where a reused occluder is rejected by first-hit validation for the current receiver.
To suppress these artifacts in the practical biased implementation, we add a conservative guard: if all light samples in the shared group are blocked, we directly set the shadow value to zero for that target.

\section{Implementation}
\label{sec:implementation}

\paragraph{System}
We build all AO and shadow variants in Falcor~\cite{Falcor} with the Vulkan backend. All experiments are conducted on an Intel Core i7-12700F CPU and an NVIDIA RTX 3080 GPU. Our reuse passes use inline ray queries rather than raygen shaders, which simplifies control over coherent thread execution during reuse.

\paragraph{Reuse Implementation}
Our current implementation uses fixed reuse patterns: neighboring pixels are partitioned into shared groups (see \appref{app:shared-group-patterns} for more details), and reuse is performed only within each group. In practice, we map each shared group in \eqref{eq:reuse-group} to threads inside a 32-thread warp and evaluate $N\in\{2,4,8\}$ group sizes. First-hit occluder data are exchanged with wave/subgroup operations~\cite{VulkanSubgroup}, so the exchange overhead remains small compared with ray traversal and intersection.

\paragraph{Blue Noise}
Blue-noise sampling makes low-sample-count Monte Carlo error less objectionable to human observers by pushing noise away from low spatial and temporal frequencies. Therefore, for all RT-based sampling in our low-spp experiments, we use precomputed spatiotemporal blue-noise masks from STBN~\cite{STBN22}. In our comparisons, blue-noise sampling is enabled consistently for all compared RT-based methods unless otherwise stated, and each comparison uses the same sampling sequence family.

\section{Results and Discussion}
\label{sec:results}

\paragraph{Experimental Setup}
All experiments are conducted on the system described in \secref{sec:implementation}. Unless otherwise specified, we render at 1080p and report the GPU runtime of the corresponding AO or shadow pass only. Reference images are generated with high-spp ray tracing using the same visibility definition as the evaluated effect. Low-spp quality comparisons use blue-noise sampling, while convergence tests use white-noise sampling to expose standard Monte Carlo behavior.
We report relative mean squared error (RelMSE, $\epsilon=10^{-3}$) against the high-spp reference as the primary quantitative metric.

We use a compact naming convention in figures and tables. The number in parentheses denotes the shared-group size used for reuse; for example, Occ~(8) denotes the biased occlusion-point reuse method with eight-pixel reuse, while Occ~(8U) denotes its unbiased counterpart with the same group size.
The method prefix indicates the baseline family: Filter denotes final-value filtering for AO, and Light denotes light-sample reuse for shadows. The detailed settings are introduced in the corresponding experiments.

\subsection{Ambient Occlusion}

\begin{figure*}[htbp]
	\centering
	\includegraphics[width=0.87\linewidth]{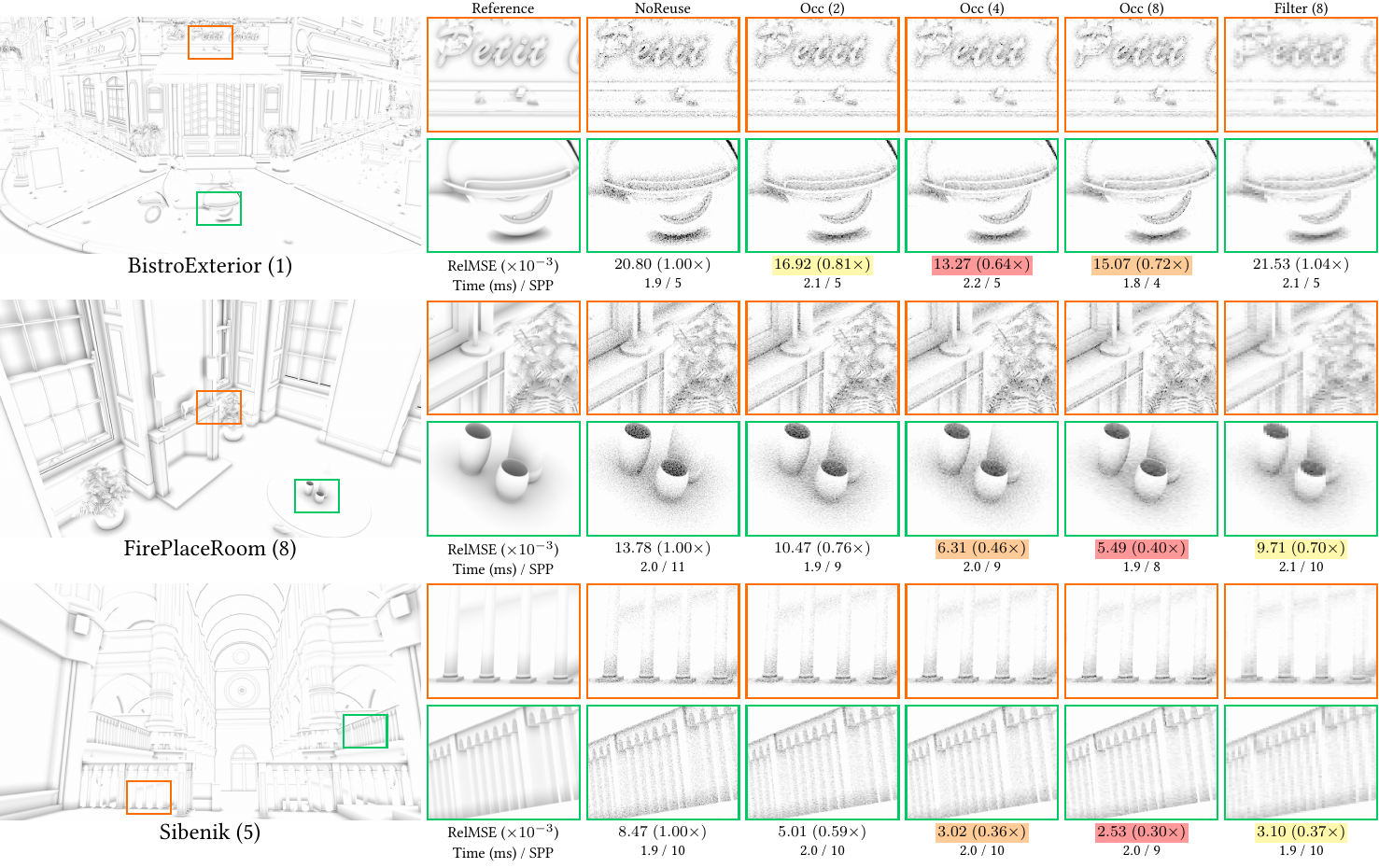}
	\caption{Equal-time AO comparison under a 2~ms budget. The panels compare no-reuse RTAO, final-value filtering, and our occlusion-point reuse variants; error maps report RelMSE against a high-spp ray-traced reference.}
	\label{fig:ao-equal-time-2ms}
	\Description{ao equal time 2 ms comparison.}
\end{figure*}

\paragraph{Equal-Time Comparison}
We first compare ray-traced AO methods under approximately equal runtime, with a target budget of 2~ms at 1080p. The baseline is standard RTAO without reuse. We compare it with final-value filtering, denoted as Filter~(8), and our biased occlusion-point reuse with shared-group sizes $N\in\{2,4,8\}$, denoted as Occ~(2), Occ~(4), and Occ~(8). \figref{fig:ao-equal-time-2ms} shows results for one representative view from each of three scenes. \tabref{tab:ao-equal-time-2ms} summarizes the quantitative result: for each scene, we evaluate several views, compute the RelMSE ratio with respect to the RTAO baseline for each view, and report the average ratio. Compared with no-reuse RTAO, our method produces smoother AO with lower noise. Compared with final-value filtering, it maintains smoothness while preserving more high-frequency details near thin geometry, contacts, and occlusion boundaries.

\begin{table}[tb]
\centering
\caption{Equal-time RTAO comparison. Each value is the average RelMSE ratio to no-reuse RTAO over multiple views of the same scene; lower is better. Configuration: 2~ms budget with blue-noise sampling. We color code the {\fboxsep1pt\colorbox[RGB]{255,153,153}{first}}, {\fboxsep1pt\colorbox[RGB]{255,204,153}{second}}, and {\fboxsep1pt\colorbox[RGB]{255,248,173}{third}} lowest numbers.}
\begin{adjustbox}{max width=\columnwidth}
\begin{tabular}{lcccc}
\toprule
Scene & Occ (2) & Occ (4) & Occ (8) & Filter (8) \\
\midrule
Airport & \cellcolor[RGB]{255,248,173}0.84x & \cellcolor[RGB]{255,204,153}0.66x & \cellcolor[RGB]{255,153,153}0.64x & 2.76x \\
Bathroom & \cellcolor[RGB]{255,248,173}0.71x & \cellcolor[RGB]{255,204,153}0.54x & \cellcolor[RGB]{255,153,153}0.46x & 1.53x \\
BistroExterior & \cellcolor[RGB]{255,248,173}0.83x & \cellcolor[RGB]{255,204,153}0.69x & \cellcolor[RGB]{255,153,153}0.63x & 1.00x \\
BistroInterior & \cellcolor[RGB]{255,248,173}0.85x & \cellcolor[RGB]{255,204,153}0.84x & \cellcolor[RGB]{255,153,153}0.78x & 1.57x \\
CartoonRoom & \cellcolor[RGB]{255,248,173}0.86x & \cellcolor[RGB]{255,204,153}0.67x & \cellcolor[RGB]{255,153,153}0.57x & 0.91x \\
FirePlaceRoom & \cellcolor[RGB]{255,248,173}0.74x & \cellcolor[RGB]{255,204,153}0.49x & \cellcolor[RGB]{255,153,153}0.45x & 1.17x \\
Room & \cellcolor[RGB]{255,248,173}0.78x & \cellcolor[RGB]{255,204,153}0.58x & \cellcolor[RGB]{255,153,153}0.52x & 0.89x \\
Sibenik & 0.59x & \cellcolor[RGB]{255,204,153}0.35x & \cellcolor[RGB]{255,153,153}0.27x & \cellcolor[RGB]{255,248,173}0.36x \\
\midrule
Average & \cellcolor[RGB]{255,248,173}0.77x & \cellcolor[RGB]{255,204,153}0.60x & \cellcolor[RGB]{255,153,153}0.54x & 1.27x \\
\bottomrule
\end{tabular}
\end{adjustbox}
\label{tab:ao-equal-time-2ms}
\end{table}

\paragraph{Efficiency}
We report the runtime overhead of different methods relative to the original no-reuse RTAO baseline in \tabref{tab:ao-efficiency}. The biased Occ~(8) variant introduces only a small extra cost ($\sim$20\%) while providing clear quality gains, making it practical for real-time AO.
A full per-scene runtime breakdown, including absolute timings, is provided in \appref{app:per-scene-efficiency}.

\paragraph{Equal-SPP Comparison}
We also evaluate a fixed-spp setting at 10 spp, where all methods use the same number of original rays per pixel.
Since our method has slightly higher per-sample cost than the compared methods (see \tabref{tab:ao-efficiency}), its relative gains over the baselines are more pronounced here than in the equal-time setting (\tabref{tab:ao-equal-time-2ms}).
Across all scenes, the average RelMSE ratios (relative to no-reuse RTAO) are 0.46x for Occ~(8) and 1.40x for Filter~(8).
Detailed quantitative results and visual comparisons are provided in \appref{app:ao-equal-spp-all}.

\begin{table}[tb]
\centering
\caption{Average runtime ratio of different methods relative to original RTAO across all scenes.}
\begin{adjustbox}{max width=\columnwidth}
\begin{tabular}{lcccc}
\toprule
Type & Occ (2) & Occ (4) & Occ (8) & Filter (8) \\
\midrule
Biased & 1.076x & 1.103x & 1.186x & 1.046x \\
Unbiased & 1.386x & 1.772x & 2.575x & -- \\
\bottomrule
\end{tabular}
\end{adjustbox}
\label{tab:ao-efficiency}
\end{table}

\begin{figure}[htbp]
	\centering
	\includegraphics[width=\linewidth]{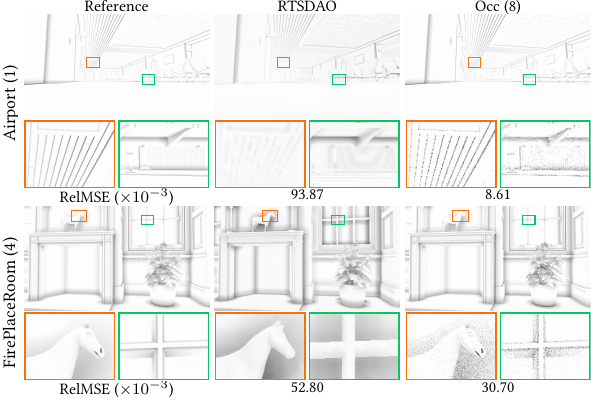}
	\caption{Screen Space AO comparison. In these two scenes, RTSDAO takes about 0.55~ms, which corresponds to Occ~(8) at 5 spp and 4 spp, respectively.}
	\label{fig:ao-screen-space}
	\Description{comparison of screen-space AO}
\end{figure}

\paragraph{Screen-Space Method Comparison}
We also compare against the state-of-the-art screen-space method RTSDAO~\cite{RTSDAO24}. We set the rendering budget to the measured runtime of RTSDAO and render our method under the same budget.
\figref{fig:ao-screen-space} shows that RTSDAO can appear visually smooth, but it misses substantial occlusion information that is not available from the current depth buffer. As a result, its image can deviate noticeably from the reference and yields a larger RelMSE error.

\begin{figure}[htbp]
	\centering
	\includegraphics[width=0.93\linewidth]{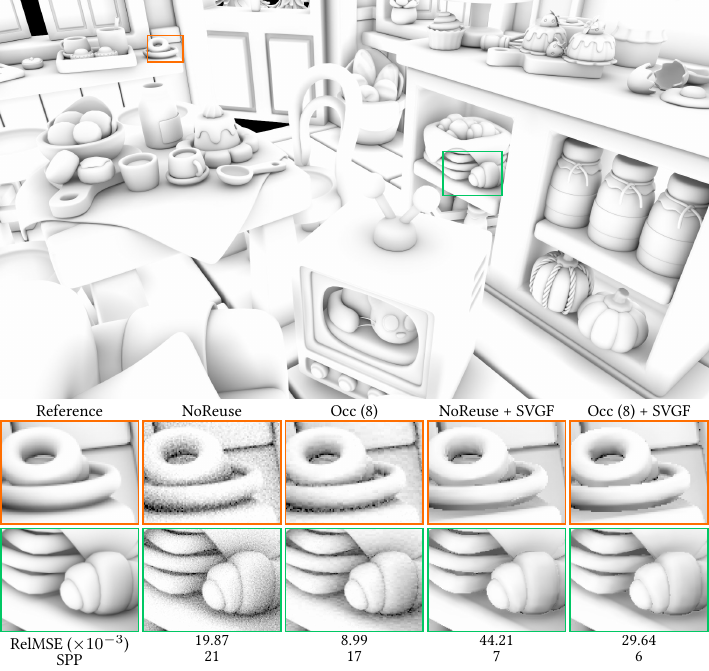
	}
	\caption{SVGF comparison for CartoonRoom (4). In this scene, SVGF costs about 4~ms, so the equal-time budget is about 6~ms in total (2~ms AO + SVGF).}
	\label{fig:ao-svgf-comparison}
	\Description{comparison of SVGF and our method for AO denoising.}
\end{figure}

\paragraph{Denoising Comparison}
We further compare with SVGF~\cite{SVGF17} with its temporal component disabled and a unified total budget of 2~ms AO + SVGF cost.
As shown in \figref{fig:ao-svgf-comparison}, SVGF effectively suppresses noise and produces visually smooth AO results. However, this smoothness may come with estimation errors, particularly near object seams, where AO tends to be over-brightened and thus leads to a large RelMSE.
Per-scene values and the overall average across all scenes are provided in \appref{app:ao-all-scenes-svgf-comparison}.

\paragraph{Convergence}
We verify that our unbiased AO reuse estimator is indeed unbiased. The convergence curves are provided in \appref{app:ao-convergence}.


\subsection{Area-Light Shadows}

For all shadow results, RelMSE is computed directly from the raw estimator output without clamping, even when the practical biased estimator produces rare negative values in very dark regions. These dark-region outlier pixels are sparse (below $1\%$ in our diagnostic example); we provide the corresponding visualization in \appref{app:shadow-negative-estimates}.

\begin{figure*}[htbp]
	\centering
	\includegraphics[width=0.87\linewidth]{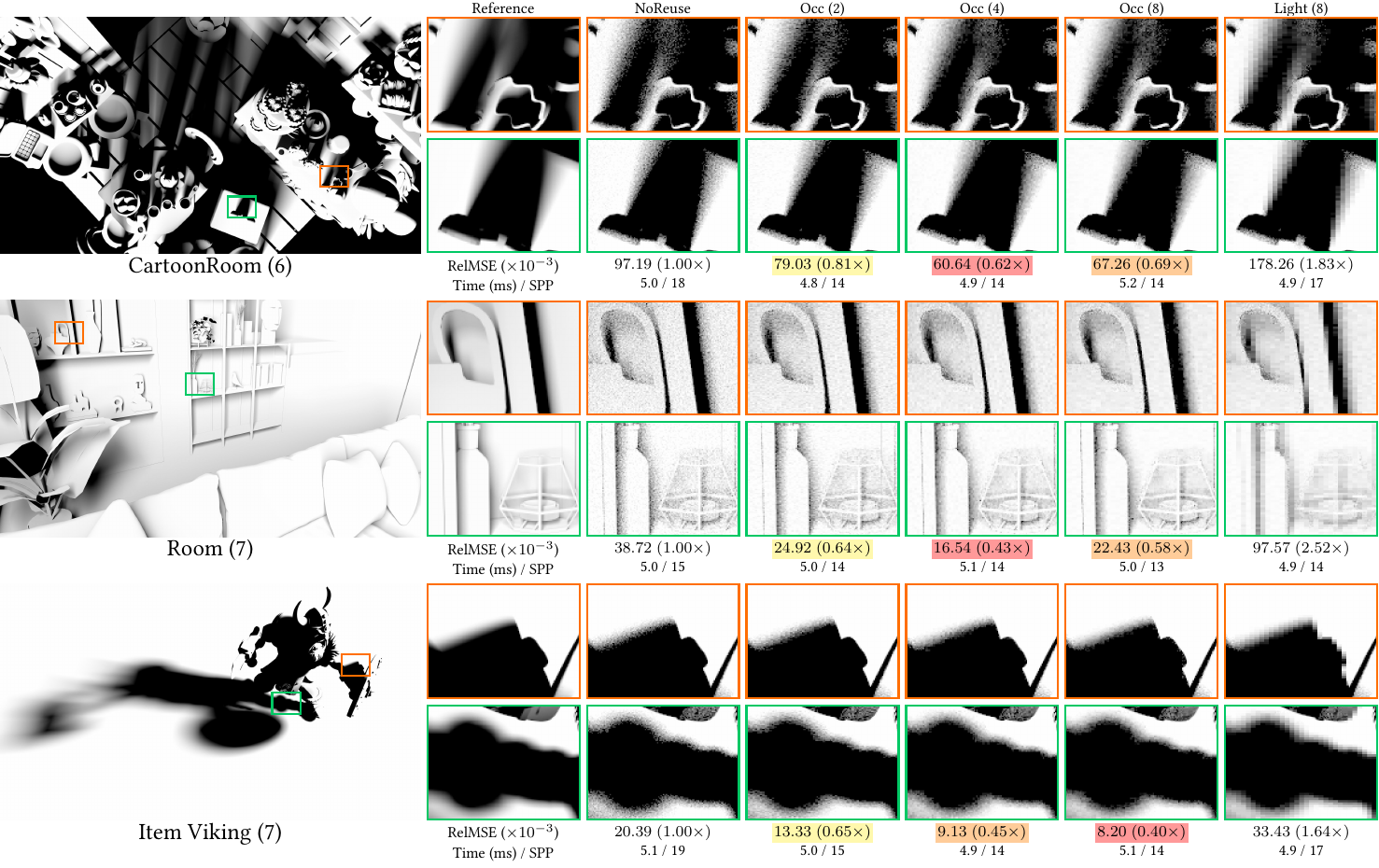}
	\caption{Equal-time area-light shadow comparison under a 5~ms budget. Compared with no reuse and light-sample reuse, occlusion-point reuse produces smoother shadows while preserving boundaries more robustly in penumbrae.}
	\label{fig:shadow-main-equal-5-ms-relmse}
	\Description{shadow equal time 5 ms comparison.}
\end{figure*}

\paragraph{Equal-Time Comparison}
Shadow estimation is more expensive than AO, so we use a unified 5~ms budget at 1080p for all methods. Following the AO naming convention, we compare no-reuse ray-traced shadows, Light~(8), and biased occlusion-point reuse with shared-group sizes $N\in\{2,4,8\}$. Light~(8) reuses light samples without additional visibility validation. Under our shadow formulation, this effectively degenerates into an averaging filter. \figref{fig:shadow-main-equal-5-ms-relmse} shows representative visual results and error maps, and \tabref{tab:shadow-equal-time-5ms} reports average RelMSE ratios over multiple views per asset. Compared with no reuse, Occ produces smoother shadows with lower noise. Compared with Light~(8), Occ better preserves object boundaries because first-hit occluder consistency is generally a more accurate approximation than visibility consistency. In complex scenes, Occ~(8) is not always better than Occ~(4); we discuss this behavior in the shared-group-size discussion below. Additional equal-time comparisons at 2~ms and 10~ms are provided in \appref{app:shadow-equal-time-2ms-and-10ms}.

\begin{table}[tb]
\centering
\caption{Equal-time area-light shadow comparison. Each value is the average RelMSE ratio to no-reuse ray-traced shadows over multiple views of the same asset; lower is better. Configuration: 5~ms budget with blue-noise sampling. We color code the {\fboxsep1pt\colorbox[RGB]{255,153,153}{first}}, {\fboxsep1pt\colorbox[RGB]{255,204,153}{second}}, and {\fboxsep1pt\colorbox[RGB]{255,248,173}{third}} lowest numbers.}
\begin{adjustbox}{max width=\columnwidth}
\begin{tabular}{lcccc}
\toprule
Scene & Occ (2) & Occ (4) & Occ (8) & Light (8) \\
\midrule
Bike & 0.67x & \cellcolor[RGB]{255,204,153}0.43x & \cellcolor[RGB]{255,153,153}0.30x & \cellcolor[RGB]{255,248,173}0.51x \\
BistroInterior & \cellcolor[RGB]{255,204,153}0.75x & \cellcolor[RGB]{255,153,153}0.71x & \cellcolor[RGB]{255,248,173}1.10x & 5.81x \\
CartoonRoom & \cellcolor[RGB]{255,204,153}0.80x & \cellcolor[RGB]{255,153,153}0.67x & \cellcolor[RGB]{255,248,173}0.82x & 3.01x \\
Room & \cellcolor[RGB]{255,248,173}0.80x & \cellcolor[RGB]{255,153,153}0.70x & \cellcolor[RGB]{255,204,153}0.74x & 1.57x \\
SunTemple & \cellcolor[RGB]{255,248,173}0.66x & \cellcolor[RGB]{255,204,153}0.41x & \cellcolor[RGB]{255,153,153}0.40x & 1.17x \\
Item Crab & \cellcolor[RGB]{255,248,173}0.64x & \cellcolor[RGB]{255,153,153}0.44x & \cellcolor[RGB]{255,204,153}0.48x & 3.91x \\
Item Hookah & \cellcolor[RGB]{255,248,173}0.62x & \cellcolor[RGB]{255,204,153}0.40x & \cellcolor[RGB]{255,153,153}0.35x & 1.16x \\
Item Plum & \cellcolor[RGB]{255,248,173}0.64x & \cellcolor[RGB]{255,204,153}0.43x & \cellcolor[RGB]{255,153,153}0.42x & 1.57x \\
Item Tree & \cellcolor[RGB]{255,248,173}0.65x & \cellcolor[RGB]{255,204,153}0.44x & \cellcolor[RGB]{255,153,153}0.40x & 1.54x \\
Item Viking & \cellcolor[RGB]{255,248,173}0.67x & \cellcolor[RGB]{255,204,153}0.43x & \cellcolor[RGB]{255,153,153}0.40x & 1.78x \\
\midrule
Average & \cellcolor[RGB]{255,248,173}0.69x & \cellcolor[RGB]{255,153,153}0.50x & \cellcolor[RGB]{255,204,153}0.54x & 2.20x \\
\bottomrule
\end{tabular}
\end{adjustbox}
\label{tab:shadow-equal-time-5ms}
\end{table}

\paragraph{Efficiency}
We report runtime overhead relative to no-reuse RT shadow in \tabref{tab:shadow-efficiency}. The biased Occ variants remain practical, with only modest overhead as shared-group size increases. In contrast, unbiased reuse is substantially more expensive due to exact per-candidate density evaluation. In the unbiased setting, Occ~(8U) is faster than Light~(8U): replacing per-candidate shading-point-to-light visibility checks with occluder-to-light checks shortens many traced rays and reduces total cost.
A full per-scene runtime breakdown, including absolute timings, is provided in \appref{app:shadow-per-scene-efficiency}.

\paragraph{Equal-SPP Comparison}
We also evaluate a fixed-spp setting at 10 spp.
Similar to AO, the relative gains over the baselines are more pronounced here than in the equal-time setting (\tabref{tab:shadow-equal-time-5ms}).
Across all scenes, the average RelMSE ratios (relative to no-reuse RTShadow) are 0.44x for Occ~(8) and 2.54x for Light~(8).
Detailed quantitative results and visual comparisons are provided in \appref{app:shadow-equal-spp-10}.

\begin{figure}[htbp]
	\centering
	\includegraphics[width=0.93\linewidth]{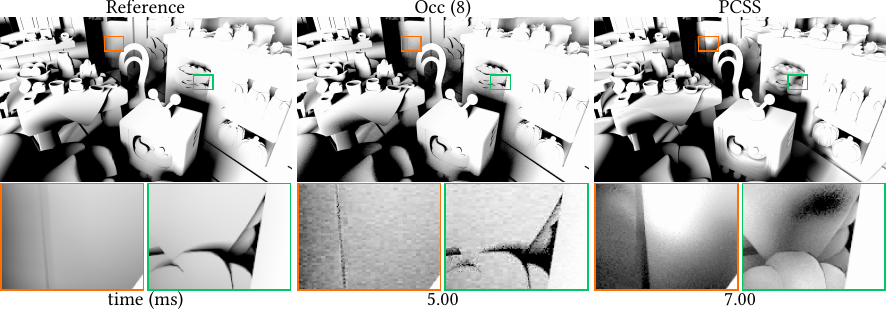}
	\caption{PCSS comparison for CartoonRoom (4). PCSS uses 64 PCF samples with a $6\times4096\times4096$ cubemap shadow map and runs at about 7~ms; our method runs at 5~ms.}
	\label{fig:shadow-pcss-comparison}
	\Description{Comparison between PCSS and our ray-traced shadow reuse method under a large area light.}
\end{figure}

\paragraph{Filtering-Based Method Comparison}
We compare against PCSS~\cite{PCSS05}, a widely used real-time approximation for hard/soft shadow transitions from area lights. For the selected view, we carefully tune PCSS parameters and report the best result we obtained in \figref{fig:shadow-pcss-comparison}. Although PCSS produces visually smooth shadows, the result remains overly hard and does not match the wide penumbra behavior of a large area light. Consequently, this approximation yields a large error relative to the ray-traced reference.

\begin{figure}[htbp]
	\centering
	\includegraphics[width=0.93\linewidth]{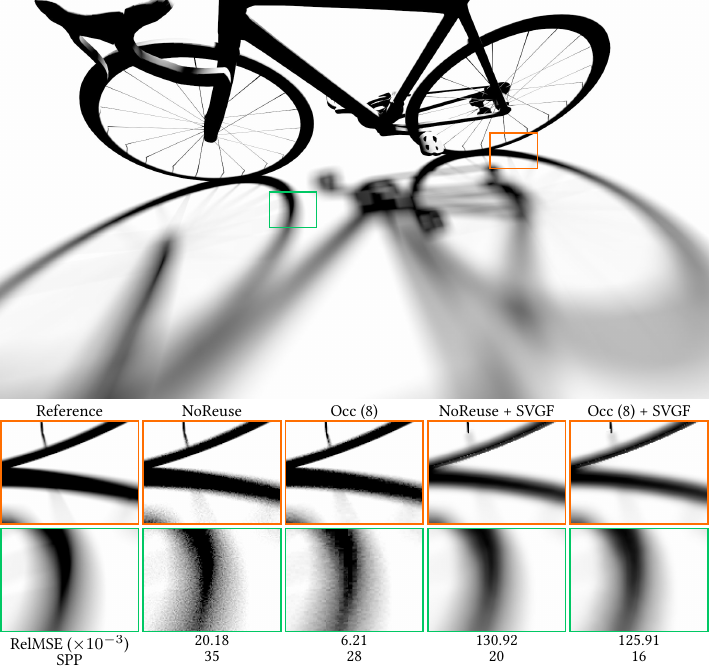}
	\caption{SVGF comparison for Bike (1). In this scene, SVGF costs about 4~ms, so the equal-time budget is about 9~ms in total (5~ms Shadow + SVGF).}
	\label{fig:shadow-svgf-comparison}
	\Description{comparison of SVGF and our method for shadow denoising.}
\end{figure}

\paragraph{Denoising Comparison}
We further compare with SVGF~\cite{SVGF17} with its temporal component disabled.
We use the same setup style as AO: total time is set to 5~ms plus the measured SVGF post-processing time.
As shown in \figref{fig:shadow-svgf-comparison}, the result is similar to AO. SVGF suppresses noise and produces smoother shadows, but it can over-smooth the estimate and deviate from the ground truth, resulting in a larger RelMSE.
Per-scene values and the overall average across all scenes are provided in \appref{app:shadow-all-scenes-svgf-comparison}.

\begin{figure}[htbp]
	\centering
	\includegraphics[width=\linewidth]{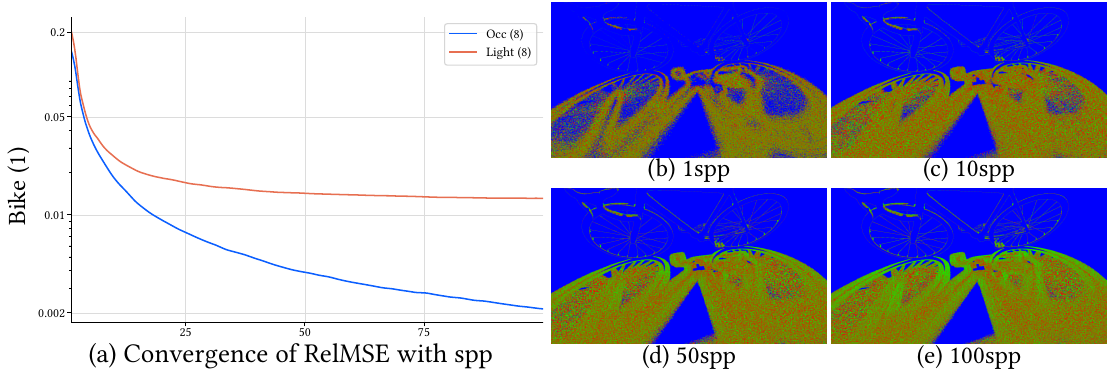}
	\caption{Biased shadow convergence comparison on Bike~(1). (a) RelMSE versus spp. (b)--(e) local winner maps at different spp, comparing Occ~(8) and Light~(8): green means lower local error for Occ~(8), red means lower local error for Light~(8), and blue means equal. As spp increases, Occ exhibits lower residual bias than Light, with clearer advantages in penumbra regions.}
	\label{fig:shadow-convergence-errors}
	\Description{shadow error-map.}
\end{figure}

\paragraph{Convergence}
We verify that our unbiased shadow reuse estimator is indeed unbiased; detailed convergence curves are provided in \appref{app:shadow-convergence}. We also compare biased estimators across increasing spp. As shown in \figref{fig:shadow-convergence-errors}, Occ is consistently more accurate than Light at higher spp, indicating lower residual bias under the first-hit occluder consistency assumption.

\begin{figure}[htbp]
	\centering
	\includegraphics[width=0.8\linewidth]{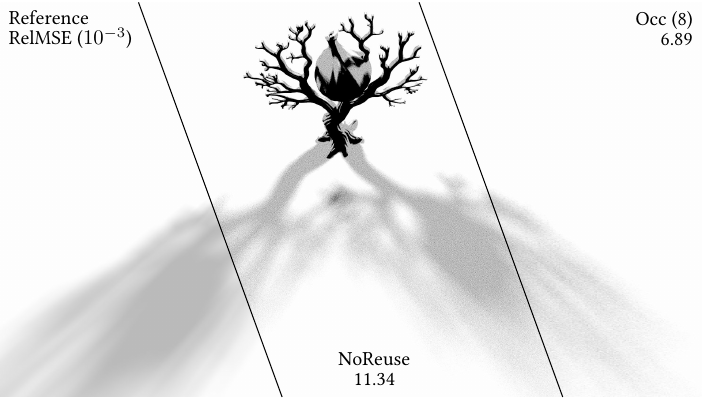}
	\caption{Two-area-light shadow comparison at 5~ms. Left: reference. Middle: no reuse. Right: Occ~(8).}
	\Description{TreeWithTwoLights.}
	\label{fig:shadow-multiple-lights}
\end{figure}

\paragraph{Multiple Lights}
Our occlusion-point reuse shadow algorithm also supports multiple area lights. \figref{fig:shadow-multiple-lights} shows a two-light example under a 5~ms rendering budget.

\begin{figure}[htbp]
	\centering
	\includegraphics[width=\linewidth]{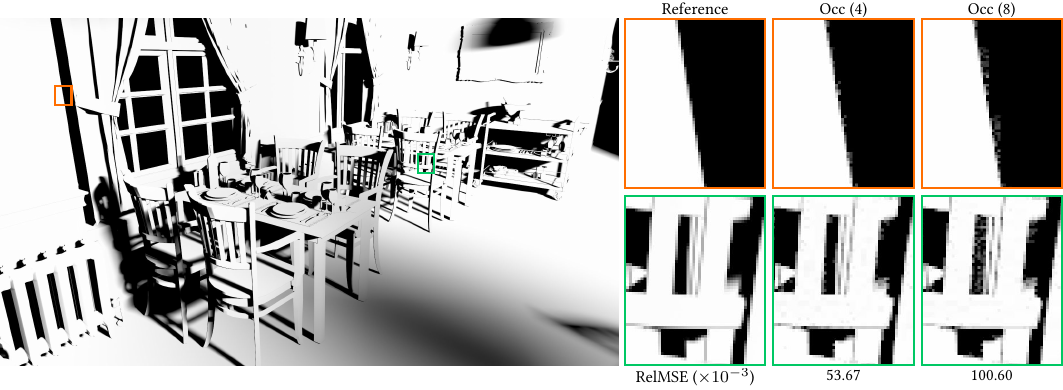}
	\caption{Failure cases of large-group biased reuse in BistroInterior~(7). Left: reference. Right-top: reuse expands into a fully dark region, causing bright outlier artifacts. Right-bottom: complex occlusion breaks local first-hit consistency, increasing bias.}
	\Description{TreeWithTwoLights.}
	\label{fig:shadow-occ-8-fail}
\end{figure}

\paragraph{Discussion of Shared-Group Size}
In complex scenes, a larger shared group is not always better for biased reuse.
We observe two main failure sources.
First, the local first-hit occluder consistency approximation can break down, especially in complex visibility configurations (see \figref{fig:rtshadow-reuse}(d)).
Second, in very dark regions, reuse may amplify bright outlier samples and produce visible bright noise spikes (see \figref{fig:rtshadow-reuse}(c)).
When a shared group spans such regions, Occ~(8) can become worse than a smaller group.
Therefore, increasing shared-group size does not monotonically improve quality.
\figref{fig:shadow-occ-8-fail} shows a representative case.


\begin{table}[tb]
\centering
\caption{Average runtime ratio of different methods relative to original
RTShadow across all scenes.}
\label{tab:shadow-efficiency}
\begin{adjustbox}{max width=\columnwidth}
\begin{tabular}{lcccc}
\toprule
Type & Occ (2) & Occ (4) & Occ (8) & Light (8) \\
\midrule
Biased & 1.188x & 1.226x & 1.304x & 1.053x \\
Unbiased & 1.494x & 2.048x & 3.124x & 7.607x \\
\bottomrule
\end{tabular}
\end{adjustbox}
\end{table}

\section{Conclusion, Limitations, and Future Work}
\label{sec:conclusion}

\paragraph{Conclusion}

We presented occlusion-point reuse, a ray-reuse framework that reuses first-hit occlusion points across neighboring receivers. By transforming visibility into an occluder-domain formulation, our method enables geometry-aware reuse for both AO and area-light shadows. This yields unbiased reference estimators as well as practical biased variants based on local first-hit consistency. Overall, occlusion-point reuse improves the quality-cost trade-off for both RTAO and ray-traced area-light shadows.

More broadly, our formulation suggests that visibility reuse can be performed through shared geometric occlusion events rather than only through sampling parameters. Because the resulting reuse mechanism is lightweight, local, and inexpensive to evaluate, we believe occlusion-point reuse is particularly well suited for high-performance real-time rendering, including resource-constrained and mobile ray-tracing platforms.

\paragraph{Limitations}
Our implementation and experiments focus on AO and normalized area-light shadow visibility. Although the appendix extends the formulation to solid-angle shadow visibility, general emitter-area sampling densities, and full direct illumination, these extensions are theoretical in this paper and are not accompanied by a complete direct-illumination implementation or systematic performance evaluation. A separate practical limitation is that the unbiased reuse estimators are too expensive for real-time use, because every reused candidate requires exact density evaluation and first-hit validation. The biased variants remove this cost with a local first-hit consistency assumption, but the reuse pattern can introduce block-like artifacts when that assumption fails.

\paragraph{Future Work}
On the implementation side, the current framework could benefit from finer adaptive control. Different warps could use different shared-group sizes according to local geometric coherence, and adaptive sampling could assign different spp budgets based on estimated variance or reuse confidence. Since our current implementation uses fixed shared-group reuse patterns, warp-specific grouping or pattern-selection strategies could further mitigate block artifacts.
A natural future direction is to extend the current implementation from AO and normalized shadow visibility to full direct illumination. Occlusion-point reuse could also be combined with light-sample reuse through MIS, since light-sample proposals may remain preferable in some regions while occluder proposals are better in others. Another promising direction is efficient occlusion-point sampling: instead of only reusing occlusion points produced by the original ray samplers, future methods could design direct proposals in the occluder domain to further improve sampling efficiency.

%
%
%
%
\clearpage
\bibliographystyle{ACM-Reference-Format}
\bibliography{ref.bib}

@online{SSAO08,
  author  = {Bavoil, Louis and Sainz, Miguel},
  title   = {Screen Space Ambient Occlusion},
  year    = {2008},
  url     = {https://developer.download.nvidia.com/SDK/10.5/direct3d/Source/ScreenSpaceAO/doc/ScreenSpaceAO.pdf},
  urldate = {2026-04-22}
}

@inproceedings{HBAO08,
  author    = {Bavoil, Louis and Sainz, Miguel and Dimitrov, Rouslan},
  title     = {Image-space horizon-based ambient occlusion},
  year      = {2008},
  isbn      = {9781605583433},
  publisher = {Association for Computing Machinery},
  address   = {New York, NY, USA},
  url       = {https://doi.org/10.1145/1401032.1401061},
  doi       = {10.1145/1401032.1401061},
  booktitle = {ACM SIGGRAPH 2008 Talks},
  articleno = {22},
  numpages  = {1},
  location  = {Los Angeles, California},
  series    = {SIGGRAPH '08}
}

@article{SDAO21,
  author     = {Vermeer, Jop and Scandolo, Leonardo and Eisemann, Elmar},
  title      = {Stochastic-Depth Ambient Occlusion},
  year       = {2021},
  issue_date = {April 2021},
  publisher  = {Association for Computing Machinery},
  address    = {New York, NY, USA},
  volume     = {4},
  number     = {1},
  url        = {https://doi.org/10.1145/3451268},
  doi        = {10.1145/3451268},
  journal    = {Proc. ACM Comput. Graph. Interact. Tech.},
  month      = apr,
  articleno  = {3},
  numpages   = {15}
}

@inproceedings{RTSDAO24,
  booktitle = {Eurographics Symposium on Rendering},
  editor    = {Haines, Eric and Garces, Elena},
  title     = {{Ray Traced Stochastic Depth Map for Ambient Occlusion}},
  author    = {Brüll, Felix and Kern, René and Grosch, Thorsten},
  year      = {2024},
  publisher = {The Eurographics Association},
  issn      = {1727-3463},
  isbn      = {978-3-03868-262-2},
  doi       = {10.2312/sr.20241160}
}

@inproceedings{SAO12,
  author    = {McGuire, Morgan and Mara, Michael and Luebke, David},
  title     = {Scalable ambient obscurance},
  year      = {2012},
  isbn      = {9783905674415},
  publisher = {Eurographics Association},
  address   = {Goslar, DEU},
  booktitle = {Proceedings of the Fourth ACM SIGGRAPH / Eurographics Conference on High-Performance Graphics},
  pages     = {97–103},
  numpages  = {7},
  location  = {Paris, France},
  series    = {EGGH-HPG'12}
}

@techreport{GTAO16,
  author    = {Hill, Stephen and McAuley, Stephen and Jover, Cyril and Lachambre, S\'{e}bastien and Pesce, Angelo and Wu, Xian-Chun and Cordes, Roger and Hery, Christophe and Hillaire, S\'{e}bastien and Hoffman, Naty and Jim\'{e}nez, Jorge and Karis, Brian and Lagarde, S\'{e}bastien and Lobl, Dan and Villemin, Ryusuke},
  title     = {Physically based shading in theory and practice},
  year      = {2016},
  isbn      = {9781450342896},
  publisher = {Association for Computing Machinery},
  address   = {New York, NY, USA},
  url       = {https://doi.org/10.1145/2897826.2927353},
  doi       = {10.1145/2897826.2927353},
  booktitle = {ACM SIGGRAPH 2016 Courses},
  articleno = {21},
  location  = {Anaheim, California},
  series    = {SIGGRAPH '16}
}

@article{LaineK10,
  author   = {Laine, Samuli and Karras, Tero},
  title    = {Two Methods for Fast Ray-Cast Ambient Occlusion},
  journal  = {Computer Graphics Forum},
  volume   = {29},
  number   = {4},
  pages    = {1325-1333},
  doi      = {https://doi.org/10.1111/j.1467-8659.2010.01728.x},
  url      = {https://onlinelibrary.wiley.com/doi/abs/10.1111/j.1467-8659.2010.01728.x},
  eprint   = {https://onlinelibrary.wiley.com/doi/pdf/10.1111/j.1467-8659.2010.01728.x},
  year     = {2010}
}

@inproceedings{AO98,
  author    = {Zhukov, S.
               and Iones, A.
               and Kronin, G.},
  editor    = {Drettakis, George
               and Max, Nelson},
  title     = {An ambient light illumination model},
  booktitle = {Rendering Techniques '98},
  year      = {1998},
  publisher = {Springer Vienna},
  address   = {Vienna},
  pages     = {45--55},
  isbn      = {978-3-7091-6453-2}
}

@inproceedings{NNAO16,
  author    = {Holden, Daniel and Saito, Jun and Komura, Taku},
  title     = {Neural network ambient occlusion},
  year      = {2016},
  isbn      = {9781450345415},
  publisher = {Association for Computing Machinery},
  address   = {New York, NY, USA},
  url       = {https://doi.org/10.1145/3005358.3005387},
  doi       = {10.1145/3005358.3005387},
  booktitle = {SIGGRAPH ASIA 2016 Technical Briefs},
  articleno = {9},
  numpages  = {4},
  location  = {Macau},
  series    = {SA '16}
}

@article{DeepShading17,
  author   = {Nalbach, O. and Arabadzhiyska, E. and Mehta, D. and Seidel, H.-P. and Ritschel, T.},
  title    = {Deep Shading: Convolutional Neural Networks for Screen Space Shading},
  journal  = {Computer Graphics Forum},
  volume   = {36},
  number   = {4},
  pages    = {65-78},
  doi      = {https://doi.org/10.1111/cgf.13225},
  url      = {https://onlinelibrary.wiley.com/doi/abs/10.1111/cgf.13225},
  eprint   = {https://onlinelibrary.wiley.com/doi/pdf/10.1111/cgf.13225},
  year     = {2017}
}

@inproceedings{SM78,
  author    = {Williams, Lance},
  title     = {Casting curved shadows on curved surfaces},
  year      = {1978},
  isbn      = {9781450379083},
  publisher = {Association for Computing Machinery},
  address   = {New York, NY, USA},
  url       = {https://doi.org/10.1145/800248.807402},
  doi       = {10.1145/800248.807402},
  booktitle = {Proceedings of the 5th Annual Conference on Computer Graphics and Interactive Techniques},
  pages     = {270–274},
  numpages  = {5},
  series    = {SIGGRAPH '78}
}

@inproceedings{PCF87,
  author    = {Reeves, William T. and Salesin, David H. and Cook, Robert L.},
  title     = {Rendering antialiased shadows with depth maps},
  year      = {1987},
  isbn      = {0897912276},
  publisher = {Association for Computing Machinery},
  address   = {New York, NY, USA},
  url       = {https://doi.org/10.1145/37401.37435},
  doi       = {10.1145/37401.37435},
  booktitle = {Proceedings of the 14th Annual Conference on Computer Graphics and Interactive Techniques},
  pages     = {283-291},
  numpages  = {9},
  series    = {SIGGRAPH '87}
}

@inproceedings{PCSS05,
  author    = {Fernando, Randima},
  title     = {Percentage-closer soft shadows},
  year      = {2005},
  isbn      = {9781450378277},
  publisher = {Association for Computing Machinery},
  address   = {New York, NY, USA},
  url       = {https://doi.org/10.1145/1187112.1187153},
  doi       = {10.1145/1187112.1187153},
  booktitle = {ACM SIGGRAPH 2005 Sketches},
  pages     = {35-es},
  location  = {Los Angeles, California},
  series    = {SIGGRAPH '05}
}

@inproceedings{VSM06,
  author    = {Donnelly, William and Lauritzen, Andrew},
  title     = {Variance shadow maps},
  year      = {2006},
  isbn      = {159593295X},
  publisher = {Association for Computing Machinery},
  address   = {New York, NY, USA},
  url       = {https://doi.org/10.1145/1111411.1111440},
  doi       = {10.1145/1111411.1111440},
  booktitle = {Proceedings of the 2006 Symposium on Interactive 3D Graphics and Games},
  pages     = {161-165},
  numpages  = {5},
  location  = {Redwood City, California},
  series    = {I3D '06}
}

@inproceedings{MSM15,
  author    = {Peters, Christoph and Klein, Reinhard},
  title     = {Moment shadow mapping},
  year      = {2015},
  isbn      = {9781450333924},
  publisher = {Association for Computing Machinery},
  address   = {New York, NY, USA},
  url       = {https://doi.org/10.1145/2699276.2699277},
  doi       = {10.1145/2699276.2699277},
  booktitle = {Proceedings of the 19th Symposium on Interactive 3D Graphics and Games},
  pages     = {7–14},
  numpages  = {8},
  location  = {San Francisco, California},
  series    = {i3D '15}
}

@inproceedings{VSSM10,
  author    = {Dong, Zhao and Yang, Baoguang},
  title     = {Variance soft shadow mapping},
  year      = {2010},
  isbn      = {9781605589398},
  publisher = {Association for Computing Machinery},
  address   = {New York, NY, USA},
  url       = {https://doi.org/10.1145/1730804.1730990},
  doi       = {10.1145/1730804.1730990},
  booktitle = {Proceedings of the 2010 ACM SIGGRAPH Symposium on Interactive 3D Graphics and Games},
  articleno = {18},
  numpages  = {1},
  location  = {Washington, D.C.},
  series    = {I3D '10}
}

@article{ESSM13,
  journal   = {Computer Graphics Forum},
  title     = {{Exponential Soft Shadow Mapping}},
  author    = {Shen, Li and Feng, Jieqing and Yang, Baoguang},
  year      = {2013},
  publisher = {The Eurographics Association and Blackwell Publishing Ltd.},
  issn      = {1467-8659},
  doi       = {10.1111/cgf.12156}
}

@article{RESTIRSM25,
  author   = {Zhang, Song and Lin, Daqi and Wyman, Chris and Yuksel, Cem},
  title    = {Many-Light Rendering Using ReSTIR-Sampled Shadow Maps},
  journal  = {Computer Graphics Forum},
  volume   = {44},
  number   = {2},
  pages    = {e70059},
  doi      = {https://doi.org/10.1111/cgf.70059},
  url      = {https://onlinelibrary.wiley.com/doi/abs/10.1111/cgf.70059},
  eprint   = {https://onlinelibrary.wiley.com/doi/pdf/10.1111/cgf.70059},
  year     = {2025}
}

@article{ReSTIR20,
  author     = {Bitterli, Benedikt and Wyman, Chris and Pharr, Matt and Shirley, Peter and Lefohn, Aaron and Jarosz, Wojciech},
  title      = {Spatiotemporal reservoir resampling for real-time ray tracing with dynamic direct lighting},
  year       = {2020},
  issue_date = {August 2020},
  publisher  = {Association for Computing Machinery},
  address    = {New York, NY, USA},
  volume     = {39},
  number     = {4},
  issn       = {0730-0301},
  url        = {https://doi.org/10.1145/3386569.3392481},
  doi        = {10.1145/3386569.3392481},
  journal    = {ACM Trans. Graph.},
  month      = aug,
  articleno  = {148},
  numpages   = {17}
}

@article{GRIS22,
  author     = {Lin, Daqi and Kettunen, Markus and Bitterli, Benedikt and Pantaleoni, Jacopo and Yuksel, Cem and Wyman, Chris},
  title      = {Generalized resampled importance sampling: foundations of ReSTIR},
  year       = {2022},
  issue_date = {July 2022},
  publisher  = {Association for Computing Machinery},
  address    = {New York, NY, USA},
  volume     = {41},
  number     = {4},
  issn       = {0730-0301},
  url        = {https://doi.org/10.1145/3528223.3530158},
  doi        = {10.1145/3528223.3530158},
  journal    = {ACM Trans. Graph.},
  month      = jul,
  articleno  = {75},
  numpages   = {23}
}

@article{ReSTIRGI21,
  author   = {Ouyang, Y. and Liu, S. and Kettunen, M. and Pharr, M. and Pantaleoni, J.},
  title    = {ReSTIR GI: Path Resampling for Real-Time Path Tracing},
  journal  = {Computer Graphics Forum},
  volume   = {40},
  number   = {8},
  pages    = {17-29},
  doi      = {https://doi.org/10.1111/cgf.14378},
  url      = {https://onlinelibrary.wiley.com/doi/abs/10.1111/cgf.14378},
  eprint   = {https://onlinelibrary.wiley.com/doi/pdf/10.1111/cgf.14378},
  year     = {2021}
}

@inproceedings{MIS95,
  author    = {Veach, Eric and Guibas, Leonidas J.},
  title     = {Optimally combining sampling techniques for Monte Carlo rendering},
  year      = {1995},
  isbn      = {0897917014},
  publisher = {Association for Computing Machinery},
  address   = {New York, NY, USA},
  url       = {https://doi.org/10.1145/218380.218498},
  doi       = {10.1145/218380.218498},
  booktitle = {Proceedings of the 22nd Annual Conference on Computer Graphics and Interactive Techniques},
  pages     = {419-428},
  numpages  = {10},
  series    = {SIGGRAPH '95}
}

@inproceedings{SVGF17,
  author    = {Schied, Christoph and Kaplanyan, Anton and Wyman, Chris and Patney, Anjul and Chaitanya, Chakravarty R. Alla and Burgess, John and Liu, Shiqiu and Dachsbacher, Carsten and Lefohn, Aaron and Salvi, Marco},
  title     = {Spatiotemporal variance-guided filtering: real-time reconstruction for path-traced global illumination},
  year      = {2017},
  isbn      = {9781450351010},
  publisher = {Association for Computing Machinery},
  address   = {New York, NY, USA},
  url       = {https://doi.org/10.1145/3105762.3105770},
  doi       = {10.1145/3105762.3105770},
  booktitle = {Proceedings of High Performance Graphics},
  articleno = {2},
  numpages  = {12},
  location  = {Los Angeles, California},
  series    = {HPG '17}
}

@article{TAAU20,
  author   = {Yang, Lei and Liu, Shiqiu and Salvi, Marco},
  title    = {A Survey of Temporal Antialiasing Techniques},
  journal  = {Computer Graphics Forum},
  volume   = {39},
  number   = {2},
  pages    = {607-621},
  doi      = {https://doi.org/10.1111/cgf.14018},
  url      = {https://onlinelibrary.wiley.com/doi/abs/10.1111/cgf.14018},
  eprint   = {https://onlinelibrary.wiley.com/doi/pdf/10.1111/cgf.14018},
  year     = {2020}
}

@misc{Falcor,
  author = {Simon Kallweit and Petrik Clarberg and Craig Kolb and Tom{'a}{\v s} Davidovi{\v c} and Kai-Hwa Yao and Theresa Foley and Yong He and Lifan Wu and Lucy Chen and Tomas Akenine-M{\"o}ller and Chris Wyman and Cyril Crassin and Nir Benty},
  title  = {The {Falcor} Rendering Framework},
  year   = {2022},
  month  = {8},
  url    = {https://github.com/NVIDIAGameWorks/Falcor},
  note   = {\url{https://github.com/NVIDIAGameWorks/Falcor}}
}

@misc{VulkanSubgroup,
  author       = {Neil Henning},
  title        = {Vulkan Subgroup Tutorial},
  howpublished = {Khronos Blog},
  year         = {2018},
  month        = {mar},
  day          = {13},
  url          = {https://www.khronos.org/blog/vulkan-subgroup-tutorial},
  note         = {Accessed: 2026-05-01}
}

@inproceedings{STBN22,
  booktitle = {Eurographics Symposium on Rendering},
  editor    = {Ghosh, Abhijeet and Wei, Li-Yi},
  title     = {{Spatiotemporal Blue Noise Masks}},
  author    = {Wolfe, Alan and Morrical, Nathan and Akenine-Möller, Tomas and Ramamoorthi, Ravi},
  year      = {2022},
  publisher = {The Eurographics Association},
  issn      = {1727-3463},
  isbn      = {978-3-03868-187-8},
  doi       = {10.2312/sr.20221161}
}

@article{LTC16,
  author     = {Heitz, Eric and Dupuy, Jonathan and Hill, Stephen and Neubelt, David},
  title      = {Real-time polygonal-light shading with linearly transformed cosines},
  year       = {2016},
  issue_date = {July 2016},
  publisher  = {Association for Computing Machinery},
  address    = {New York, NY, USA},
  volume     = {35},
  number     = {4},
  issn       = {0730-0301},
  url        = {https://doi.org/10.1145/2897824.2925895},
  doi        = {10.1145/2897824.2925895},
  journal    = {ACM Trans. Graph.},
  month      = jul,
  articleno  = {41},
  numpages   = {8}
}
\clearpage

\clearpage
\appendix
\twocolumn

\section{Derivation of Area-Light Shadow Visibility}
\label{app:shadow-visibility}

The direct illumination from area lights can be written in area form as
\begin{equation}
	\begin{aligned}
		L_{\mathrm{DI}}(\mathrm{x},\omega_o)
		 & =
		\int_{\mathcal{A}}
		L_i(\mathrm{x},\mathrm{y})
		f_r(\mathrm{x},\omega_{\mathrm{x}\to\mathrm{y}},\omega_o)
		V(\mathrm{x},\mathrm{y}) \\
		 & \quad
		\frac{
		\cosangle{\mathbf{n}_{\mathrm{x}}}{\omega_{\mathrm{x}\to\mathrm{y}}}
		\cosangle{\mathbf{n}_{\mathrm{y}}}{\omega_{\mathrm{y}\to\mathrm{x}}}
		}{
		\|\mathrm{x}-\mathrm{y}\|^2
		}
		\,\mathrm{d}\mu(\mathrm{y}),
	\end{aligned}
	\label{eq:app-direct-area}
\end{equation}
where the two cosine factors are the projected-area factors at the receiver and light, respectively. Let $g(\mathrm{x},\mathrm{y})$ collect all non-visibility factors in \eqref{eq:app-direct-area}, including emission, BSDF, geometry, and cosine terms. The direct-light integral can then be written as
\begin{equation}
	L_{\mathrm{DI}}(\mathrm{x},\omega_o)
	=
	\int_{\mathcal{A}}
	V(\mathrm{x},\mathrm{y})
	g(\mathrm{x},\mathrm{y})
	\,\mathrm{d}\mu(\mathrm{y}).
	\label{eq:app-direct-area-compact}
\end{equation}

To obtain the scalar shadow term used in the main paper, we approximate this shadowed direct-light integral as an unshadowed direct-light integral multiplied by the average visibility over the emitter:
\begin{equation}
	\int_{\mathcal{A}} V(\mathrm{x},\mathrm{y})g(\mathrm{x},\mathrm{y})\,\mathrm{d}\mu(\mathrm{y})
	\approx
	\frac{\int_{\mathcal{A}} V(\mathrm{x},\mathrm{y})\,\mathrm{d}\mu(\mathrm{y})}
	{\int_{\mathcal{A}}\mathrm{d}\mu(\mathrm{y})}
	\int_{\mathcal{A}} g(\mathrm{x},\mathrm{y})\,\mathrm{d}\mu(\mathrm{y}).
	\label{eq:app-shadow-factorization}
\end{equation}
This approximation follows the generic factorization
\begin{equation}
	\int_{\Omega} f(\xi)g(\xi)\,\mathrm{d}\xi
	\approx
	\frac{\int_{\Omega}f(\xi)\,\mathrm{d}\xi}
	{\int_{\Omega}\mathrm{d}\xi}
	\int_{\Omega}g(\xi)\,\mathrm{d}\xi,
	\label{eq:app-factorization-generic}
\end{equation}
which is widely used in real-time rendering to decouple the visibility term from the remaining lighting factors. Applying this approximation to area-light direct illumination gives the normalized shadow term
\begin{equation}
	\mathrm{Shadow}(\mathrm{x})
	=
	\frac{1}{\mathrm{Area}(\mathcal{A})}
	\int_{\mathcal{A}}
	V(\mathrm{x},\mathrm{y})
	\,\mathrm{d}\mu(\mathrm{y}).
	\label{eq:app-shadow-area}
\end{equation}
With uniform area sampling on the light, $p_{\mathcal{A}}(\mathrm{y})=1/\mathrm{Area}(\mathcal{A})$, the standard estimator is the average of binary visibility samples.

\paragraph{Notes}
In the direct-lighting setting, besides the area-domain definition in \eqref{eq:app-shadow-area}, shadow can also be defined on the emitter-visible solid-angle domain:
\begin{equation}
	\mathrm{Shadow}_{\Omega}(\mathrm{x})
	=
	\frac{1}{\mathrm{Area}(\Omega_{\mathcal{A}}(\mathrm{x}))}
	\int_{\Omega_{\mathcal{A}}(\mathrm{x})}
	V(\mathrm{x},\omega)\,\mathrm{d}\omega,
	\label{eq:app-shadow-definition-solid-angle}
\end{equation}
where $\Omega_{\mathcal{A}}(\mathrm{x})$ denotes the set of directions from $\mathrm{x}$ that intersect the emitter.
Under this definition, the occluder-domain transformation, induced-density construction, and reuse estimator follow the similar derivation pattern as in the area-domain case, leading to a similar algorithmic pipeline; we provide the full derivation in \appref{app:shadow-occluder-derivation-solid-angle}.
In this paper, we use the area-domain definition for both theoretical presentation and implementation.

\section{Additional Direct-Illumination Decompositions}
\label{app:di-two-decompositions}

For clarity, we first summarize the two direct-illumination decomposition forms relevant to the following discussion. Here we only need two facts: direct illumination admits both area-domain and solid-angle-domain visibility decompositions, and after factoring out $V$, the remaining parts represent the same lighting contribution under two different measures. The area-domain expression is
\begin{equation}
	L_{\mathrm{DI}}(\mathrm{x},\omega_o)
	=
	\int_{\mathcal{A}}
	V(\mathrm{x},\mathrm{y})
	g(\mathrm{x},\mathrm{y},\omega_o)
	\,\mathrm{d}\mu(\mathrm{y}),
	\label{eq:reapp-di-area-compact}
\end{equation}
and the solid-angle-domain expression is
\begin{equation}
	L_{\mathrm{DI}}(\mathrm{x},\omega_o)
	=
	\int_{\Omega_{\mathcal{A}}(\mathrm{x})}
	V(\mathrm{x},\omega)
	g_{\Omega}(\mathrm{x},\omega,\omega_o)
	\,\mathrm{d}\omega,
	\label{eq:reapp-di-solid-angle-compact}
\end{equation}
where $g$ and $g_{\Omega}$ collect all non-visibility factors under the corresponding measures. After factoring out $V$, the remaining parts are the same unshadowed direct-illumination contribution written in two different domains, and satisfy
\begin{equation}
	g(\mathrm{x},\mathrm{y},\omega_o)\,\mathrm{d}\mu(\mathrm{y})
	=
	g_{\Omega}(\mathrm{x},\omega_{\mathrm{x}\to\mathrm{y}},\omega_o)\,\mathrm{d}\omega,
	\label{eq:reapp-di-integrand-measure-relation}
\end{equation}
namely, the light-side geometry factor is explicit in the area-domain form and absorbed into the solid-angle measure in the directional form.

\section{Derivation of Occluder-Domain Ambient Occlusion}
\label{app:ao-occluder-derivation}

AO starts from the cosine-weighted hemisphere integral in \eqref{eq:ao-hemisphere}. Ignoring the finite AO radius for the moment, the normal hemisphere can be split into directions that miss the scene surface $\mathcal{S}$ and directions that intersect it:
\begin{equation}
	\begin{aligned}
		\Omega(\mathrm{x})
		 & =
		\Omega_{\mathrm{vis}}(\mathrm{x})
		\cup
		\Omega_{\mathrm{occ}}(\mathrm{x}),                                            \\
		\Omega_{\mathrm{vis}}(\mathrm{x})
		 & =
		\{\omega\in\Omega(\mathrm{x})\mid \omega\ \text{does not hit}\ \mathcal{S}\}, \\
		\Omega_{\mathrm{occ}}(\mathrm{x})
		 & =
		\{\omega\in\Omega(\mathrm{x})\mid \omega\ \text{has a first hit on}\ \mathcal{S}\}.
	\end{aligned}
	\label{eq:app-ao-domain-split}
\end{equation}
The complement of AO, $1-\mathrm{AO}(\mathrm{x})$, then measures the finite-radius occluded part of this hemisphere:
\begin{equation}
	\begin{aligned}
		1-\mathrm{AO}(\mathrm{x})
		 & =
		\frac{1}{\pi}
		\int_{\Omega(\mathrm{x})}
		\left(1-V(\mathrm{x},\omega,R)\right)
		\\
		 & \quad
		\cosangle{\mathbf{n}_{\mathrm{x}}}{\omega}
		\,\mathrm{d}\omega \\
		 & =
		\frac{1}{\pi}
		\bigl(
		\int_{\Omega_{\mathrm{vis}}(\mathrm{x})}
		0\cdot\cosangle{\mathbf{n}_{\mathrm{x}}}{\omega}
		\,\mathrm{d}\omega
		+
		\\
		 & \quad
		\int_{\Omega_{\mathrm{occ}}(\mathrm{x})}
		\mathbb{I}_{r(\mathrm{x},\omega)\le R}
		\cosangle{\mathbf{n}_{\mathrm{x}}}{\omega}
		\,\mathrm{d}\omega
		\bigr)             \\
		 & =
		\frac{1}{\pi}
		\int_{\Omega_{\mathrm{occ}}(\mathrm{x})}
		\mathbb{I}_{r(\mathrm{x},\omega)\le R}
		\cosangle{\mathbf{n}_{\mathrm{x}}}{\omega}
		\,\mathrm{d}\omega.
	\end{aligned}
	\label{eq:app-ao-blocked-direction}
\end{equation}
where $r(\mathrm{x},\omega)$ is the distance from $\mathrm{x}$ to the first hit along $\omega$. For each direction in $\Omega_{\mathrm{occ}}(\mathrm{x})$, let $\mathrm{z}\in\mathcal{S}$ be this first hit. The solid-angle to area-measure conversion is
\begin{equation}
	\mathrm{d}\omega
	=
	\frac{\cosangle{\mathbf{n}_{\mathrm{z}}}{\omega_{\mathrm{z}\to\mathrm{x}}}}
	{\|\mathrm{x}-\mathrm{z}\|^2}
	\,\mathrm{d}\mu(\mathrm{z}),
	\label{eq:app-ao-solid-angle}
\end{equation}
where the numerator is the projected-area factor at $\mathrm{z}$. The change of variables is one-to-one only between directions in $\Omega_{\mathrm{occ}}(\mathrm{x})$ and their first-hit surface points. When we rewrite the result as an integral over the full scene surface $\mathcal{S}$, $\mathrm{Occ}(\mathrm{x},\mathrm{z})$ extends this mapping by zero outside the first-hit set, while the radius indicator removes first hits outside the AO support:
\begin{equation*}
	J_{\Omega\!\to\!\mathcal{S}}(\mathrm{x},\mathrm{z})
	=
	\frac{\mathrm{d}\omega}{\mathrm{d}\mu(\mathrm{z})}
	=
	\frac{\cosangle{\mathbf{n}_{\mathrm{z}}}{\omega_{\mathrm{z}\to\mathrm{x}}}}
	{\|\mathrm{x}-\mathrm{z}\|^2}.
\end{equation*}
The receiver cosine remains the original AO integrand term, while $J_{\Omega\!\to\!\mathcal{S}}$ comes only from the domain conversion. Therefore,
\begin{equation}
	1-\mathrm{AO}(\mathrm{x})
	=
	\frac{1}{\pi}
	\int_{\mathcal{S}}
	\mathrm{Occ}(\mathrm{x},\mathrm{z})
	\cosangle{\mathbf{n}_{\mathrm{x}}}{\omega_{\mathrm{x}\to\mathrm{z}}}
	J_{\Omega\!\to\!\mathcal{S}}(\mathrm{x},\mathrm{z})
	\mathbb{I}_{\|\mathrm{x}-\mathrm{z}\|\le R}
	\,\mathrm{d}\mu(\mathrm{z}).
	\label{eq:app-ao-blocked}
\end{equation}
Rearranging \eqref{eq:app-ao-blocked} gives \eqref{eq:ao-occluder-domain}.

The same change of variables gives the finite-support density used in the main paper. The directional density is
\begin{equation*}
	p_{\Omega}(\omega\mid\mathrm{x})
	=
	\frac{\cosangle{\mathbf{n}_{\mathrm{x}}}{\omega}}{\pi}.
\end{equation*}
On the valid one-to-one mapping, the induced surface density is the directional density multiplied by the Jacobian $\mathrm{d}\omega/\mathrm{d}\mu(\mathrm{z})$. To define this density over arbitrary candidate surface points, we again multiply by the finite-radius support and by the first-hit predicate:
\begin{equation}
	\begin{aligned}
		p_{\mathrm{AO}}(\mathrm{z}\mid\mathrm{x})
		 & =
		p_{\Omega}(\omega\mid\mathrm{x})
		\frac{\mathrm{d}\omega}{\mathrm{d}\mu(\mathrm{z})}
		\mathbb{I}_{\|\mathrm{x}-\mathrm{z}\|\le R}
		\mathrm{Occ}(\mathrm{x},\mathrm{z}) \\
		 & =
		\frac{1}{\pi}
		\cosangle{\mathbf{n}_{\mathrm{x}}}{\omega_{\mathrm{x}\to\mathrm{z}}}
		J_{\Omega\!\to\!\mathcal{S}}(\mathrm{x},\mathrm{z})
		\mathbb{I}_{\|\mathrm{x}-\mathrm{z}\|\le R}
		\mathrm{Occ}(\mathrm{x},\mathrm{z}),
	\end{aligned}
	\label{eq:app-ao-pdf}
\end{equation}
for arbitrary candidate surface points; the indicator and first-hit predicate make the density zero outside the valid finite-support mapping. The first-hit predicate is part of the exact induced density, which is why cross-pixel pdf evaluation requires a first-hit validation.

\section{AO Reuse Algorithm}
\label{app:ao-reuse-algorithm}
\algref{alg:ao-unbiased} gives the unbiased implementation of the AO estimator in \eqref{eq:ao-unbiased-estimator}. Each target thread evaluates exact induced densities for the occluders shared by its local shared group, including the first-hit validation, then combines the contributions with a MIS formulation.

\begin{algorithm}[t]
	\caption{Unbiased AO reuse for target~$\mathrm{x}_i$}
	\label{alg:ao-unbiased}
	$B_i\leftarrow 0$\;
	\ForEach{AO sample}{
		\tcp*[l]{Step 1: all threads in the shared group generate candidates}
		Sample a cosine-weighted direction $\omega_i$ at $\mathrm{x}_i$\;
		\tcp*[l]{Use a finite ray because the AO radius is part of the sampling support}
		Trace along $\omega_i$ with $t_{\max}=R$ to obtain first hit $\mathrm{z}_i$\;
		Store hit position and oriented hit normal if the ray hits $\mathcal{S}$\;
		\;\tcp*[l]{Step 2: expose all candidates $\{\mathrm{z}_j\}$ in the shared group}
		Initialize $p_i[j]\leftarrow0$\;
		\tcp*[l]{$p_i[j]$: exact density $p_{\mathrm{AO}}(\mathrm{z}_j\mid\mathrm{x}_i)$}
		\ForEach{source thread $j\in\mathcal{N}_i$}{
			Read $\mathrm{z}_j$ and its hit normal from source thread $j$\;
			\If{$\mathrm{z}_j$ is a valid source hit}{
				Evaluate temporary density $q$ from the support and geometric terms of \eqref{eq:ao-pdf}\;
				\If{$\mathrm{z}_j$ is the first hit from $\mathrm{x}_i$}{
					$p_i[j]\leftarrow q$\;
				}
			}
		}
		\;\tcp*[l]{Step 3: this thread evaluates MIS contributions for target $\mathrm{x}_i$}
		\ForEach{candidate $j\in\mathcal{N}_i$}{
			Read $p_k[j]$ from source thread $k$ and accumulate $P_j\leftarrow\sum_{k=1}^{N}p_k[j]$\;
			\If{$p_i[j]>0 \land P_j>0$}{
				$B_i\leftarrow B_i+p_i[j]/P_j$\;
			}
		}
	}
	\Return{$1-B_i/M$}\tcp*[l]{$M$: number of AO samples accumulated for target $\mathrm{x}_i$}
\end{algorithm}

\section{Derivation of Occluder-Domain Area-Light Shadows}
\label{app:shadow-occluder-derivation}

We start from the normalized shadow term in \eqref{eq:shadow-area},
\begin{equation}
	\mathrm{Shadow}(\mathrm{x})
	=
	\frac{1}{\mathrm{Area}(\mathcal{A})}
	\int_{\mathcal{A}}
	V(\mathrm{x},\mathrm{y})
	\,\mathrm{d}\mu(\mathrm{y}).
	\label{eq:app-shadow-start}
\end{equation}

For a fixed receiver $\mathrm{x}$, split the emitter domain into visible and blocked parts:
\begin{equation}
	\begin{aligned}
		\mathcal{A}
		 & =
		\mathcal{A}_{\mathrm{vis}}(\mathrm{x})
		\cup
		\mathcal{A}_{\mathrm{occ}}(\mathrm{x}),                      \\
		\mathcal{A}_{\mathrm{vis}}(\mathrm{x})
		 & =
		\{\mathrm{y}\in\mathcal{A}\mid V(\mathrm{x},\mathrm{y})=1\}, \\
		\mathcal{A}_{\mathrm{occ}}(\mathrm{x})
		 & =
		\{\mathrm{y}\in\mathcal{A}\mid V(\mathrm{x},\mathrm{y})=0\}.
	\end{aligned}
	\label{eq:app-shadow-domain-split}
\end{equation}
Therefore, the visible emitter area can be written as the total emitter area minus the blocked emitter area:
\begin{equation}
	\begin{aligned}
		\int_{\mathcal{A}}
		V(\mathrm{x},\mathrm{y})
		\,\mathrm{d}\mu(\mathrm{y})
		 & =
		\int_{\mathcal{A}_{\mathrm{vis}}(\mathrm{x})}
		\,\mathrm{d}\mu(\mathrm{y}) \\
		 & =
		\mathrm{Area}(\mathcal{A})
		-
		\int_{\mathcal{A}_{\mathrm{occ}}(\mathrm{x})}
		\,\mathrm{d}\mu(\mathrm{y}).
	\end{aligned}
	\label{eq:app-shadow-visible-minus-blocked}
\end{equation}

For each blocked light point $\mathrm{y}\in\mathcal{A}_{\mathrm{occ}}(\mathrm{x})$, let $\mathrm{z}$ be the first occluder along the segment from $\mathrm{x}$ to $\mathrm{y}$. In the following domain conversion, $\mathrm{y}$ is the original emitter point being mapped to its first occluder.

The light point $\mathrm{y}$ and occlusion point $\mathrm{z}$ lie on the same ray from $\mathrm{x}$, and therefore subtend the same differential solid angle at $\mathrm{x}$:
\begin{equation}
	\frac{
	\cosangle{\mathbf{n}_{\mathrm{y}}}{\omega_{\mathrm{y}\to\mathrm{x}}}
	}{
	\|\mathrm{x}-\mathrm{y}\|^2
	}
	\,\mathrm{d}\mu(\mathrm{y})
	=
	\frac{
	\cosangle{\mathbf{n}_{\mathrm{z}}}{\omega_{\mathrm{z}\to\mathrm{x}}}
	}{
	\|\mathrm{x}-\mathrm{z}\|^2
	}
	\,\mathrm{d}\mu(\mathrm{z}).
	\label{eq:app-shadow-solid-angle}
\end{equation}
Rearranging \eqref{eq:app-shadow-solid-angle} gives the projected-area measure conversion
\begin{equation}
	\mathrm{d}\mu(\mathrm{y})
	=
	J(\mathrm{x},\mathrm{z})\,\mathrm{d}\mu(\mathrm{z}),
	\label{eq:app-shadow-measure}
\end{equation}
where
\begin{equation}
	J(\mathrm{x},\mathrm{z})
	=
	\frac{
	\cosangle{\mathbf{n}_{\mathrm{z}}}{\omega_{\mathrm{z}\to\mathrm{x}}}
	\|\mathrm{x}-\mathrm{y}\|^2
	}{
	\cosangle{\mathbf{n}_{\mathrm{y}}}{\omega_{\mathrm{y}\to\mathrm{x}}}
	\|\mathrm{x}-\mathrm{z}\|^2
	}.
	\label{eq:app-shadow-jacobian}
\end{equation}

We now map the blocked emitter-area integral in \eqref{eq:app-shadow-visible-minus-blocked} to a surface integral. The local measure conversion above is valid for a one-to-one correspondence between a blocked emitter point and its first occluder. When the integral is written over the full scene surface $\mathcal{S}$, $\mathrm{Occ}(\mathrm{x},\mathrm{z})$ keeps only first occluders, and $\chi_{\mathcal{A}}(\mathrm{x},\mathrm{z})$ keeps only occluders whose ray from $\mathrm{x}$ reaches a valid point on $\mathcal{A}$. Thus,
\begin{equation}
	\int_{\mathcal{A}_{\mathrm{occ}}(\mathrm{x})}
	\,\mathrm{d}\mu(\mathrm{y})
	=
	\int_{\mathcal{S}}
	\mathrm{Occ}(\mathrm{x},\mathrm{z})
	J(\mathrm{x},\mathrm{z})
	\chi_{\mathcal{A}}(\mathrm{x},\mathrm{z})
	\,\mathrm{d}\mu(\mathrm{z}).
	\label{eq:app-shadow-blocked-area}
\end{equation}
Substituting \eqref{eq:app-shadow-blocked-area} into \eqref{eq:app-shadow-visible-minus-blocked} yields
\begin{equation}
	\int_{\mathcal{A}}V(\mathrm{x},\mathrm{y})\,\mathrm{d}\mu(\mathrm{y})
	=
	\mathrm{Area}(\mathcal{A})
	-
	\int_{\mathcal{S}}
	\mathrm{Occ}(\mathrm{x},\mathrm{z})
	J(\mathrm{x},\mathrm{z})
	\chi_{\mathcal{A}}(\mathrm{x},\mathrm{z})
	\,\mathrm{d}\mu(\mathrm{z}).
	\label{eq:app-shadow-visible-area}
\end{equation}
Finally, substituting \eqref{eq:app-shadow-visible-area} into the normalized shadow definition \eqref{eq:app-shadow-start} gives \eqref{eq:shadow-occluder-domain}.

During reuse, an occlusion point $\mathrm{z}$ may have been generated from a different receiver, so the original sampled light point $\mathrm{y}$ is not generally the emitter point associated with the current receiver $\mathrm{x}$. We therefore define $\mathrm{y}'(\mathrm{x},\mathrm{z})$ as the intersection between the ray from $\mathrm{x}$ through $\mathrm{z}$ and the emitter domain, when it exists. The same Jacobian is then evaluated with $\mathrm{y}$ replaced by this light-intersection point $\mathrm{y}'$.

If a light point is sampled from an area density $p_{\mathcal{A}}(\mathrm{y})$, the corresponding first-hit occluder density is obtained by the same measure conversion on the valid one-to-one mapping. To evaluate the density for arbitrary reused candidates, we multiply by the same emitter hit support and first-hit predicate:
\begin{equation}
	\begin{aligned}
		p_{\mathrm{Sh}}(\mathrm{z}\mid\mathrm{x})
		 & =
		p_{\mathcal{A}}(\mathrm{y}')
		\frac{\mathrm{d}\mu(\mathrm{y}')}{\mathrm{d}\mu(\mathrm{z})}
		\chi_{\mathcal{A}}(\mathrm{x},\mathrm{z})
		\mathrm{Occ}(\mathrm{x},\mathrm{z}) \\
		 & =
		\frac{1}{\mathrm{Area}(\mathcal{A})}
		J(\mathrm{x},\mathrm{z})
		\chi_{\mathcal{A}}(\mathrm{x},\mathrm{z})
		\mathrm{Occ}(\mathrm{x},\mathrm{z}),
	\end{aligned}
	\label{eq:app-shadow-pdf}
\end{equation}
The second line uses the uniform area density
\begin{equation*}
	p_{\mathcal{A}}(\mathrm{y}')=\frac{1}{\mathrm{Area}(\mathcal{A})}.
\end{equation*}
The density is zero if $\mathrm{z}$ is not the first hit from $\mathrm{x}$, the ray through $\mathrm{z}$ does not hit the emitter surface, $\mathrm{y}'(\mathrm{x},\mathrm{z})$ lies outside $\mathcal{A}$, or any projected-area factor is invalid.

\section{Derivation of Occluder-Domain Area-Light Shadows under the Solid-Angle Definition}
\label{app:shadow-occluder-derivation-solid-angle}

As discussed in \appref{app:shadow-visibility}, the same framework also applies to area-light shadow estimation under the emitter-visible solid-angle definition. Here, for completeness, we provide the full derivation. In this case, we only need to transform the blocked solid-angle contribution into the first-hit occluder domain, while keeping the actual sampling density on the light area.
Starting from the solid-angle definition,
\begin{equation}
	\mathrm{Shadow}_{\Omega}(\mathrm{x})
	=
	\frac{1}{\mathrm{Area}(\Omega_{\mathcal{A}}(\mathrm{x}))}
	\int_{\Omega_{\mathcal{A}}(\mathrm{x})}
	V(\mathrm{x},\omega)\,\mathrm{d}\omega,
	\label{eq:reapp-shadow-definition-solid-angle}
\end{equation}
where $\Omega_{\mathcal{A}}(\mathrm{x})$ denotes the set of directions from $\mathrm{x}$ that intersect the emitter domain $\mathcal{A}$.
The normalization term $\mathrm{Area}(\Omega_{\mathcal{A}}(\mathrm{x}))$, namely the solid angle subtended by the emitter at $\mathrm{x}$, depends only on the receiver--light configuration and can be computed analytically.

For a fixed receiver $\mathrm{x}$, split $\Omega_{\mathcal{A}}(\mathrm{x})$ into visible and blocked directions:
\begin{equation}
	\begin{aligned}
		\Omega_{\mathcal{A}}(\mathrm{x})
		 & =
		\Omega_{\mathcal{A},\mathrm{vis}}(\mathrm{x})
		\cup
		\Omega_{\mathcal{A},\mathrm{occ}}(\mathrm{x}),                            \\[2pt]
		\Omega_{\mathcal{A},\mathrm{vis}}(\mathrm{x})
		 & =
		\{\omega\in\Omega_{\mathcal{A}}(\mathrm{x})\mid V(\mathrm{x},\omega)=1\}, \\[2pt]
		\Omega_{\mathcal{A},\mathrm{occ}}(\mathrm{x})
		 & =
		\{\omega\in\Omega_{\mathcal{A}}(\mathrm{x})\mid V(\mathrm{x},\omega)=0\}.
	\end{aligned}
	\label{eq:reapp-shadow-solid-angle-domain-split}
\end{equation}
Therefore, the visible solid angle equals the total emitter solid angle minus the blocked part:
\begin{equation}
	\begin{aligned}
		\int_{\Omega_{\mathcal{A}}(\mathrm{x})}
		V(\mathrm{x},\omega)\,\mathrm{d}\omega
		 & =
		\int_{\Omega_{\mathcal{A},\mathrm{vis}}(\mathrm{x})}
		\,\mathrm{d}\omega \\
		 & =
		\mathrm{Area}(\Omega_{\mathcal{A}}(\mathrm{x}))
		-
		\int_{\Omega_{\mathcal{A},\mathrm{occ}}(\mathrm{x})}
		\,\mathrm{d}\omega.
	\end{aligned}
	\label{eq:reapp-shadow-solid-angle-visible-minus-blocked}
\end{equation}

For each blocked direction $\omega\in\Omega_{\mathcal{A},\mathrm{occ}}(\mathrm{x})$, let $\mathrm{z}$ be the first occluder hit along the ray from $\mathrm{x}$.
The corresponding differential solid angle and surface area satisfy
\begin{equation}
	\mathrm{d}\omega
	=
	\frac{
	\cosangle{\mathbf{n}_{\mathrm{z}}}{\omega_{\mathrm{z}\to\mathrm{x}}}
	}{
	\|\mathrm{x}-\mathrm{z}\|^2
	}
	\,\mathrm{d}\mu(\mathrm{z}),
	\label{eq:reapp-shadow-solid-angle-to-surface}
\end{equation}
or equivalently,
\begin{equation*}
	J_{\Omega\!\to\!\mathcal{S}}(\mathrm{x},\mathrm{z})
	=
	\frac{\mathrm{d}\omega}{\mathrm{d}\mu(\mathrm{z})}
	=
	\frac{
	\cosangle{\mathbf{n}_{\mathrm{z}}}{\omega_{\mathrm{z}\to\mathrm{x}}}
	}{
	\|\mathrm{x}-\mathrm{z}\|^2
	}.
\end{equation*}
This change of variables is one-to-one only between a blocked direction and its first-hit occluder.
When the integral is written over the full scene surface $\mathcal{S}$, $\mathrm{Occ}(\mathrm{x},\mathrm{z})$ keeps only first occluders, and $\chi_{\mathcal{A}}(\mathrm{x},\mathrm{z})$ keeps only directions whose ray from $\mathrm{x}$ through $\mathrm{z}$ reaches a valid point on the emitter.
Thus,
\begin{equation}
	\int_{\Omega_{\mathcal{A},\mathrm{occ}}(\mathrm{x})}
	\,\mathrm{d}\omega
	=
	\int_{\mathcal{S}}
	\mathrm{Occ}(\mathrm{x},\mathrm{z})
	J_{\Omega\!\to\!\mathcal{S}}(\mathrm{x},\mathrm{z})
	\chi_{\mathcal{A}}(\mathrm{x},\mathrm{z})
	\,\mathrm{d}\mu(\mathrm{z}).
	\label{eq:reapp-shadow-solid-angle-blocked}
\end{equation}
Substituting \eqref{eq:reapp-shadow-solid-angle-blocked} into \eqref{eq:reapp-shadow-solid-angle-visible-minus-blocked} gives
\begin{equation}
	\begin{aligned}
		\int_{\Omega_{\mathcal{A}}(\mathrm{x})}
		V(\mathrm{x},\omega)\,\mathrm{d}\omega
		 & =
		\mathrm{Area}(\Omega_{\mathcal{A}}(\mathrm{x}))
		-
		\\
		 & \quad
		\int_{\mathcal{S}}
		\mathrm{Occ}(\mathrm{x},\mathrm{z})
		J_{\Omega\!\to\!\mathcal{S}}(\mathrm{x},\mathrm{z})
		\chi_{\mathcal{A}}(\mathrm{x},\mathrm{z})
		\,\mathrm{d}\mu(\mathrm{z}).
	\end{aligned}
	\label{eq:reapp-shadow-solid-angle-visible}
\end{equation}
Finally, substituting \eqref{eq:reapp-shadow-solid-angle-visible} into \eqref{eq:reapp-shadow-definition-solid-angle} yields the occluder-domain form
\begin{equation}
	\begin{aligned}
		\mathrm{Shadow}_{\Omega}(\mathrm{x})
		 & =
		1-
		\frac{1}{\mathrm{Area}(\Omega_{\mathcal{A}}(\mathrm{x}))}
		\\
		 & \quad
		\int_{\mathcal{S}}
		\mathrm{Occ}(\mathrm{x},\mathrm{z})
		J_{\Omega\!\to\!\mathcal{S}}(\mathrm{x},\mathrm{z})
		\chi_{\mathcal{A}}(\mathrm{x},\mathrm{z})
		\,\mathrm{d}\mu(\mathrm{z}).
	\end{aligned}
	\label{eq:reapp-shadow-solid-angle-occluder-domain}
\end{equation}

Although the shadow definition above is written on the solid-angle domain, our implementation still samples emitter points uniformly on the light area:
\begin{equation*}
	p_{\mathcal{A}}(\mathrm{y}')
	=
	\frac{1}{\mathrm{Area}(\mathcal{A})}.
\end{equation*}
Therefore, the first-hit occluder density is still induced by the area-domain sampling density:
\begin{equation}
	\begin{aligned}
		p_{\mathrm{Sh}}(\mathrm{z}\mid\mathrm{x})
		 & =
		p_{\mathcal{A}}(\mathrm{y}')
		\frac{\mathrm{d}\mu(\mathrm{y}')}{\mathrm{d}\mu(\mathrm{z})}
		\chi_{\mathcal{A}}(\mathrm{x},\mathrm{z})
		\mathrm{Occ}(\mathrm{x},\mathrm{z}) \\
		 & =
		\frac{1}{\mathrm{Area}(\mathcal{A})}
		J(\mathrm{x},\mathrm{z})
		\chi_{\mathcal{A}}(\mathrm{x},\mathrm{z})
		\mathrm{Occ}(\mathrm{x},\mathrm{z}).
	\end{aligned}
	\label{eq:reapp-shadow-solid-angle-pdf}
\end{equation}
Here $J(\mathrm{x},\mathrm{z})=\mathrm{d}\mu(\mathrm{y}')/\mathrm{d}\mu(\mathrm{z})$ is the same area-to-occluder Jacobian as in the area-domain derivation, evaluated using the light-intersection point $\mathrm{y}'(\mathrm{x},\mathrm{z})$, namely
\begin{equation*}
	J(\mathrm{x},\mathrm{z})
	=
	\frac{
	\cosangle{\mathbf{n}_{\mathrm{z}}}{\omega_{\mathrm{z}\to\mathrm{x}}}
	\|\mathrm{x}-\mathrm{y}'\|^2
	}{
	\cosangle{\mathbf{n}_{\mathrm{y}'}}{\omega_{\mathrm{y}'\to\mathrm{x}}}
	\|\mathrm{x}-\mathrm{z}\|^2
	}.
\end{equation*}
Combining this density with the solid-angle-domain integrand in \eqref{eq:reapp-shadow-solid-angle-occluder-domain}, the corresponding one-sample Monte Carlo estimator of the blocked term is
\begin{equation}
	\begin{aligned}
		\left\langle B_{\Omega}(\mathrm{x}) \right\rangle
		 & =
		\frac{
			\frac{1}{\mathrm{Area}(\Omega_{\mathcal{A}}(\mathrm{x}))}
			J_{\Omega\!\to\!\mathcal{S}}(\mathrm{x},\mathrm{z})
			\chi_{\mathcal{A}}(\mathrm{x},\mathrm{z})
			\mathrm{Occ}(\mathrm{x},\mathrm{z})
		}{
			p_{\mathrm{Sh}}(\mathrm{z}\mid\mathrm{x})
		}    \\
		 & =
		\frac{\mathrm{Area}(\mathcal{A})}
		{\mathrm{Area}(\Omega_{\mathcal{A}}(\mathrm{x}))}
		\frac{
			J_{\Omega\!\to\!\mathcal{S}}(\mathrm{x},\mathrm{z})
		}{
			J(\mathrm{x},\mathrm{z})
		}    \\
		 & =
		\frac{\mathrm{Area}(\mathcal{A})}
		{\mathrm{Area}(\Omega_{\mathcal{A}}(\mathrm{x}))}
		\frac{
		\cosangle{\mathbf{n}_{\mathrm{y}'}}{\omega_{\mathrm{y}'\to\mathrm{x}}}
		}{
		\|\mathrm{x}-\mathrm{y}'\|^2
		},
	\end{aligned}
	\label{eq:reapp-shadow-solid-angle-contribution}
\end{equation}
for valid reused candidates.

If one candidate occluder $\mathrm{z}_j$ is provided by each source receiver $\mathrm{x}_j$ and the balance heuristic is used for MIS, the blocked-solid-angle term is estimated as
\begin{equation}
	\left\langle B_{\Omega}(\mathrm{x}_i) \right\rangle_{\mathrm{MIS}}
	=
	\sum_{j=1}^{N}
	\frac{
		\mathrm{Occ}(\mathrm{x}_i,\mathrm{z}_j)
		F_{\mathrm{Sh},\Omega}(\mathrm{x}_i,\mathrm{z}_j)
	}{
		\sum_{k=1}^{N} p_{\mathrm{Sh}}(\mathrm{z}_j\mid\mathrm{x}_k)
	},
	\label{eq:reapp-shadow-solid-angle-mis-blocked}
\end{equation}
where
\begin{equation}
	F_{\mathrm{Sh},\Omega}(\mathrm{x},\mathrm{z})
	=
	\frac{1}{\mathrm{Area}(\Omega_{\mathcal{A}}(\mathrm{x}))}
	J_{\Omega\!\to\!\mathcal{S}}(\mathrm{x},\mathrm{z})
	\chi_{\mathcal{A}}(\mathrm{x},\mathrm{z}).
	\label{eq:reapp-shadow-solid-angle-integrand}
\end{equation}
Using \eqref{eq:reapp-shadow-solid-angle-contribution}, the numerator can be written as
\begin{equation}
	\begin{aligned}
		     & \mathrm{Occ}(\mathrm{x}_i,\mathrm{z}_j)
		F_{\mathrm{Sh},\Omega}(\mathrm{x}_i,\mathrm{z}_j) \\
		= {} & C_{\Omega}(\mathrm{x}_i,\mathrm{z}_j)\,
		p_{\mathrm{Sh}}(\mathrm{z}_j\mid\mathrm{x}_i),
	\end{aligned}
	\label{eq:reapp-shadow-solid-angle-mis-numerator}
\end{equation}
with
\begin{equation}
	C_{\Omega}(\mathrm{x},\mathrm{z})
	=
	\frac{\mathrm{Area}(\mathcal{A})}
	{\mathrm{Area}(\Omega_{\mathcal{A}}(\mathrm{x}))}
	\frac{
	\cosangle{\mathbf{n}_{\mathrm{y}'}}{\omega_{\mathrm{y}'\to\mathrm{x}}}
	}{
	\|\mathrm{x}-\mathrm{y}'\|^2
	}.
	\label{eq:reapp-shadow-solid-angle-mis-factor}
\end{equation}
Therefore, the unbiased MIS estimator under the solid-angle definition becomes
\begin{equation}
	\left\langle \mathrm{Shadow}_{\Omega}(\mathrm{x}_i) \right\rangle_{\mathrm{MIS}}
	=
	1-
	\sum_{j=1}^{N}
	C_{\Omega}(\mathrm{x}_i,\mathrm{z}_j)\,
	\frac{
		p_{\mathrm{Sh}}(\mathrm{z}_j\mid\mathrm{x}_i)
	}{
		\sum_{k=1}^{N} p_{\mathrm{Sh}}(\mathrm{z}_j\mid\mathrm{x}_k)
	}.
	\label{eq:reapp-shadow-solid-angle-estimator}
\end{equation}
Unlike the area-domain formulation, the numerator does not collapse directly to the induced density itself; an additional geometric factor $C_{\Omega}$ remains because the target quantity is normalized in solid angle, while the actual sampling is still performed uniformly on the emitter area.

\subsection{General Area-Domain Sampling Density}

We also note that the same derivation is not restricted to uniform light-area sampling. Once a general emitter-area pdf is given, we only need to replace the induced first-hit occluder density and the corresponding estimator factors accordingly.

The derivation above assumes uniform area sampling on the emitter only to obtain the explicit factor $1/\mathrm{Area}(\mathcal{A})$.
More generally, if the light sample is drawn from an arbitrary area-domain density
\begin{equation}
	p_{\mathcal{A}}(\mathrm{y}'\mid\mathrm{x}),
	\label{eq:reapp-shadow-general-area-pdf}
\end{equation}
defined on the emitter surface, then the corresponding area-induced first-hit occluder density becomes
\begin{equation}
	\begin{aligned}
		p_{\mathrm{Sh}}^{\mathrm{gen}}(\mathrm{z}\mid\mathrm{x})
		 & =
		p_{\mathcal{A}}(\mathrm{y}'\mid\mathrm{x})
		\frac{\mathrm{d}\mu(\mathrm{y}')}{\mathrm{d}\mu(\mathrm{z})}
		\chi_{\mathcal{A}}(\mathrm{x},\mathrm{z})
		\mathrm{Occ}(\mathrm{x},\mathrm{z}) \\
		 & =
		p_{\mathcal{A}}(\mathrm{y}'\mid\mathrm{x})
		J(\mathrm{x},\mathrm{z})
		\chi_{\mathcal{A}}(\mathrm{x},\mathrm{z})
		\mathrm{Occ}(\mathrm{x},\mathrm{z}).
	\end{aligned}
	\label{eq:reapp-shadow-general-solid-angle-pdf}
\end{equation}
The corresponding one-sample Monte Carlo estimator of the blocked solid-angle term is
\begin{equation}
	\begin{aligned}
		\left\langle B_{\Omega}^{\mathrm{gen}}(\mathrm{x}) \right\rangle
		 & =
		\frac{
			\frac{1}{\mathrm{Area}(\Omega_{\mathcal{A}}(\mathrm{x}))}
			J_{\Omega\!\to\!\mathcal{S}}(\mathrm{x},\mathrm{z})
			\chi_{\mathcal{A}}(\mathrm{x},\mathrm{z})
			\mathrm{Occ}(\mathrm{x},\mathrm{z})
		}{
			p_{\mathrm{Sh}}^{\mathrm{gen}}(\mathrm{z}\mid\mathrm{x})
		}    \\
		 & =
		\frac{
			J_{\Omega\!\to\!\mathcal{S}}(\mathrm{x},\mathrm{z})
		}{
			\mathrm{Area}(\Omega_{\mathcal{A}}(\mathrm{x}))
			p_{\mathcal{A}}(\mathrm{y}'\mid\mathrm{x})
			J(\mathrm{x},\mathrm{z})
		}.
	\end{aligned}
	\label{eq:reapp-shadow-general-solid-angle-contribution}
\end{equation}

Under the same balance-heuristic MIS construction, the numerator is
\begin{equation}
	\begin{aligned}
		     & \mathrm{Occ}(\mathrm{x}_i,\mathrm{z}_j)
		F_{\mathrm{Sh},\Omega}(\mathrm{x}_i,\mathrm{z}_j)             \\
		= {} & C_{\Omega}^{\mathrm{gen}}(\mathrm{x}_i,\mathrm{z}_j)\,
		p_{\mathrm{Sh}}^{\mathrm{gen}}(\mathrm{z}_j\mid\mathrm{x}_i),
	\end{aligned}
	\label{eq:reapp-shadow-general-solid-angle-mis-numerator}
\end{equation}
with
\begin{equation}
	\begin{aligned}
		C_{\Omega}^{\mathrm{gen}}(\mathrm{x},\mathrm{z})
		 & =
		\frac{
			J_{\Omega\!\to\!\mathcal{S}}(\mathrm{x},\mathrm{z})
		}{
			\mathrm{Area}(\Omega_{\mathcal{A}}(\mathrm{x}))
			p_{\mathcal{A}}(\mathrm{y}'\mid\mathrm{x})
			J(\mathrm{x},\mathrm{z})
		} \\
		 & =
		\frac{
		\cosangle{\mathbf{n}_{\mathrm{y}'}}{\omega_{\mathrm{y}'\to\mathrm{x}}}
		}{
		\mathrm{Area}(\Omega_{\mathcal{A}}(\mathrm{x}))
		p_{\mathcal{A}}(\mathrm{y}'\mid\mathrm{x})
		\|\mathrm{x}-\mathrm{y}'\|^2
		}.
	\end{aligned}
	\label{eq:reapp-shadow-general-solid-angle-mis-factor}
\end{equation}
Therefore, the corresponding MIS estimator is
\begin{equation}
	\left\langle \mathrm{Shadow}_{\Omega}(\mathrm{x}_i) \right\rangle_{\mathrm{MIS}}^{\mathrm{gen}}
	=
	1-
	\sum_{j=1}^{N}
	C_{\Omega}^{\mathrm{gen}}(\mathrm{x}_i,\mathrm{z}_j)\,
	\frac{
		p_{\mathrm{Sh}}^{\mathrm{gen}}(\mathrm{z}_j\mid\mathrm{x}_i)
	}{
		\sum_{k=1}^{N} p_{\mathrm{Sh}}^{\mathrm{gen}}(\mathrm{z}_j\mid\mathrm{x}_k)
	}.
	\label{eq:reapp-shadow-general-solid-angle-estimator}
\end{equation}
The uniform-area result above is recovered by substituting
\begin{equation*}
	p_{\mathcal{A}}(\mathrm{y}'\mid\mathrm{x})=\frac{1}{\mathrm{Area}(\mathcal{A})}.
\end{equation*}

\section{Occluder-Domain Direct Illumination}
\label{app:di-occluder-derivation}

For completeness, we further derive the occluder-domain form for direct illumination itself, rather than only for the normalized shadow quantity. In this case, we move the blocked emitter contribution to the first-hit occluder domain while leaving the unshadowed term separate. Starting from the area-domain visibility decomposition in \appref{app:di-two-decompositions}, specifically \eqref{eq:reapp-di-area-compact}, we give the corresponding formulation below.

For a fixed receiver $\mathrm{x}$, split the emitter into visible and blocked parts as in the shadow derivation:
\begin{equation}
	\mathcal{A}
	=
	\mathcal{A}_{\mathrm{vis}}(\mathrm{x})
	\cup
	\mathcal{A}_{\mathrm{occ}}(\mathrm{x}).
	\label{eq:reapp-di-domain-split}
\end{equation}
Therefore,
\begin{equation}
	\begin{aligned}
		L_{\mathrm{DI}}(\mathrm{x},\omega_o)
		 & =
		\int_{\mathcal{A}_{\mathrm{vis}}(\mathrm{x})}
		g(\mathrm{x},\mathrm{y},\omega_o)
		\,\mathrm{d}\mu(\mathrm{y}) \\
		 & =
		\int_{\mathcal{A}}
		g(\mathrm{x},\mathrm{y},\omega_o)
		\,\mathrm{d}\mu(\mathrm{y})
		-
		\int_{\mathcal{A}_{\mathrm{occ}}(\mathrm{x})}
		g(\mathrm{x},\mathrm{y},\omega_o)
		\,\mathrm{d}\mu(\mathrm{y}).
	\end{aligned}
	\label{eq:reapp-di-visible-minus-blocked}
\end{equation}

For each blocked emitter point $\mathrm{y}\in\mathcal{A}_{\mathrm{occ}}(\mathrm{x})$, let $\mathrm{z}$ be the first occluder along the segment from $\mathrm{x}$ to $\mathrm{y}$.
The same projected-area Jacobian as in the area-domain shadow derivation applies:
\begin{equation}
	\mathrm{d}\mu(\mathrm{y})
	=
	J(\mathrm{x},\mathrm{z})\,\mathrm{d}\mu(\mathrm{z}),
	\label{eq:reapp-di-area-jacobian}
\end{equation}
where $\mathrm{y}'(\mathrm{x},\mathrm{z})$ denotes the intersection between the ray from $\mathrm{x}$ through $\mathrm{z}$ and the emitter, and
\begin{equation}
	J(\mathrm{x},\mathrm{z})
	=
	\frac{
	\cosangle{\mathbf{n}_{\mathrm{z}}}{\omega_{\mathrm{z}\to\mathrm{x}}}
	\|\mathrm{x}-\mathrm{y}'\|^2
	}{
	\cosangle{\mathbf{n}_{\mathrm{y}'}}{\omega_{\mathrm{y}'\to\mathrm{x}}}
	\|\mathrm{x}-\mathrm{z}\|^2
	}.
	\label{eq:reapp-di-jacobian}
\end{equation}
Writing the blocked-emitter integral over the full scene surface then gives
\begin{equation}
	\begin{aligned}
		  & \int_{\mathcal{A}_{\mathrm{occ}}(\mathrm{x})}
		g(\mathrm{x},\mathrm{y},\omega_o)
		\,\mathrm{d}\mu(\mathrm{y})                       \\
		= &
		\int_{\mathcal{S}}
		\mathrm{Occ}(\mathrm{x},\mathrm{z})
		g(\mathrm{x},\mathrm{y}'(\mathrm{x},\mathrm{z}),\omega_o)
		J(\mathrm{x},\mathrm{z})
		\chi_{\mathcal{A}}(\mathrm{x},\mathrm{z})
		\,\mathrm{d}\mu(\mathrm{z}).
	\end{aligned}
	\label{eq:reapp-di-blocked-occ}
\end{equation}
Substituting \eqref{eq:reapp-di-blocked-occ} into \eqref{eq:reapp-di-visible-minus-blocked} yields the occluder-domain direct-illumination expression
\begin{equation}
	\begin{aligned}
		L_{\mathrm{DI}}(\mathrm{x},\omega_o)
		 & =
		\int_{\mathcal{A}}
		g(\mathrm{x},\mathrm{y},\omega_o)
		\,\mathrm{d}\mu(\mathrm{y})
		\\
		 & \quad
		-
		\int_{\mathcal{S}}
		\mathrm{Occ}(\mathrm{x},\mathrm{z})
		g(\mathrm{x},\mathrm{y}'(\mathrm{x},\mathrm{z}),\omega_o)
		J(\mathrm{x},\mathrm{z})
		\chi_{\mathcal{A}}(\mathrm{x},\mathrm{z})
		\,\mathrm{d}\mu(\mathrm{z}).
	\end{aligned}
	\label{eq:reapp-di-occluder-domain}
\end{equation}
Here the first term is simply the unshadowed direct-illumination integral over the emitter domain.
In our current setting, the emitter is a rectangular area light, so this term can be evaluated efficiently using LTCs (Linearly Transformed Cosines)~\cite{LTC16}.
Therefore, the remaining occluder-domain term only needs to account for the blocked contribution transported to the first-hit occluder domain.
Since the unshadowed term in \eqref{eq:reapp-di-occluder-domain} can be evaluated directly, we denote it by
\begin{equation}
	U(\mathrm{x},\omega_o)
	=
	\int_{\mathcal{A}}
	g(\mathrm{x},\mathrm{y},\omega_o)
	\,\mathrm{d}\mu(\mathrm{y}).
	\label{eq:reapp-di-unshadowed-term}
\end{equation}
Then \eqref{eq:reapp-di-occluder-domain} becomes
\begin{equation}
	\begin{aligned}
		L_{\mathrm{DI}}(\mathrm{x},\omega_o)
		 & =
		U(\mathrm{x},\omega_o)
		\\
		 & \quad
		-
		\int_{\mathcal{S}}
		\mathrm{Occ}(\mathrm{x},\mathrm{z})
		g(\mathrm{x},\mathrm{y}'(\mathrm{x},\mathrm{z}),\omega_o)
		J(\mathrm{x},\mathrm{z})
		\chi_{\mathcal{A}}(\mathrm{x},\mathrm{z})
		\,\mathrm{d}\mu(\mathrm{z}).
	\end{aligned}
	\label{eq:reapp-di-occluder-domain-compact}
\end{equation}
For a general light-area sampling density
\begin{equation}
	p_{\mathcal{A}}(\mathrm{y}'\mid\mathrm{x}).
	\label{eq:reapp-di-general-area-pdf}
\end{equation}
The corresponding area-induced first-hit occluder density is
\begin{equation}
	\begin{aligned}
		p_{\mathrm{DI}}^{\mathrm{gen}}(\mathrm{z}\mid\mathrm{x})
		 & =
		p_{\mathcal{A}}(\mathrm{y}'\mid\mathrm{x})
		\frac{\mathrm{d}\mu(\mathrm{y}')}{\mathrm{d}\mu(\mathrm{z})}
		\chi_{\mathcal{A}}(\mathrm{x},\mathrm{z})
		\mathrm{Occ}(\mathrm{x},\mathrm{z}) \\
		 & =
		p_{\mathcal{A}}(\mathrm{y}'\mid\mathrm{x})
		J(\mathrm{x},\mathrm{z})
		\chi_{\mathcal{A}}(\mathrm{x},\mathrm{z})
		\mathrm{Occ}(\mathrm{x},\mathrm{z}).
	\end{aligned}
	\label{eq:reapp-di-general-occ-pdf}
\end{equation}
Therefore, the corresponding one-sample Monte Carlo estimator of the blocked term is
\begin{equation}
	\begin{aligned}
		\left\langle B_{\mathrm{DI}}(\mathrm{x},\omega_o) \right\rangle^{\mathrm{gen}}
		 & =
		\frac{
			\mathrm{Occ}(\mathrm{x},\mathrm{z})
			g(\mathrm{x},\mathrm{y}'(\mathrm{x},\mathrm{z}),\omega_o)
			J(\mathrm{x},\mathrm{z})
			\chi_{\mathcal{A}}(\mathrm{x},\mathrm{z})
		}{
			p_{\mathrm{DI}}^{\mathrm{gen}}(\mathrm{z}\mid\mathrm{x})
		}
		\\
		 & =
		\frac{
			g(\mathrm{x},\mathrm{y}'(\mathrm{x},\mathrm{z}),\omega_o)
		}{
			p_{\mathcal{A}}(\mathrm{y}'\mid\mathrm{x})
		}.
	\end{aligned}
	\label{eq:reapp-di-general-one-sample-blocked}
\end{equation}
and the corresponding direct-illumination estimator is
\begin{equation}
	\left\langle L_{\mathrm{DI}}(\mathrm{x},\omega_o) \right\rangle^{\mathrm{gen}}
	=
	U(\mathrm{x},\omega_o)
	-
	\left\langle B_{\mathrm{DI}}(\mathrm{x},\omega_o) \right\rangle^{\mathrm{gen}}.
	\label{eq:reapp-di-general-one-sample-estimator}
\end{equation}

If one candidate occluder $\mathrm{z}_j$ is provided by each source receiver $\mathrm{x}_j$ and the balance heuristic is used for MIS, the blocked term is estimated as
\begin{equation}
	\left\langle B_{\mathrm{DI}}(\mathrm{x}_i,\omega_o) \right\rangle_{\mathrm{MIS}}^{\mathrm{gen}}
	=
	\sum_{j=1}^{N}
	\frac{
		\mathrm{Occ}(\mathrm{x}_i,\mathrm{z}_j)
		g(\mathrm{x}_i,\mathrm{y}'(\mathrm{x}_i,\mathrm{z}_j),\omega_o)
		J(\mathrm{x}_i,\mathrm{z}_j)
		\chi_{\mathcal{A}}(\mathrm{x}_i,\mathrm{z}_j)
	}{
		\sum_{k=1}^{N} p_{\mathrm{DI}}^{\mathrm{gen}}(\mathrm{z}_j\mid\mathrm{x}_k)
	}.
	\label{eq:reapp-di-general-mis-blocked}
\end{equation}
Its numerator can be written as
\begin{equation}
	\begin{aligned}
		     & \mathrm{Occ}(\mathrm{x}_i,\mathrm{z}_j)
		g(\mathrm{x}_i,\mathrm{y}'(\mathrm{x}_i,\mathrm{z}_j),\omega_o)
		J(\mathrm{x}_i,\mathrm{z}_j)
		\chi_{\mathcal{A}}(\mathrm{x}_i,\mathrm{z}_j) \\
		= {} & C_{\mathrm{DI}}^{\mathrm{gen}}(\mathrm{x}_i,\mathrm{z}_j,\omega_o)\,
		p_{\mathrm{DI}}^{\mathrm{gen}}(\mathrm{z}_j\mid\mathrm{x}_i),
	\end{aligned}
	\label{eq:reapp-di-general-mis-numerator}
\end{equation}
with
\begin{equation}
	C_{\mathrm{DI}}^{\mathrm{gen}}(\mathrm{x},\mathrm{z},\omega_o)
	=
	\frac{
		g(\mathrm{x},\mathrm{y}'(\mathrm{x},\mathrm{z}),\omega_o)
	}{
		p_{\mathcal{A}}(\mathrm{y}'\mid\mathrm{x})
	}.
	\label{eq:reapp-di-general-mis-factor}
\end{equation}
Therefore, the corresponding MIS estimator of direct illumination is
\begin{equation}
	\left\langle L_{\mathrm{DI}}(\mathrm{x}_i,\omega_o) \right\rangle_{\mathrm{MIS}}^{\mathrm{gen}}
	=
	U(\mathrm{x}_i,\omega_o)
	-
	\sum_{j=1}^{N}
	C_{\mathrm{DI}}^{\mathrm{gen}}(\mathrm{x}_i,\mathrm{z}_j,\omega_o)\,
	\frac{
		p_{\mathrm{DI}}^{\mathrm{gen}}(\mathrm{z}_j\mid\mathrm{x}_i)
	}{
		\sum_{k=1}^{N} p_{\mathrm{DI}}^{\mathrm{gen}}(\mathrm{z}_j\mid\mathrm{x}_k)
	}.
	\label{eq:reapp-di-general-mis-estimator}
\end{equation}

\section{Shadow Reuse Algorithm}
\label{app:shadow-reuse-algorithm}

\algref{alg:shadow-unbiased} gives the unbiased implementation of the shadow estimator in \eqref{eq:shadow-estimator}. Each target thread evaluates exact induced densities for the occluders shared by its local shared group, including the emitter hit test and the first-hit validation, then combines the contributions with a MIS formulation.

\begin{algorithm}[t]
	\caption{Unbiased shadow reuse for target~$\mathrm{x}_i$}
	\label{alg:shadow-unbiased}
	$B_i\leftarrow 0$\;
	\ForEach{shadow sample}{
		\tcp*[l]{Step 1: all threads in the shared group generate candidates}
		Sample a point $\mathrm{y}_i$ uniformly on the emitter domain $\mathcal{A}$\;
		Trace a closest-hit ray from $\mathrm{x}_i$ toward $\mathrm{y}_i$ with $t_{\max}=\|\mathrm{y}_i-\mathrm{x}_i\|$\;
		Store the first hit $\mathrm{z}_i$ and its oriented normal if the segment is blocked\;
		\;\tcp*[l]{Step 2: expose blocked candidates $\{\mathrm{z}_j\}$ in the shared group}
		Initialize $p_i[j]\leftarrow0$\;
		\tcp*[l]{$p_i[j]$: exact density $p_{\mathrm{Sh}}(\mathrm{z}_j\mid\mathrm{x}_i)$}
		\ForEach{source thread $j\in\mathcal{N}_i$}{
			Read $\mathrm{z}_j$ and its hit normal from source thread $j$\;
			\If{$\mathrm{z}_j$ is a valid blocked source hit}{
				Intersect the ray from $\mathrm{x}_i$ through $\mathrm{z}_j$ with the emitter surface\;
				\If{the intersection point $\mathrm{y}'(\mathrm{x}_i,\mathrm{z}_j)$ lies on $\mathcal{A}$}{
					Evaluate temporary density $q$ from the support and geometric terms of \eqref{eq:shadow-pdf}\;
					\If{$\mathrm{z}_j$ is the first hit from $\mathrm{x}_i$}{
						$p_i[j]\leftarrow q$\;
					}
				}
			}
		}
		\;\tcp*[l]{Step 3: this thread evaluates MIS contributions for target $\mathrm{x}_i$}
		\ForEach{candidate $j\in\mathcal{N}_i$}{
			Read $p_k[j]$ from source thread $k$ and accumulate $P_j\leftarrow\sum_{k=1}^{N}p_k[j]$\;
			\If{$p_i[j]>0 \land P_j>0$}{
				$B_i\leftarrow B_i+p_i[j]/P_j$\;
			}
		}
	}
	\Return{$1-B_i/M$}\tcp*[l]{$M$: number of shadow samples accumulated for target $\mathrm{x}_i$}
\end{algorithm}

The biased shadow variant follows the same data flow, but replaces the exact density with $\widetilde{p}_{\mathrm{Sh}}$ by assuming local first-hit consistency. This removes the first-hit validation inside Step 2 while retaining the emitter hit test and Jacobian evaluation.

\section{Implementation Details}
\label{app:implementation-details}

\subsection{Shared-Group Patterns}
\label{app:shared-group-patterns}
\begin{figure}[ht]
	\centering
	\includegraphics[width=0.85\linewidth]{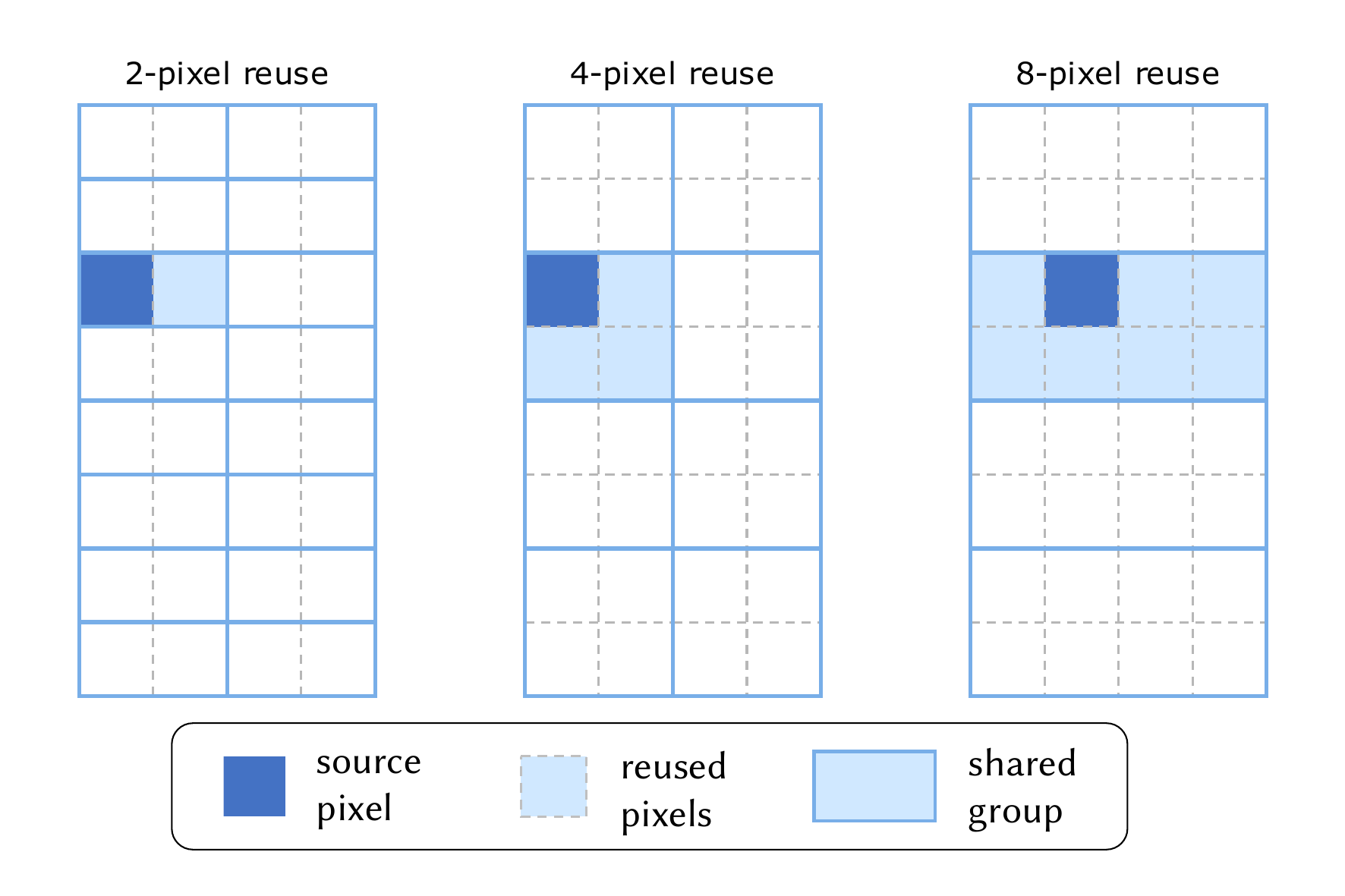}
	\caption{Illustration of fixed shared-group reuse patterns for $N=2,4,8$.}
	\label{fig:shared-group-patterns}
	\Description{Illustration of fixed shared-group reuse patterns for N=2,4,8.}
\end{figure}

\figref{fig:shared-group-patterns} illustrates the fixed shared-group reuse patterns for $N=2,4,8$.

\section{Additional Results for AO}
\label{app:additional-results-ao}

\subsection{Equal-spp Comparison}
\label{app:ao-equal-spp-all}
\begin{figure*}[t]
	\centering
	\includegraphics[width=\linewidth]{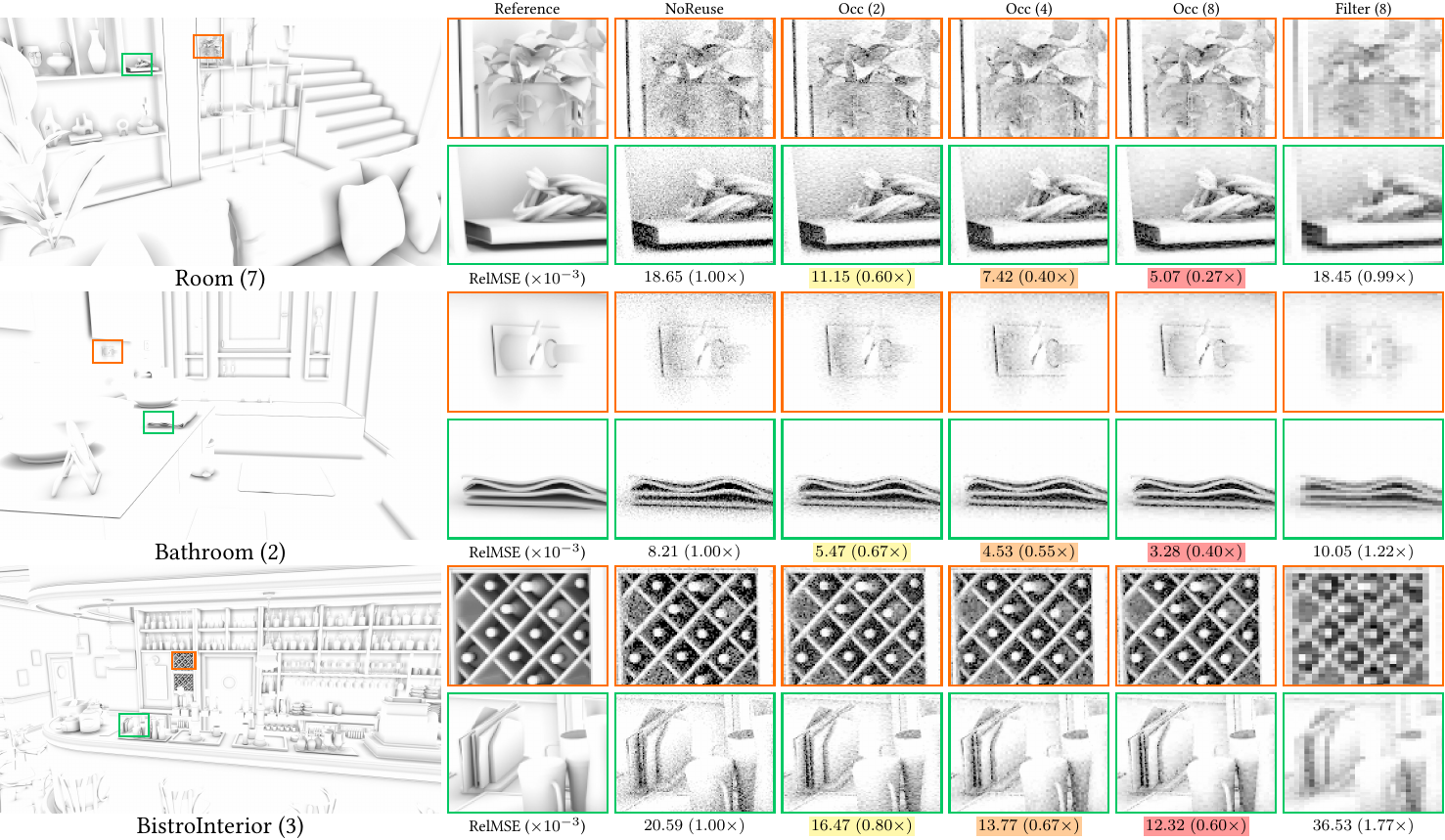}
	\caption{Equal-spp AO comparison at 10 spp. The figure shows that occlusion-point reuse reduces variance while preserving local occlusion detail better than final-value filtering.}
	\label{fig:ao-main-equal-10-spp-relmse}
	\Description{ao equal 10 spp comparison.}
\end{figure*}

We report AO fixed-spp (10 spp) results in this subsection.
Table~\ref{tab:ao-equal-spp-10} summarizes per-scene averaged RelMSE ratios, and \figref{fig:ao-main-equal-10-spp-relmse} shows the corresponding all-scene visual comparisons.
Here, Occ denotes the practical biased reuse variants (e.g., Occ~(2), Occ~(4), and Occ~(8)).
Across scenes, Occ reuse is consistently better than final-value filtering in both detail fidelity and error.
The key reason is that our reuse is built on occlusion-point transport and therefore naturally preserves local geometric detail, while final-value filtering tends to smooth across thin structures and contact regions, which can increase bias and RelMSE.

\begin{table}[tb]
\centering
\caption{Equal-spp RTAO comparison. Each value is the average RelMSE ratio to no-reuse RTAO over multiple views of the same scene; lower is better. Configuration: 10 spp with blue-noise sampling. We color code the {\fboxsep1pt\colorbox[RGB]{255,153,153}{first}}, {\fboxsep1pt\colorbox[RGB]{255,204,153}{second}}, and {\fboxsep1pt\colorbox[RGB]{255,248,173}{third}} lowest numbers.}
\begin{adjustbox}{max width=\columnwidth}
\begin{tabular}{lcccc}
\toprule
Scene & Occ (2) & Occ (4) & Occ (8) & Filter (8) \\
\midrule
Airport & \cellcolor[RGB]{255,248,173}0.81x & \cellcolor[RGB]{255,204,153}0.65x & \cellcolor[RGB]{255,153,153}0.56x & 2.66x \\
Bathroom & \cellcolor[RGB]{255,248,173}0.67x & \cellcolor[RGB]{255,204,153}0.50x & \cellcolor[RGB]{255,153,153}0.40x & 1.45x \\
BistroExterior & \cellcolor[RGB]{255,248,173}0.79x & \cellcolor[RGB]{255,204,153}0.66x & \cellcolor[RGB]{255,153,153}0.59x & 1.79x \\
BistroInterior & \cellcolor[RGB]{255,248,173}0.83x & \cellcolor[RGB]{255,204,153}0.77x & \cellcolor[RGB]{255,153,153}0.68x & 1.95x \\
CartoonRoom & \cellcolor[RGB]{255,248,173}0.76x & \cellcolor[RGB]{255,204,153}0.56x & \cellcolor[RGB]{255,153,153}0.47x & 0.97x \\
FirePlaceRoom & \cellcolor[RGB]{255,248,173}0.67x & \cellcolor[RGB]{255,204,153}0.44x & \cellcolor[RGB]{255,153,153}0.35x & 1.13x \\
Room & \cellcolor[RGB]{255,248,173}0.70x & \cellcolor[RGB]{255,204,153}0.52x & \cellcolor[RGB]{255,153,153}0.43x & 0.88x \\
Sibenik & 0.58x & \cellcolor[RGB]{255,204,153}0.34x & \cellcolor[RGB]{255,153,153}0.23x & \cellcolor[RGB]{255,248,173}0.35x \\
\midrule
Average & \cellcolor[RGB]{255,248,173}0.72x & \cellcolor[RGB]{255,204,153}0.55x & \cellcolor[RGB]{255,153,153}0.46x & 1.40x \\
\bottomrule
\end{tabular}
\end{adjustbox}
\label{tab:ao-equal-spp-10}
\end{table}

\subsection{Per-Scene Efficiency Results}
\label{app:per-scene-efficiency}
\tabref{tab:app-ao-per-scene-efficiency} reports per-scene runtime ratios to no-reuse RTAO for AO reuse variants.
For biased Occ variants, the trend is consistent across scenes: they add only a small overhead, and the overhead increases only slightly as shared-group size grows, which keeps these settings practical.
For unbiased variants, the overhead increase is much larger because each reused candidate requires exact first-hit validation.

\begin{table}[tb]
\centering
\caption{Per-scene runtime ratio to no-reuse RTAO. Values are averaged over tested views of each scene, with biased and unbiased variants reported separately (lower is faster).}
\label{tab:app-ao-per-scene-efficiency}
\begin{adjustbox}{max width=\columnwidth}
\begin{tabular}{llcccc}
\toprule
Scene & Type & Occ (2) & Occ (4) & Occ (8) & Filter (8) \\
\midrule
Bathroom & Biased & 1.064x & 1.090x & 1.207x & 1.024x \\
 & Unbiased & 1.323x & 1.594x & 2.203x & -- \\
\midrule
BistroExterior & Biased & 1.041x & 1.056x & 1.088x & 1.036x \\
 & Unbiased & 1.379x & 1.914x & 3.064x & -- \\
\midrule
BistroInterior & Biased & 1.060x & 1.095x & 1.156x & 1.037x \\
 & Unbiased & 1.350x & 1.710x & 2.453x & -- \\
\midrule
CartoonRoom & Biased & 1.136x & 1.162x & 1.237x & 1.091x \\
 & Unbiased & 1.548x & 2.063x & 3.122x & -- \\
\midrule
FirePlaceRoom & Biased & 1.151x & 1.192x & 1.297x & 1.115x \\
 & Unbiased & 1.564x & 2.006x & 2.940x & -- \\
\midrule
Room & Biased & 1.085x & 1.112x & 1.213x & 1.045x \\
 & Unbiased & 1.414x & 1.814x & 2.613x & -- \\
\midrule
Sibenik & Biased & 1.035x & 1.058x & 1.150x & 1.010x \\
 & Unbiased & 1.276x & 1.589x & 2.197x & -- \\
\midrule
Airport & Biased & 1.034x & 1.058x & 1.138x & 1.014x \\
 & Unbiased & 1.237x & 1.487x & 2.005x & -- \\
\bottomrule
\end{tabular}
\end{adjustbox}
\end{table}

\paragraph{Absolute runtime.}
For completeness, we also report the absolute runtime of each AO method. To obtain stable measurements, we run each method at 200 spp and repeat the measurement 100 times, then divide the total time by the accumulated sample count to estimate the runtime cost of 1 spp. The reported values are in milliseconds and averaged over the evaluated variants of each scene; \tabref{tab:re-ao-absolute-runtime-per-scene} summarizes the AO results.
\begin{table}[tb]
\centering
\caption{Per-scene estimated 1-spp RTAO pass time in milliseconds. Each value is obtained by timing the method at 200 spp, repeating the measurement 100 times, and dividing by the accumulated sample count; results are then averaged over the evaluated variants of each scene, with biased and unbiased implementations reported separately.}
\label{tab:re-ao-absolute-runtime-per-scene}
\begin{adjustbox}{max width=\columnwidth}
\begin{tabular}{llccccc}
\toprule
Scene & Type & NoReuse & Occ (2) & Occ (4) & Occ (8) & Filter (8) \\
\midrule
Bathroom & Biased & -- & 0.192 & 0.196 & 0.217 & 0.185 \\
 & Unbiased & 0.180 & 0.238 & 0.287 & 0.396 & -- \\
\midrule
BistroExterior & Biased & -- & 0.398 & 0.403 & 0.416 & 0.396 \\
 & Unbiased & 0.382 & 0.529 & 0.737 & 1.182 & -- \\
\midrule
BistroInterior & Biased & -- & 0.258 & 0.266 & 0.281 & 0.252 \\
 & Unbiased & 0.243 & 0.328 & 0.417 & 0.598 & -- \\
\midrule
CartoonRoom & Biased & -- & 0.267 & 0.273 & 0.289 & 0.257 \\
 & Unbiased & 0.237 & 0.359 & 0.477 & 0.723 & -- \\
\midrule
FirePlaceRoom & Biased & -- & 0.209 & 0.216 & 0.235 & 0.202 \\
 & Unbiased & 0.181 & 0.284 & 0.365 & 0.535 & -- \\
\midrule
Room & Biased & -- & 0.214 & 0.219 & 0.239 & 0.206 \\
 & Unbiased & 0.197 & 0.279 & 0.358 & 0.516 & -- \\
\midrule
Sibenik & Biased & -- & 0.193 & 0.198 & 0.215 & 0.189 \\
 & Unbiased & 0.187 & 0.238 & 0.297 & 0.411 & -- \\
\midrule
Airport & Biased & -- & 0.197 & 0.202 & 0.217 & 0.193 \\
 & Unbiased & 0.191 & 0.236 & 0.284 & 0.382 & -- \\
\bottomrule
\end{tabular}
\end{adjustbox}
\end{table}

\subsection{Results of All scenes in SVGF Comparison}
\label{app:ao-all-scenes-svgf-comparison}
\tabref{tab:app-ao-svgf} reports the per-scene RelMSE ratios for the SVGF comparison and their average across all scenes.
The AO budget follows the same setup as the main text: a 2~ms AO pass, with optional SVGF as an additional post-process.
Compared with the 2~ms equal-time AO table in the main paper (\tabref{tab:ao-equal-time-2ms}), the Occ~(8) numbers here are slightly worse.
This is expected because Occ~(8) is a biased estimator: increasing the budget does not guarantee convergence to the same reference as the unbiased/no-reuse estimator, so a small residual error can remain.

\begin{table}[tb]
\centering
\caption{AO denoising comparison across all scenes. Values are RelMSE ratios to the no-reuse RTAO baseline (lower is better), evaluated with total time set to 2~ms + scene-dependent SVGF time, using blue-noise sampling. The {\fboxsep1pt\colorbox[RGB]{255,153,153}{first}}, {\fboxsep1pt\colorbox[RGB]{255,204,153}{second}}, and {\fboxsep1pt\colorbox[RGB]{255,248,173}{third}} lowest values in each row are highlighted.}
\label{tab:app-ao-svgf}
\begin{adjustbox}{max width=\columnwidth}
\begin{tabular}{lccc}
\toprule
Scene & Occ (8) & NoReuse + SVGF & Occ (8) + SVGF \\
\midrule
Airport & \cellcolor[RGB]{255,153,153}0.74x & \cellcolor[RGB]{255,248,173}7.76x & \cellcolor[RGB]{255,204,153}7.34x \\
Bathroom & \cellcolor[RGB]{255,153,153}0.51x & \cellcolor[RGB]{255,248,173}8.14x & \cellcolor[RGB]{255,204,153}7.82x \\
BistroExterior & \cellcolor[RGB]{255,153,153}0.64x & \cellcolor[RGB]{255,204,153}3.67x & \cellcolor[RGB]{255,248,173}3.79x \\
BistroInterior & \cellcolor[RGB]{255,153,153}0.83x & \cellcolor[RGB]{255,204,153}5.53x & \cellcolor[RGB]{255,248,173}5.73x \\
CartoonRoom & \cellcolor[RGB]{255,153,153}0.59x & \cellcolor[RGB]{255,248,173}1.71x & \cellcolor[RGB]{255,204,153}1.58x \\
FirePlaceRoom & \cellcolor[RGB]{255,153,153}0.50x & \cellcolor[RGB]{255,248,173}2.96x & \cellcolor[RGB]{255,204,153}2.83x \\
Room & \cellcolor[RGB]{255,153,153}0.54x & \cellcolor[RGB]{255,248,173}3.75x & \cellcolor[RGB]{255,204,153}3.62x \\
Sibenik & \cellcolor[RGB]{255,153,153}0.27x & \cellcolor[RGB]{255,248,173}1.55x & \cellcolor[RGB]{255,204,153}1.38x \\
\midrule
Average & \cellcolor[RGB]{255,153,153}0.58x & \cellcolor[RGB]{255,248,173}4.38x & \cellcolor[RGB]{255,204,153}4.26x \\
\bottomrule
\end{tabular}
\end{adjustbox}
\end{table}

\subsection{Convergence Results}
\label{app:ao-convergence}
\begin{figure*}[htbp]
	\centering
	\includegraphics[width=0.9\linewidth]{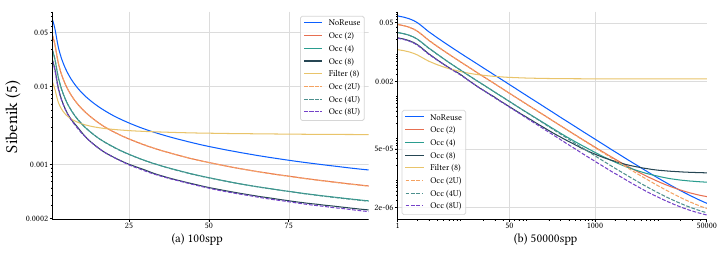}
	\caption{AO convergence on one representative scene (white-noise sampling, RelMSE vs spp), shown with log-scaled axes. (a) Low-spp view up to 100 spp: Under biased settings, filter is competitive only at very low spp, while biased Occ becomes better as spp increases. (b) Full-range view: Occ unbiased converges to the reference, and larger reuse groups converge faster than no reuse.}
	\label{fig:ao-convergence}
	\Description{AO convergence curves with a low-spp zoom and full-range view, comparing filter, biased Occ, unbiased Occ, and no reuse.}
\end{figure*}

We report AO convergence by plotting RelMSE versus spp under white-noise sampling.
\figref{fig:ao-convergence}(a) shows the low-spp regime (up to 100 spp). At very low spp ($<5$), Filter can look better because strong smoothing suppresses dominant noise. As spp increases, the biased Occ variants consistently outperform the Filter-based biased result.
\figref{fig:ao-convergence}(b) provides two complementary observations: first, Occ unbiased converges to the reference, confirming that the unbiased reuse estimator is indeed unbiased; second, reuse improves convergence speed compared with no reuse, and larger reuse groups further accelerate convergence.

\section{Additional Results for Shadows}
\label{app:additional-results-shadows}

\subsection{More Results for Equal-Time Comparison}
\label{app:shadow-equal-time-2ms-and-10ms}

\begin{table}[tb]
\centering
\caption{Equal-time area-light shadow comparison. Each value is the average RelMSE ratio to no-reuse ray-traced shadows over multiple views of the same asset; lower is better. Configuration: 2~ms budget with blue-noise sampling. We color code the {\fboxsep1pt\colorbox[RGB]{255,153,153}{first}}, {\fboxsep1pt\colorbox[RGB]{255,204,153}{second}}, and {\fboxsep1pt\colorbox[RGB]{255,248,173}{third}} lowest numbers.}
\begin{adjustbox}{max width=\columnwidth}
\begin{tabular}{lcccc}
\toprule
Scene & Occ (2) & Occ (4) & Occ (8) & Light (8) \\
\midrule
Bike & 0.73x & \cellcolor[RGB]{255,248,173}0.38x & \cellcolor[RGB]{255,153,153}0.31x & \cellcolor[RGB]{255,204,153}0.33x \\
BistroInterior & \cellcolor[RGB]{255,204,153}0.80x & \cellcolor[RGB]{255,153,153}0.73x & \cellcolor[RGB]{255,248,173}0.99x & 2.62x \\
CartoonRoom & \cellcolor[RGB]{255,248,173}0.81x & \cellcolor[RGB]{255,153,153}0.67x & \cellcolor[RGB]{255,204,153}0.78x & 1.45x \\
Room & \cellcolor[RGB]{255,248,173}0.82x & \cellcolor[RGB]{255,153,153}0.71x & \cellcolor[RGB]{255,204,153}0.74x & 0.97x \\
SunTemple & 0.65x & \cellcolor[RGB]{255,204,153}0.38x & \cellcolor[RGB]{255,153,153}0.36x & \cellcolor[RGB]{255,248,173}0.57x \\
Item Crab & \cellcolor[RGB]{255,248,173}0.63x & \cellcolor[RGB]{255,153,153}0.44x & \cellcolor[RGB]{255,204,153}0.46x & 1.92x \\
Item Hookah & 0.62x & \cellcolor[RGB]{255,204,153}0.40x & \cellcolor[RGB]{255,153,153}0.35x & \cellcolor[RGB]{255,248,173}0.56x \\
Item Plum & \cellcolor[RGB]{255,248,173}0.63x & \cellcolor[RGB]{255,153,153}0.41x & \cellcolor[RGB]{255,204,153}0.43x & 0.75x \\
Item Tree & \cellcolor[RGB]{255,248,173}0.65x & \cellcolor[RGB]{255,204,153}0.44x & \cellcolor[RGB]{255,153,153}0.37x & 0.70x \\
Item Viking & \cellcolor[RGB]{255,248,173}0.67x & \cellcolor[RGB]{255,204,153}0.41x & \cellcolor[RGB]{255,153,153}0.39x & 0.80x \\
\midrule
Average & \cellcolor[RGB]{255,248,173}0.70x & \cellcolor[RGB]{255,153,153}0.50x & \cellcolor[RGB]{255,204,153}0.52x & 1.07x \\
\bottomrule
\end{tabular}
\end{adjustbox}
\label{tab:shadow-equal-time-2ms}
\end{table}


\begin{table}[tb]
\centering
\caption{Equal-time area-light shadow comparison. Each value is the average RelMSE ratio to no-reuse ray-traced shadows over multiple views of the same asset; lower is better. Configuration: 10~ms budget with blue-noise sampling. We color code the {\fboxsep1pt\colorbox[RGB]{255,153,153}{first}}, {\fboxsep1pt\colorbox[RGB]{255,204,153}{second}}, and {\fboxsep1pt\colorbox[RGB]{255,248,173}{third}} lowest numbers.}
\begin{adjustbox}{max width=\columnwidth}
\begin{tabular}{lcccc}
\toprule
Scene & Occ (2) & Occ (4) & Occ (8) & Light (8) \\
\midrule
Bike & \cellcolor[RGB]{255,248,173}0.68x & \cellcolor[RGB]{255,204,153}0.39x & \cellcolor[RGB]{255,153,153}0.30x & 0.84x \\
BistroInterior & \cellcolor[RGB]{255,153,153}0.75x & \cellcolor[RGB]{255,204,153}0.75x & \cellcolor[RGB]{255,248,173}1.30x & 11.37x \\
CartoonRoom & \cellcolor[RGB]{255,204,153}0.80x & \cellcolor[RGB]{255,153,153}0.68x & \cellcolor[RGB]{255,248,173}0.92x & 5.57x \\
Room & \cellcolor[RGB]{255,248,173}0.81x & \cellcolor[RGB]{255,153,153}0.72x & \cellcolor[RGB]{255,204,153}0.78x & 2.64x \\
SunTemple & \cellcolor[RGB]{255,248,173}0.66x & \cellcolor[RGB]{255,153,153}0.42x & \cellcolor[RGB]{255,204,153}0.47x & 2.15x \\
Item Crab & \cellcolor[RGB]{255,248,173}0.65x & \cellcolor[RGB]{255,153,153}0.46x & \cellcolor[RGB]{255,204,153}0.49x & 8.11x \\
Item Hookah & \cellcolor[RGB]{255,248,173}0.63x & \cellcolor[RGB]{255,204,153}0.41x & \cellcolor[RGB]{255,153,153}0.36x & 2.12x \\
Item Plum & \cellcolor[RGB]{255,248,173}0.64x & \cellcolor[RGB]{255,204,153}0.43x & \cellcolor[RGB]{255,153,153}0.42x & 2.96x \\
Item Tree & \cellcolor[RGB]{255,248,173}0.67x & \cellcolor[RGB]{255,204,153}0.44x & \cellcolor[RGB]{255,153,153}0.40x & 2.83x \\
Item Viking & \cellcolor[RGB]{255,248,173}0.68x & \cellcolor[RGB]{255,204,153}0.43x & \cellcolor[RGB]{255,153,153}0.42x & 3.37x \\
\midrule
Average & \cellcolor[RGB]{255,248,173}0.70x & \cellcolor[RGB]{255,153,153}0.51x & \cellcolor[RGB]{255,204,153}0.59x & 4.20x \\
\bottomrule
\end{tabular}
\end{adjustbox}
\label{tab:shadow-equal-time-10ms}
\end{table}

\tabref{tab:shadow-equal-time-2ms} and \tabref{tab:shadow-equal-time-10ms} provide additional equal-time comparisons.
The overall trend is consistent with the main text: occlusion-point reuse is generally more robust than Light~(8), because its approximation is built on first-hit occluder consistency, which is more reliable than direct visibility consistency in many regions.

As the time budget increases, bias affects RelMSE more clearly. For Light~(8), the underlying approximation often breaks in penumbrae and around visibility transitions, so the error degradation is much larger at higher budgets.
For Occ, the approximation is more accurate in most cases, so the degradation is noticeably smaller. However, in complex scenes such as BistroInterior, the bias of Occ~(8) can still accumulate with higher sampling budgets, and its RelMSE may become worse than no reuse.

\subsection{Equal-spp Comparison}
\label{app:shadow-equal-spp-10}
\begin{figure*}[t]
	\centering
	\includegraphics[width=\linewidth]{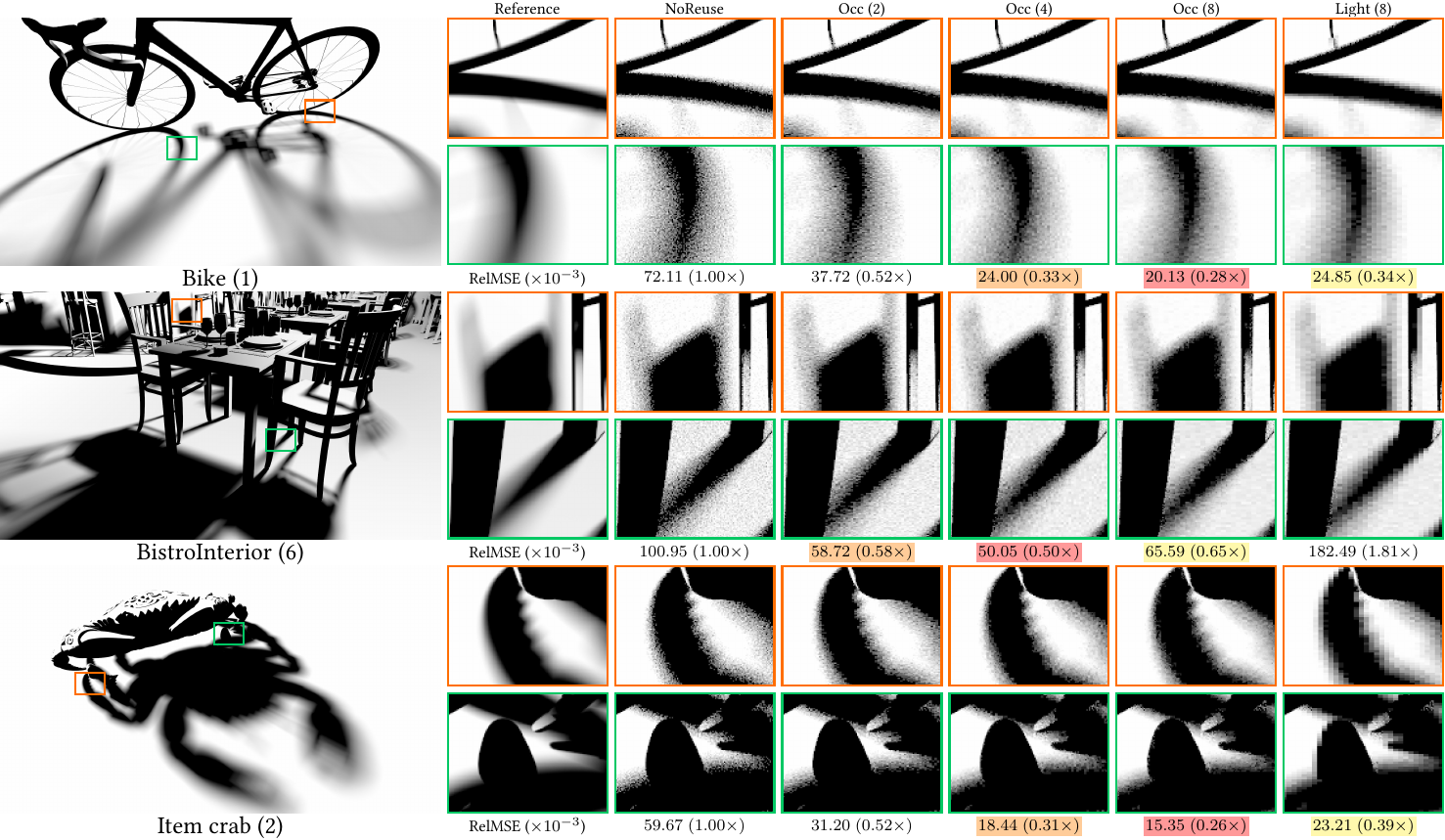}
	\caption{Equal-spp area-light shadow comparison at 10 spp. Error maps compare no reuse, light-sample reuse, and occlusion-point reuse against a high-spp ray-traced reference.}
	\label{fig:shadow-main-equal-10-spp-relmse}
	\Description{shadow equal 10 spp comparison.}
\end{figure*}

We also report shadow fixed-spp (10 spp) results in this subsection.
Table~\ref{tab:shadow-equal-spp-10} summarizes per-scene averaged RelMSE ratios, and \figref{fig:shadow-main-equal-10-spp-relmse} presents representative visual comparisons.
Here, Occ denotes the practical biased reuse variants (Occ~(2), Occ~(4), and Occ~(8)).
The trend is consistent across scenes: Occ reuse significantly improves over no reuse.

Compared with Light~(8), Occ generally preserves boundaries and penumbra transitions better.
The main reason is that Occ relies on first-hit occluder consistency, which is usually a more reliable local approximation than directly assuming binary visibility consistency for reused light samples.

As in the equal-time setting, shared-group size also introduces a trade-off in complex scenes: larger groups can reduce variance more aggressively, but they may increase bias when local occluder consistency is weaker.
This explains why Occ~(8) is not always better than Occ~(4) in difficult scene configurations.

\begin{table}[tb]
\centering
\caption{Equal-spp ray-traced shadow comparison. Each value is the average RelMSE ratio to no-reuse ray-traced shadows over multiple views of the same scene; lower is better. Configuration: 10 spp with blue-noise sampling. We color code the {\fboxsep1pt\colorbox[RGB]{255,153,153}{first}}, {\fboxsep1pt\colorbox[RGB]{255,204,153}{second}}, and {\fboxsep1pt\colorbox[RGB]{255,248,173}{third}} lowest numbers.}
\begin{adjustbox}{max width=\columnwidth}
\begin{tabular}{lcccc}
\toprule
Scene & Occ (2) & Occ (4) & Occ (8) & Light (8) \\
\midrule
Bike & 0.56x & \cellcolor[RGB]{255,204,153}0.35x & \cellcolor[RGB]{255,153,153}0.25x & \cellcolor[RGB]{255,248,173}0.49x \\
BistroInterior & \cellcolor[RGB]{255,204,153}0.65x & \cellcolor[RGB]{255,153,153}0.63x & \cellcolor[RGB]{255,248,173}1.01x & 7.98x \\
CartoonRoom & \cellcolor[RGB]{255,248,173}0.68x & \cellcolor[RGB]{255,153,153}0.56x & \cellcolor[RGB]{255,204,153}0.66x & 3.09x \\
Room & \cellcolor[RGB]{255,248,173}0.74x & \cellcolor[RGB]{255,153,153}0.63x & \cellcolor[RGB]{255,204,153}0.65x & 1.82x \\
SunTemple & \cellcolor[RGB]{255,248,173}0.54x & \cellcolor[RGB]{255,153,153}0.33x & \cellcolor[RGB]{255,204,153}0.34x & 1.27x \\
Item Crab & \cellcolor[RGB]{255,248,173}0.56x & \cellcolor[RGB]{255,204,153}0.38x & \cellcolor[RGB]{255,153,153}0.37x & 4.53x \\
Item Hookah & \cellcolor[RGB]{255,248,173}0.54x & \cellcolor[RGB]{255,204,153}0.34x & \cellcolor[RGB]{255,153,153}0.28x & 1.16x \\
Item Plum & \cellcolor[RGB]{255,248,173}0.54x & \cellcolor[RGB]{255,204,153}0.33x & \cellcolor[RGB]{255,153,153}0.30x & 1.62x \\
Item Tree & \cellcolor[RGB]{255,248,173}0.55x & \cellcolor[RGB]{255,204,153}0.35x & \cellcolor[RGB]{255,153,153}0.29x & 1.53x \\
Item Viking & \cellcolor[RGB]{255,248,173}0.53x & \cellcolor[RGB]{255,204,153}0.33x & \cellcolor[RGB]{255,153,153}0.30x & 1.89x \\
\midrule
Average & \cellcolor[RGB]{255,248,173}0.59x & \cellcolor[RGB]{255,153,153}0.42x & \cellcolor[RGB]{255,204,153}0.44x & 2.54x \\
\bottomrule
\end{tabular}
\end{adjustbox}
\label{tab:shadow-equal-spp-10}
\end{table}

\subsection{Per-Scene Efficiency Results}
\label{app:shadow-per-scene-efficiency}
\tabref{tab:app-shadow-per-scene-efficiency} reports per-scene runtime ratios to no-reuse RT shadow.
For biased variants, Occ introduces only modest overhead and remains practical across scenes.
For unbiased variants, all methods are significantly more expensive, and Occ~(8U) is faster than Light~(8U).
This gap is consistent with our implementation path: replacing shading-point-to-light visibility checks with occluder-to-light checks yields shorter traced segments in many reused candidates, reducing total tracing cost.

\begin{table}[tb]
\centering
\caption{Per-scene runtime ratio to no-reuse RT shadow. Values are averaged over tested views of each scene, with biased and unbiased variants reported separately (lower is faster).}
\label{tab:app-shadow-per-scene-efficiency}
\begin{adjustbox}{max width=\columnwidth}
\begin{tabular}{llcccc}
\toprule
Scene & Type & Occ (2) & Occ (4) & Occ (8) & Light (8) \\
\midrule
Bike & Biased & 1.233x & 1.166x & 1.265x & 1.038x \\
 & Unbiased & 1.369x & 1.843x & 2.808x & 7.258x \\
\midrule
BistroInterior & Biased & 1.164x & 1.196x & 1.221x & 1.015x \\
 & Unbiased & 1.488x & 2.195x & 3.654x & 7.558x \\
\midrule
CartoonRoom & Biased & 1.171x & 1.205x & 1.289x & 1.026x \\
 & Unbiased & 1.533x & 2.248x & 3.642x & 7.069x \\
\midrule
Room & Biased & 1.088x & 1.121x & 1.178x & 1.041x \\
 & Unbiased & 1.308x & 1.732x & 2.528x & 6.813x \\
\midrule
SunTemple & Biased & 1.214x & 1.230x & 1.286x & 1.027x \\
 & Unbiased & 1.753x & 2.882x & 5.147x & 7.736x \\
\midrule
Item Crab & Biased & 1.172x & 1.219x & 1.276x & 1.062x \\
 & Unbiased & 1.435x & 1.927x & 2.847x & 7.829x \\
\midrule
Item Hookah & Biased & 1.160x & 1.201x & 1.295x & 1.062x \\
 & Unbiased & 1.358x & 1.646x & 2.197x & 7.938x \\
\midrule
Item Plum & Biased & 1.197x & 1.293x & 1.396x & 1.095x \\
 & Unbiased & 1.535x & 2.010x & 2.874x & 7.983x \\
\midrule
Item Tree & Biased & 1.211x & 1.309x & 1.429x & 1.093x \\
 & Unbiased & 1.520x & 1.873x & 2.519x & 7.905x \\
\midrule
Item Viking & Biased & 1.270x & 1.321x & 1.411x & 1.073x \\
 & Unbiased & 1.640x & 2.128x & 3.023x & 7.977x \\
\bottomrule
\end{tabular}
\end{adjustbox}
\end{table}

\paragraph{Absolute runtime.}
For completeness, we also report the absolute runtime of each shadow method using the same protocol: each method is run at 200 spp and repeated 100 times, and the total time is divided by the accumulated sample count to estimate the runtime cost of 1 spp. The reported values are in milliseconds and averaged over the evaluated variants of each scene; \tabref{tab:re-shadow-absolute-runtime-per-scene} summarizes the shadow results.
\begin{table}[tb]
\centering
\caption{Per-scene estimated 1-spp ray-traced shadow pass time in milliseconds. Each value is obtained by timing the method at 200 spp, repeating the measurement 100 times, and dividing by the accumulated sample count; results are then averaged over the evaluated variants of each scene, with biased and unbiased implementations reported separately.}
\label{tab:re-shadow-absolute-runtime-per-scene}
\begin{adjustbox}{max width=\columnwidth}
\begin{tabular}{llccccc}
\toprule
Scene & Type & NoReuse & Occ (2) & Occ (4) & Occ (8) & Light (8) \\
\midrule
Bike & Biased & -- & 0.309 & 0.292 & 0.317 & 0.260 \\
 & Unbiased & 0.251 & 0.343 & 0.462 & 0.704 & 1.820 \\
\midrule
BistroInterior & Biased & -- & 0.389 & 0.399 & 0.407 & 0.340 \\
 & Unbiased & 0.335 & 0.495 & 0.729 & 1.211 & 2.540 \\
\midrule
CartoonRoom & Biased & -- & 0.324 & 0.333 & 0.355 & 0.284 \\
 & Unbiased & 0.277 & 0.424 & 0.622 & 1.010 & 1.971 \\
\midrule
Room & Biased & -- & 0.329 & 0.339 & 0.355 & 0.314 \\
 & Unbiased & 0.302 & 0.395 & 0.522 & 0.762 & 2.062 \\
\midrule
SunTemple & Biased & -- & 0.364 & 0.369 & 0.385 & 0.308 \\
 & Unbiased & 0.299 & 0.533 & 0.882 & 1.582 & 2.335 \\
\midrule
Item Crab & Biased & -- & 0.327 & 0.340 & 0.356 & 0.297 \\
 & Unbiased & 0.279 & 0.401 & 0.538 & 0.794 & 2.187 \\
\midrule
Item Hookah & Biased & -- & 0.293 & 0.303 & 0.327 & 0.268 \\
 & Unbiased & 0.252 & 0.342 & 0.415 & 0.554 & 2.002 \\
\midrule
Item Plum & Biased & -- & 0.316 & 0.341 & 0.368 & 0.288 \\
 & Unbiased & 0.263 & 0.405 & 0.531 & 0.760 & 2.104 \\
\midrule
Item Tree & Biased & -- & 0.309 & 0.335 & 0.365 & 0.279 \\
 & Unbiased & 0.255 & 0.389 & 0.479 & 0.646 & 2.017 \\
\midrule
Item Viking & Biased & -- & 0.340 & 0.354 & 0.378 & 0.287 \\
 & Unbiased & 0.268 & 0.439 & 0.570 & 0.811 & 2.138 \\
\bottomrule
\end{tabular}
\end{adjustbox}
\end{table}

\subsection{Results of All scenes in SVGF Comparison}
\label{app:shadow-all-scenes-svgf-comparison}
\tabref{tab:app-shadow-svgf-5ms} reports per-scene RelMSE ratios for the shadow SVGF comparison and the average across all scenes.
The setup follows the main text: a 5~ms shadow pass, with optional SVGF as an additional post-process.

\begin{table}[tb]
\centering
\caption{Per-scene shadow SVGF comparison. Values are RelMSE ratios to no-reuse ray-traced shadows (lower is better), averaged over evaluated views per scene. Configuration: total time set to 5~ms + scene-dependent SVGF time, with blue-noise sampling. We color code the {\fboxsep1pt\colorbox[RGB]{255,153,153}{first}}, {\fboxsep1pt\colorbox[RGB]{255,204,153}{second}}, and {\fboxsep1pt\colorbox[RGB]{255,248,173}{third}} lowest numbers.}
\label{tab:app-shadow-svgf-5ms}
\begin{adjustbox}{max width=\columnwidth}
\begin{tabular}{lccc}
\toprule
Scene & Occ (8) & NoReuse + SVGF & Occ (8) + SVGF \\
\midrule
Bike & \cellcolor[RGB]{255,153,153}0.31x & \cellcolor[RGB]{255,248,173}6.49x & \cellcolor[RGB]{255,204,153}6.24x \\
BistroInterior & \cellcolor[RGB]{255,153,153}1.24x & \cellcolor[RGB]{255,248,173}28.16x & \cellcolor[RGB]{255,204,153}28.14x \\
CartoonRoom LargeLight & \cellcolor[RGB]{255,153,153}0.89x & \cellcolor[RGB]{255,248,173}11.90x & \cellcolor[RGB]{255,204,153}11.86x \\
Room & \cellcolor[RGB]{255,153,153}0.78x & \cellcolor[RGB]{255,248,173}8.30x & \cellcolor[RGB]{255,204,153}7.68x \\
SunTemple LargeLight & \cellcolor[RGB]{255,153,153}0.45x & \cellcolor[RGB]{255,204,153}5.49x & \cellcolor[RGB]{255,248,173}5.78x \\
Item Crab & \cellcolor[RGB]{255,153,153}0.50x & \cellcolor[RGB]{255,248,173}14.23x & \cellcolor[RGB]{255,204,153}14.05x \\
Item Hookah & \cellcolor[RGB]{255,153,153}0.35x & \cellcolor[RGB]{255,248,173}10.11x & \cellcolor[RGB]{255,204,153}9.93x \\
Item Plum & \cellcolor[RGB]{255,153,153}0.42x & \cellcolor[RGB]{255,204,153}10.83x & \cellcolor[RGB]{255,248,173}11.03x \\
Item Tree & \cellcolor[RGB]{255,153,153}0.41x & \cellcolor[RGB]{255,248,173}15.85x & \cellcolor[RGB]{255,204,153}15.19x \\
Item Viking & \cellcolor[RGB]{255,153,153}0.42x & \cellcolor[RGB]{255,204,153}14.37x & \cellcolor[RGB]{255,248,173}15.40x \\
\midrule
Average & \cellcolor[RGB]{255,153,153}0.58x & \cellcolor[RGB]{255,248,173}12.57x & \cellcolor[RGB]{255,204,153}12.53x \\
\bottomrule
\end{tabular}
\end{adjustbox}
\end{table}

\subsection{Negative Estimates}
\label{app:shadow-negative-estimates}

\begin{figure*}[htbp]
	\centering
	\includegraphics[width=\linewidth]{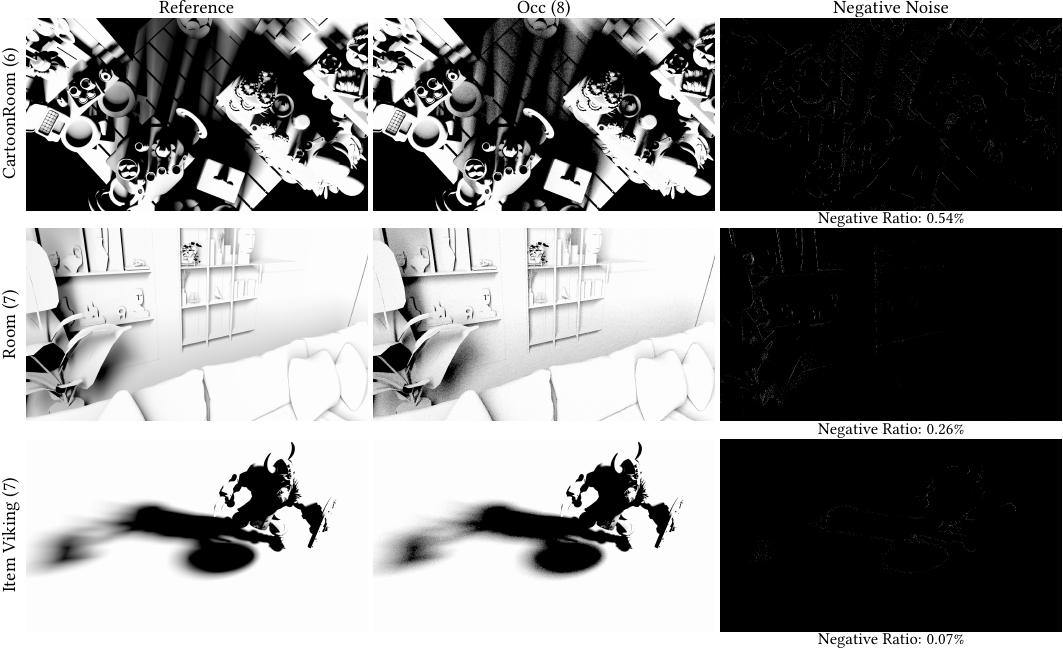}
	\caption{Visualization of negative estimates for the practical biased shadow implementation. For RelMSE evaluation throughout the paper, we use the raw estimator without clamping negative values. In this visualization, pixels with negative estimates are highlighted in white to show where such cases occur spatially; the annotated proportion of negative pixels is below $1\%$.}
	\label{fig:re-shadow-negative-visualization}
	\Description{Visualization of spatial regions where the practical biased shadow estimator becomes negative; negative pixels are shown in white.}
\end{figure*}

In our practical biased implementation, the estimator can occasionally become negative in very dark regions.
Throughout the paper, when computing RelMSE, we do not clamp these negative values; instead, we directly measure the error of the raw estimator against the reference.
To make this behavior explicit, we additionally visualize the spatial distribution of negative estimates by marking negative pixels in white, as shown in \figref{fig:re-shadow-negative-visualization}. The fraction of negative pixels is below $1\%$ in this example.

\subsection{Convergence Results}
\label{app:shadow-convergence}
We report shadow convergence with white-noise sampling by plotting RelMSE versus spp.

\begin{figure*}[t]
	\centering
	\includegraphics[width=0.9\linewidth]{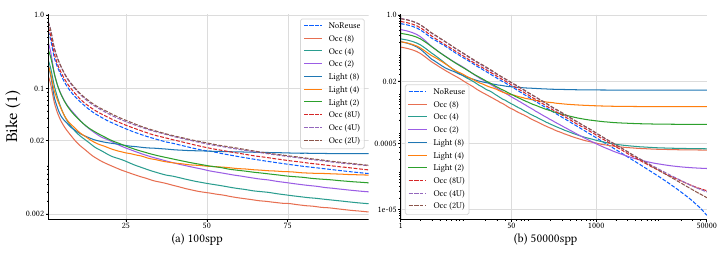}
	\caption{Shadow convergence on Bike~(1) (white-noise sampling, RelMSE vs spp), shown with log-scaled axes. (a) Low-spp view up to 100 spp: Under biased settings, Light can be competitive at very low spp ($\le5$), but as spp increases and variance decreases, residual bias dominates and Occ becomes more accurate. (b) Full-range view: the unbiased Occ estimator keeps decreasing with spp, confirming unbiasedness. In practice, it is not always better than no reuse because bright noise outliers in dark regions can dominate finite-spp error statistics (see \secref{sec:application-shadow}).}
	\label{fig:shadow-convergence-errors-50000spp}
	\Description{shadow error-map.}
\end{figure*}

\figref{fig:shadow-convergence-errors-50000spp}(a) focuses on the low-spp regime.
For the biased comparison, Light can achieve lower RelMSE at very low spp because stronger smoothing suppresses dominant variance.
As spp increases and variance is reduced, residual bias becomes the dominant error source, and Occ yields lower error than Light.

\figref{fig:shadow-convergence-errors-50000spp}(b) gives the full-range behavior.
Occ unbiased continues to converge, validating the unbiased formulation.
However, the unbiased reuse estimator can still underperform no reuse in some spp ranges, because bright outliers in dark regions may dominate finite-spp statistics and cause the image-wide RelMSE to be worse than no reuse.

\clearpage
\clearpage

\end{document}